\documentclass[]{partc}
\usepackage{graphicx}

\usepackage{rotating}
\usepackage{multirow}
\usepackage{amsmath}
\usepackage{comment}
\usepackage{tabularx}
\usepackage{subcaption}
\usepackage{xcolor}
\usepackage{amsfonts}
\usepackage{bm}
\usepackage{makecell}
\usepackage{array}

\title{An Empirical Markov Chain Car-Following (MC-CF) Model}

\author[1]{Sungyong Chung}
\author[1]{Yanlin Zhang}
\author[2]{Nachuan Li}
\author[2]{Dana Monzer}
\author[1]{Alireza Talebpour\thanks{Corresponding author. Email: ataleb@illinois.edu}}

\affil[1]{Department of Civil and Environmental Engineering, University of Illinois Urbana-Champaign}
\affil[2]{Northwestern University Transportation Center, Northwestern University}

\date{}

\begin{document}
\maketitle
\begin{abstract}
Car-following behavior is fundamental to traffic flow theory, yet traditional models often fail to capture the stochasticity of naturalistic driving. This paper proposes an empirical probabilistic sampling approach to car-following modeling that bypasses conventional parametric assumptions. Under this approach, we introduce the Markov Chain Car-Following (MC-CF) model, which represents state transitions as a Markov process and predicts behavior by randomly sampling accelerations from empirical distributions within discretized state bins. Evaluation on the Waymo Open Motion Dataset (WOMD) demonstrates that MC-CF variants significantly outperform all physics-based baselines (IDM, Gipps, FVDM, and SIDM) across both one-step and open-loop trajectory prediction metrics, and remain competitive with modern data-driven baselines including neural network and Gaussian mixture model approaches. Zero-shot generalization on the Naturalistic Phoenix (PHX) dataset further confirms cross-domain transferability. Finally, microscopic ring road simulations validate the framework's scalability: by incrementally integrating unconstrained free-flow trajectories and high-speed freeway data (TGSIM) alongside a conservative inference strategy, the model substantially reduces collisions across most tested scenarios and successfully reproduces naturalistic and stochastic shockwave propagation, though crashes persist under severe shockwave conditions. Overall, the proposed MC-CF model provides a robust and scalable foundation for simulating population-level stochastic traffic behavior that requires no behavioral parameter calibration, making it well-suited for the data-rich future of intelligent transportation.
\end{abstract}


\section{Introduction}\label{sec:intro}
Car-following behavior analysis forms a cornerstone of traffic flow theory, providing the microscopic foundation for understanding traffic dynamics, from individual vehicle interactions to large-scale congestion patterns. For decades, this understanding has relied on models developed from limited, often experimental datasets. However, the recent emergence of large-scale, high-fidelity trajectory datasets such as Waymo Open Motion Dataset (WOMD) \cite{Kan_2024_icra, ettinger2021waymo}, Lyft level-5 Open Dataset \cite{houston2021one}, and Argoverse Dataset \cite{chang2019argoverse}, has transformed this landscape. These datasets capture millions of real-world interactions in complex urban environments at high temporal resolution, encompassing both automated vehicles (AVs) and human-driven vehicles (HDVs). This data revolution presents a unique opportunity to build a new generation of car-following models that are more accurate, robust, and representative of real-world driving complexity.

Traditionally, car-following research has been dominated by parametric, physics-based formulations such as the Gipps model \cite{gipps1981behavioural}  and the Intelligent Driver Model (IDM) \cite{treiber2000congested}, along with stochastic variants like the Stochastic IDM (SIDM) \cite{treiber2017intelligent}. These models have been invaluable for their interpretability, as they frame driving behavior through well-defined rules and parameters related to safety, speed, and spacing. However, their reliance on simplified assumptions and limited number of parameters limits their capacity to exploit the richness of modern trajectory data. As a result, they often fall short in capturing behavioral heterogeneity, the stochastic nature of human driving, and the emergent dynamics in mixed traffic of AVs and HDVs.

Recent developments in data-driven modeling, powered by machine learning techniques such as neural networks and Bayesian mixture models, have brought remarkable advances \cite{papathanasopoulou2015towards, he2015simple, panwai2007neural, wang2017capturing, khodayari2012modified, ma2020sequence, sutskever2014sequence, chen2023bayesian}. These models excel at capturing complex, nonlinear relationships directly from data, frequently achieving higher predictive accuracy than traditional approaches. However, these models either function as black boxes with limited interpretability, or impose parametric distributional assumptions that may not reflect the true heterogeneity of driving behavior. Both limitations pose challenges for interpretability and safety validation in simulation environments.

To address this gap, this paper introduces an empirical probabilistic sampling approach that retains the transparency of classical state-based logic while abandoning restrictive parametric behavioral assumptions about driver behavior. Building on recent empirical evidence from \citep{li2025assessing}, which provides rigorous statistical validation for the Markov property in car-following state transitions where the state is defined by follower speed, leader speed, and spacing, this approach represents car-following dynamics through (i) empirical transition probabilities between discrete traffic states and (ii) state-conditional empirical acceleration distributions estimated directly from trajectory data. As a result, behavioral interpretation is statistically grounded. Each traffic state is associated with an observable acceleration profile, enabling direct inference of mean response and variability across regimes without invoking latent parameters.

As a concrete realization of this approach, we propose the Markov Chain Car-Following (MC-CF) model and its variants. Specifically, the MC-CF model learns state transition probabilities directly from trajectory data over a discretized state space and predicts behavior by sampling accelerations from the empirical acceleration distribution of the estimated next state. By synthesizing a statistically validated Markovian structure with empirical sampling, the MC-CF model provides a simple yet interpretable framework designed to leverage large-scale naturalistic driving data without requiring behavioral parameter calibration.

We note that the MC-CF model is particularly suited for the data-rich era ahead. As AV deployment accelerates, the volume of high-quality driving data (from human-AV and human-human interactions) collected from onboard sensors is expected to grow exponentially. Unlike parametric models that require complete recalibration when new data arrive, the inherent structure of the MC-CF model allows continuous refinement. Specifically, new trajectory data can be seamlessly incorporated by updating transition probabilities and adding acceleration samples to each state's empirical acceleration distribution. This scalability and adaptability position the MC-CF model as a practical tool for data-driven transportation analysis that can become more accurate as more data becomes available.

The remainder of this paper is organized as follows. Section~\ref{sec:literature} reviews physics-based and data-driven car-following models and positions the proposed approach within the broader modeling landscape. Section~\ref{sec:data_description} describes the trajectory dataset and preprocessing procedures used for empirical implementation. Section~\ref{sec:methodology} presents the MC-CF model, including state definition, transition probability estimation, and construction of state-conditional acceleration distributions. Section~\ref{sec:results} evaluates the proposed framework against representative car-following model baselines in terms of trajectory accuracy with empirical data. We then discuss model scalability in microscopic simulation and its implications for data-rich transportation systems. Finally, Section~\ref{sec:conclusion} concludes the paper and outlines directions for future research.

% ---------- Literature Review ----------
\section{Literature Review}\label{sec:literature}

The landscape of car-following modeling has continuously evolved in pursuit of balancing empirical realism with structural transparency. Traditionally, this domain has been dominated by parametric, physics-based equations. While these classical models offer clear theoretical insights, they often struggle to capture the full spectrum of naturalistic human driving. In response, data-driven models emerged, offering a more accurate and reliable representation of real-world behaviors by directly leveraging trajectory datasets without restrictive parametric constraints \cite{papathanasopoulou2015towards}. However, these machine learning approaches are frequently criticized for functioning as black boxes; they offer limited insight into traffic flow theory and present challenges for safety validation and transferability across different traffic conditions \cite{papathanasopoulou2015towards}. To navigate these competing challenges, the modeling landscape currently spans classical physics-based models, complex machine learning architectures, and probabilistic data-driven models. While probabilistic approaches offer richer uncertainty representation, they often impose parametric distributional assumptions or require computationally intensive inference, leaving an open need for an approach that is simultaneously empirically flexible, transparent, and assumption-free.

\subsection{Physics-Based Car-Following Models}\label{sec:physics_based_cf}

Physics-based car-following models have long served as foundational tools for simulating driver behavior and traffic dynamics by employing well-defined mathematical rules. For instance, the Gipps model \cite{gipps1981behavioural}, an extensively used framework, operates on the assumption that individuals have a limited range of desirable acceleration and deceleration rates, enforcing strict collision-avoidance constraints. However, it has been heavily criticized for its inability to replicate metastability, often resulting in unconditionally stable traffic states \cite{higgs2011analysis, krauss_1998}.

Seeking to capture smoother, more continuous car-following dynamics, the IDM \cite{treiber2000congested} became highly adopted. Despite its popularity, the IDM struggles to simulate reaction delays and lacks the psychological human-factors necessary to reproduce heterogeneous, real-world reactions, sometimes yielding unrealistic vehicle dynamics \cite{saifuzzaman2014incorporating, Albeaik_2022}. To address this deterministic rigidity, the SIDM \cite{treiber2017intelligent} introduced time-varying stochastic fluctuations to reproduce flow oscillations and driver indifference. Nevertheless, adding random noise to deterministic equations is generally criticized, as it can inadvertently cause negative vehicle speeds \cite{zheng2020analyzing}.

Parallel to the IDM family, the Full Velocity Difference Model (FVDM) \cite{jiang2001full} was introduced to improve the generalized force model (GFM) \cite{helbing1998generalized} by incorporating the impacts of both positive and negative relative speeds. Variants of the FVDM, such as those utilizing constant time headways (FVDM-CTH) or non-linear sigmoids (FVDM-sigmoid), have been used to model AVs. Yet, empirical evaluations have shown that simpler models like Gipps can sometimes outperform these FVDM variants in AV modeling \cite{punzo2021calibration}. Extending these physics-based principles to connected environments, Van Arem \cite{van2006impact} modeled cooperative adaptive cruise control systems utilizing vehicle-to-vehicle communication to allow for smaller following spacings. However, similar to its predecessors, this model enforces specific safety constraints and desired traffic variables without allowing for the flexible, heterogeneous objective prioritization seen in actual driving \cite{ames2016control}.

Recognizing the limitations of static parameters in these traditional equations, a recent evolution within physics-based modeling has employed Markovian frameworks to capture the dynamic, multi-modal nature of driving. These studies conceptualize driving as a sequence of transitions between discrete behavioral states or latent regimes. For instance, \citep{zaky2014car} applied a Markov switching regression model to classify regimes such as stable following and braking. \citep{zou2022multivariate} utilized a Coupled Hidden Markov Model (CHMM) to segment driving into primitive patterns, while \citep{yao2023identification} introduced the concept of action-chains using coupled Markov chains. More recently, \citep{zhang2025markov} proposed a Factorial Hidden Markov Model (FHMM) integrated directly with the IDM, calibrating a unique set of IDM parameters for each latent driving regime. While this study significantly enhances accuracy by modeling regime switching with a larger set of calibrated parameters, it remains fundamentally anchored in the physics-based approach, as the underlying vehicle dynamics within each state are still dictated by rigid parametric equations.

\subsection{Data-Driven Car-Following Models}

To bypass the restrictive assumptions of parametric equations, researchers have increasingly turned to data-driven techniques. \citep{papathanasopoulou2015towards} developed an improved weighted regression (loess) technique to model car-following behavior and compared it with the Gipps model to find that it outperformed the latter. The model was based on the dataset collected in Naples by \citep{punzo2005data}. However, the model needs more validation before being integrated in traffic simulation \cite{papathanasopoulou2015towards}. \citep{he2015simple} introduced a nonparametric car-following model using k-nearest neighbor (kNN) and field data, avoiding assumptions about driver behavior parameters or fundamental diagrams. Based on the idea that drivers respond to traffic stimuli in consistent, learned patterns, the model accurately reproduced empirical traffic features like stop-and-go oscillations. However, because the kNN approach predicts vehicle movement by taking the strict average of the most similar historical cases, it inherently smooths out the natural, heterogeneous variability of driver responses. Consequently, to capture stochastic driving errors and test traffic instability, the model relies on the artificial injection of white Gaussian noise, rather than deriving the true probabilistic distribution of driving behaviors directly from the empirical data.

Artificial neural networks (ANNs) have also brought remarkable advances. \citep{panwai2007neural} developed an ANN-based model that demonstrated superior performance compared to Gipps and psychophysical models. \citep{wang2017capturing} systematically evaluated deep learning architectures for car-following and demonstrated that a Gated Recurrent Unit (GRU) network conditioned on a 1-second trajectory history outperforms Feedforward Neural Network (FNN)-based models and the IDM in one-step prediction accuracy. The GRU's sequential hidden state allows it to implicitly encode short-term memory effects, an advantage not available to instantaneous-state models. Another direction focuses on sequence-to-sequence (seq2seq) approaches that account for memory effects and driver reaction delays \cite{ma2020sequence}. A common limitation of these neural network approaches is, however, that they produce deterministic point estimates of acceleration, providing no principled representation of behavioral uncertainty or driver heterogeneity. Their autoregressive rollout also tends to accumulate errors under long-horizon open-loop simulation.

To address this gap, recent work has turned to probabilistic learning frameworks. \citep{chen2023bayesian} proposed a Bayesian Gaussian Mixture Model (BayesGMM) for probabilistic car-following that represents the joint distribution of a 1-second kinematic history window and a short prediction horizon as a mixture of $K$ Gaussian components over an augmented state space. Model parameters are inferred via a Gibbs sampler under a Gaussian-inverse-Wishart prior, yielding a full posterior distribution over the acceleration rather than a point prediction. This approach provides richer uncertainty quantification than deterministic neural networks and achieves competitive one-step prediction accuracy on benchmark datasets. However, the MCMC-based inference introduces substantial computational overhead, and independently drawn step-level samples under autoregressive rollout can produce high jerk and trajectory instability, limiting applicability in long-horizon closed-loop simulation. While these data-driven models excel at capturing complex, nonlinear relationships and achieving high predictive accuracy, they frequently function as black boxes or impose parametric distributional assumptions. This opacity poses significant challenges for behavioral interpretability and safety validation in simulation environments.

\subsection{The Empirical Probabilistic Sampling Approach}

To bridge the gap between the transparency of classical physics-based logic and the flexibility of data-driven approaches, this paper introduces an empirical probabilistic sampling approach. Rather than imposing restrictive physics-based parametric assumptions, this approach derives behavioral dynamics directly from observed data distributions.

The foundation for this approach relies on the rigorous statistical validation of state transitions. A recent study by \cite{li2025assessing} confirmed that the Markov property holds for vehicle state transitions, where a state is explicitly defined by the leader vehicle's speed, the follower vehicle's speed, and the spacing between them. Building directly on this validated principle, we propose the MC-CF model.

Crucially, the MC-CF model diverges from the regime-switching approaches discussed in Section \ref{sec:physics_based_cf}. Instead of identifying latent behavioral regimes and calibrating separate parametric equations within them, the MC-CF model learns the state transition probabilities directly from trajectory data over a discretized state space. Furthermore, rather than assuming any deterministic acceleration function for a given state, our model embraces the inherent stochasticity of driving by directly sampling acceleration from an empirical distribution constructed from observed data for each discrete state. Unlike deterministic neural networks \cite{wang2017capturing} that offer no uncertainty representation and are prone to open-loop instability, or probabilistic mixture models \cite{chen2023bayesian} that impose parametric Gaussian forms and require MCMC inference, the MC-CF model constructs non-parametric empirical acceleration distributions without distributional assumptions and can be incrementally updated without retraining.

% ---------- Data Description ----------
\section{Data Description}\label{sec:data_description}

This study primarily draws on the WOMD in order to train the proposed models. WOMD captures car-following interactions near the Waymo vehicles (AVs), and provides a data volume that is unprecedented compared to other datasets commonly used for car-following analysis, thereby enabling the proposed empirical probabilistic sampling approach. The dataset includes all interaction types: AV-following-HDV, HDV-following-AV, and HDV-following-HDV. For each interaction type, 90\% of the extracted car-following pairs are used for training, and the remaining 10\% are reserved for testing to evaluate the performance of the proposed and baseline models. The split is performed at the car-following pair level via simple random sampling using a fixed seed across all models. Because WOMD pairs are extracted from over 100,000 independent 20-second scenes spanning six geographically diverse US cities, pair-level random sampling produces training and test sets with broadly comparable distributions of kinematic states, road types, and traffic conditions. In addition, we employ the Naturalistic Phoenix (PHX) dataset \cite{zhang2025waymo} to assess zero-shot generalization. Here, zero-shot generalization refers to a model's ability to generate motion predictions for time series originating from previously unseen datasets \cite{wu2024smart}. Specifically, for the zero-shot generalization analysis, the models are trained on WOMD training dataset and validated on the entire PHX dataset.

From each dataset, we extract key variables relevant to car-following modeling, including follower and leader positions (x and y coordinates), speeds, accelerations, spacings, and relative speeds. Trajectories in both datasets are sampled at 10 Hz. Preprocessing involves filtering for valid car-following pairs based on the following criteria: a minimum duration of 10 seconds, a maximum spacing of 45 m, and a minimum speed threshold of 3 m/s (that is, the maximum speed within each car-following pair must exceed 3 m/s to ensure that at least one vehicle is moving). Time steps with accelerations outside the range of -10 $m/s^2$ to 5 $m/s^2$ are removed, as they likely stem from sensor noise; these accounted for approximately 0.81\% of the WOMD and 1.26\% of the PHX. Additionally, the first and last 2 seconds of each pair are discarded to minimize non car-following behavior related to lane changes or merging. The resulting car-following pairs are then partitioned into three groups based on interaction type within each dataset.

\subsection{Waymo Open Motion Dataset}
The WOMD provides a large-scale benchmark for motion forecasting, comprising over 100,000 interactive 20 second scenes across 1,750 km of roadways in six US cities \cite{ettinger2021waymo}. The dataset provides information including scenario IDs, unique tracking IDs for objects, object types (vehicles, pedestrians, and cyclists), Waymo vehicle identifiers, and detailed object attributes including position (x, y, z), dimensions (length, width, height), heading, and velocity. In addition, the dataset offers map-related features for each scenario represented as 3D polylines, including lane centers, lane boundaries, and road boundaries. These map features also specify the travel direction associated with each lane center ID and describe its connections to other lane center IDs, such as exit lanes and adjacent left and right lanes. However, in contrast to the traditional NGSIM trajectory dataset \cite{USDOT2016}, which is one of the most widely used in car-following studies, WOMD does not include assigned lane center IDs for each vehicle, a key element for identifying leader–follower pairs.

To systematically extract all possible car-following pairs, vehicles were assigned to lanes by mapping their position to the nearest point on lane centers. As vehicles moved, lane assignments were updated based on the closest point from the (i) current lane center ID, (ii) any exit lane center IDs, or (iii) neighboring left and right lane center IDs. After assigning lane centers to each vehicle at each time step, we identified leaders as the closest vehicles to the follower in the direction of travel, either within the same lane center as the follower or in lane centers connected to it. This procedure, consistent with the methodology described in \cite{Chung2025lanechanging}, enabled us to extract all candidate car-following pairs from the dataset. Following extraction, we applied the preprocessing steps described earlier to obtain the final car-following pairs for analysis.

Table ~\ref{tab:trajectory_seconds_WOMD} summarizes the trajectory durations, in seconds, and the number of car-following pairs for each interaction type extracted from the WOMD. As noted, because each WOMD scenario has a maximum length of 20 seconds, the mean duration of car-following pairs is around 9.5 seconds across all interaction types. Specifically, 4,198 car-following pairs were extracted for AV-following-HDV interactions, while 5,440 and 32,753 pairs were extracted for HDV-following-AV and HDV-following-HDV interactions, respectively.

\begin{table}[tb!]
\centering
\caption{Descriptive statistics of trajectory durations in seconds and number of car-following (CF) pairs for different interaction types in WOMD.}
\vspace{0.1cm}
\resizebox{0.7\textwidth}{!}{
\begin{tabular}{l l r r r r r}
\hline
Dataset & Interaction Type & CF Pairs & Mean (s) & Std (s) & Max (s) & Min (s) \\
\hline
WOMD (Train) & AV-following-HDV & 3,778 & 9.28 & 2.67 & 15.7 & 5.1 \\
WOMD (Test) & AV-following-HDV & 420 & 9.34 & 2.80 & 15.8 & 5.4 \\
WOMD (Train) & HDV-following-AV & 4,896 & 9.65 & 2.74 & 15.7 & 5.8 \\
WOMD (Test) & HDV-following-AV & 544 & 9.67 & 2.78 & 15.7 & 6.0 \\
WOMD (Train) & HDV-following-HDV & 29,477 & 9.43 & 2.63 & 15.8 & 5.2 \\
WOMD (Test) & HDV-following-HDV & 3,276 & 9.48 & 2.65 & 15.7 & 5.9 \\
\hline
\end{tabular}}
\label{tab:trajectory_seconds_WOMD}
\end{table}

\subsection{Naturalistic Phoenix Dataset}

The Naturalistic PHX Dataset was developed to capture detailed Waymo vehicle trajectories and behavioral interactions in real-world traffic conditions across the Phoenix metropolitan area. High-resolution aerial videos were collected using a stabilized camera mounted on a helicopter, covering major arterial corridors with varying geometric and control features. These raw videos were processed through a multi-stage extraction pipeline \cite{ammourah2025introduction}. The resulting dataset provides precise, continuous trajectories at 0.1 second resolution. Compared with the WOMD, the PHX dataset offers longer trajectory durations, enabling the analysis of dynamic scenarios such as queue formation, dissipation, and heterogeneous car-following behaviors.

To ensure high-fidelity trajectory reconstruction, all vehicle positions in the PHX dataset were transformed into a unified, ground-fixed coordinate system. Following the data collection procedure in \citep{ammourah2025introduction}, the image coordinates of the moving aerial platform were mapped to a consistent spatial reference frame, ensuring high-precision tracking across the study area.

Object detections extracted from the aerial video were projected into this same fixed coordinate system, ensuring spatial consistency across all frames. Each detected vehicle was assigned to its corresponding lane region based on its centroid location. After establishing lane assignment, the longitudinal ordering of vehicles within each lane was used to identify leader–follower pairs dynamically over time. This procedure allowed every vehicle's immediate leader and follower at any given moment to be determined with high spatial accuracy, forming the foundation for robust model calibration and behavioral analysis. The resulting pairs were then further processed using the identical preprocessing steps described earlier to obtain the final set of car-following pairs used in the analysis.

Table~\ref{tab:trajectory_seconds_PHX} shows summary statistics of trajectory durations and the number of car-following pairs for different interaction types in the PHX dataset. Compared to WOMD, PHX provides substantially longer trajectories, with mean durations exceeding 25 seconds and greater variability in length, making it particularly suitable for evaluating zero-shot generation. Given that AV behaviors may differ between WOMD and PHX \cite{zhang2025waymo}, we restrict the zero-shot generation analysis to the 106 HDV-following-HDV pairs in PHX in order to assess whether the proposed and baseline models can reproduce heterogeneous HDV behaviors in a completely new dataset.

\begin{table}[tb!]
\centering
\caption{Descriptive statistics of trajectory durations in seconds and number of car-following pairs in PHX.}
\vspace{0.1cm}
\resizebox{0.7\textwidth}{!}{
\begin{tabular}{l l r r r r r}
\hline
Dataset & Interaction Type & CF Pairs & Mean (s) & Std (s) & Max (s) & Min (s) \\
\hline
PHX & AV-following-HDV & 16 & 26.41 & 19.28 & 74.6 & 6.1 \\
PHX & HDV-following-AV & 20 & 28.22 & 23.05 & 100.9 & 7.2 \\
PHX & HDV-following-HDV & 106 & 25.60 & 18.16 & 89.2 & 4.9 \\
\hline
\end{tabular}}
\label{tab:trajectory_seconds_PHX}
\end{table}

\subsection{Third Generation Simulation Dataset}
The Third Generation Simulation (TGSIM) dataset~\cite{ammourah2025introduction,talebpour2024third} represents a recent effort to generate high–fidelity vehicle trajectories that capture interactions among HDVs and partially AVs under naturalistic traffic conditions. In contrast to the primarily urban environments of the WOMD and PHX datasets, TGSIM includes extensive data from freeway segments, such as I--294, I--90, and I--94. In this study, we specifically utilize the I--294 dataset to evaluate the scalability of the proposed MC-CF model when augmented with high speed freeway dynamics.

The trajectories were extracted using a unified pipeline incorporating image stabilization, deep-learning-based vehicle detection, multi-frame tracking, and Kalman-filter-based smoothing. The moving–helicopter approach used to collect the dataset provides long continuous trajectories of the vehicles, and the resulting trajectories cover a wide range of operational scenarios, including free-flow, slow-and-go, congested shockwaves, forced merges, discretionary lane changes, and AV responses to surrounding traffic.

% ---------- Methodology ----------
\section{Methodology}\label{sec:methodology}
In this section, we present the MC-CF model. Instead of prescribing a deterministic acceleration function as in classical physics-based car-following models, this study adopts a probabilistic formulation that treats state transitions as a Markov chain. We also summarize the baseline models and the calibration method used for comparison.

\subsection{Markov Chain Car-Following Model}
The evolution of the follower vehicle is represented in discrete time with step size \(\Delta t\).
At each time step $t$, the state of the system is described by the triplet $s_t = \big( v_t, \Delta v_t, d_t \big)$,
where \(v_t\) is the speed of the follower, \(\Delta v_t = v_t - v_t^{\text{lead}}\) is the relative speed with respect to the leader, and \(d_t = x_t^{\text{lead}} - x_t-l\) is the spacing. Here, $l$ represents the average length of the leader and follower vehicles. The acceleration of the follower at time step $t$ is denoted by \(a_t\).

The stochastic dynamics of the system are modeled as a first-order Markov chain on the discretized state space \(\mathcal{S}\). The continuous state space is partitioned into a three-dimensional grid of bins using the Freedman--Diaconis rule for automatic bandwidth selection \cite{freedman1981histogram, masserano2024wavelet}, which adapts the number of bins to the data distribution while minimizing bias. The bin width $h$ is computed as
\begin{equation}
\label{eq:bin_width}
    h = 2 \cdot \frac{\mathrm{IQR}(X)}{n^{1/3}},
\end{equation}

\noindent where $\mathrm{IQR}(X)$ is the interquartile range of the data $X$ and $n$ is the sample size. The number of bins is then calculated as
\begin{equation}
\label{eq:num_bin}
    k = \left\lceil \frac{x_{\max} - x_{\min}}{h} \right\rceil.
\end{equation}

The ranges for each dimension, relative speed $\Delta v_t \in [-10, 10]$ m/s, spacing $d_t \in [0, 45]$ m, and follower speed $v_t \in [0, 20]$ m/s, were set based on the dataset to ensure all data samples fall within the grid and are classified into a bin. Table~\ref{tab:fd_bins} summarizes the binning results across the three interaction types. Due to the larger number of samples in HDV-following-HDV, the Freedman--Diaconis rule yields a much smaller step size, resulting in a significantly larger number of bins.

\begin{table}[tb]
\centering
\caption{Binning results for different car-following cases using the Freedman--Diaconis rule.}
\vspace{0.1cm}
\label{tab:fd_bins}
\resizebox{0.8\textwidth}{!}{
\begin{tabular}{lcrrrrrr}
\hline
Dimension & Range & \multicolumn{2}{c}{AV-following-HDV} & \multicolumn{2}{c}{HDV-following-AV} & \multicolumn{2}{c}{HDV-following-HDV} \\
 & & \# bins & Step size & \# bins & Step size & \# bins & Step size \\
\hline
Relative Speed  \\ ($m/s$) & $(-10, 10)$ & 394 & $\sim 0.05$ & 444 & $\sim 0.05$ & 1,041 & $\sim 0.02$ \\
Spacing \\ ($m$) & $(0, 45)$ & 109 & $\sim 0.41$ & 139 & $\sim 0.32$ & 357 & $\sim 0.13$ \\
Follower Speed \\ ($m/s$) & $(0, 20)$ & 89 & $\sim 0.22$ & 90 & $\sim 0.22$ & 230 & $\sim 0.09$ \\
\hline
\end{tabular}}
\end{table}

To characterize driving behavior, one could theoretically estimate transition probabilities directly between the discretized bins defined based on the Freedman-Diaconis rule (Table \ref{tab:fd_bins}). However, given the high dimensionality of the state space, many bins may contain insufficient samples for reliable statistical estimation. To address this sparsity while preserving local behavioral fidelity, we employ a spatially constrained state clustering algorithm, detailed in Algorithm \ref{alg:clustering}.

This approach structurally refines the state space during the training phase. We first define a minimum sample threshold $N_{\min}=10$. This value was selected to maintain a balance between behavioral resolution and statistical reliability: an excessively high threshold risks merging distinct states that are far apart, while an overly low threshold may have bins dominated by sensor noise. The algorithm identifies sparse bins that fall below this threshold and iteratively merges them into their nearest neighbor. Importantly, to ensure that the distance metric accounts for the different scales of the three state variables, the nearest neighbor search is performed using Euclidean distance in a normalized feature space, where each dimension is scaled by its range.

This process yields a refined set of robust state clusters $\mathcal{C}$ and a mapping function $\mathcal{M}: \mathcal{S} \rightarrow \mathcal{C}$ that assigns a raw discretized state $s \in \mathcal{S}$ to a specific cluster $C \in \mathcal{C}$. Consequently, the effective state space of the MC-CF model becomes the set of these validated clusters. By design, every cluster contains at least $N_{\min}$ acceleration samples without the need for synthetic imputation.

Empirically, this procedure significantly mitigates state space sparsity. For instance, in the HDV-following-HDV case, the Freedman-Diaconis rule-based discretization yields a theoretical grid of over 85.4 million bins (=$1041 \times 357 \times 230)$. However, observed driving behaviors occupy only 1.46 million of these bins (approximately 1.7\%). The clustering algorithm further merges these active bins into 123,235 clusters, achieving a compression ratio of 11.89. This process effectively isolates the relevant behavioral subspace, ensuring that every cluster meets the minimum acceleration sample requirement for reliable statistical estimation.

\begin{algorithm}[tb!]
\caption{Spatially Constrained State Clustering Training Procedure}
\label{alg:clustering}
\SetAlgoLined
\small
\KwIn{Trajectory Data $\mathcal{D}$, Bin Edges $B_r, B_s, B_f$, Min Samples $N_{\min}=10$}
\KwOut{Cluster Map $\mathcal{M}$, Transition Matrix $P$, Acceleration Distributions $\mathcal{A}$}

\tcp{1. Initialization}
Discretize $\mathcal{D}$ into 3D grid bins using $B_r, B_s, B_f$\;
Initialize set of clusters $\mathcal{C}$ where each unique grid bin is a cluster $C_i$\;
\For{each cluster $C_i \in \mathcal{C}$}{
    Calculate sample count $n_i$, centroid $\mu_i$, and collected accelerations $A_i$\;
    Calculate normalized centroid $\hat{\mu}_i$ based on feature ranges\;
}

\tcp{2. Iterative Merging}
Identify sparse clusters $\mathcal{S} = \{C_i \in \mathcal{C} \mid n_i < N_{\min}\}$\;
\While{$\mathcal{S} \neq \emptyset$}{
    \For{each sparse cluster $C_{src} \in \mathcal{S}$}{
        Find nearest neighbor cluster $C_{dst} \in \mathcal{C} \setminus \{C_{src}\}$ minimizing $||\hat{\mu}_{src} - \hat{\mu}_{dst}||_2$\;
        Record potential merge tuple $(C_{src}, C_{dst}, \text{dist})$\;
    }
    Sort merge tuples by source count (ascending) and distance (ascending)\;

    \For{each valid merge tuple $(C_{src}, C_{dst})$}{
        \If{$C_{src}$ or $C_{dst}$ has been removed in current batch}{
            Continue\;
        }
        \tcp{Merge Source into Destination}
        $n_{dst} \leftarrow n_{dst} + n_{src}$\;
        $\mu_{dst} \leftarrow \text{WeightedAvg}(\mu_{dst}, \mu_{src})$\;
        $\hat{\mu}_{dst} \leftarrow \text{WeightedAvg}(\hat{\mu}_{dst}, \hat{\mu}_{src})$\;
        $A_{dst} \leftarrow A_{dst} \cup A_{src}$\;
        Remove $C_{src}$ from $\mathcal{C}$\;
        Update map $\mathcal{M}: \text{original\_bins}(C_{src}) \to C_{dst}$\;
    }
    Update sparse set $\mathcal{S}$ based on new counts\;
}

\tcp{3. Transition Probability Calculation}
\For{each transition $(C_t, C_{t+1})$ in $\mathcal{D}$ using $\mathcal{M}$}{
    Increment transition count $T_{C_t \to C_{t+1}}$\;
}
Compute row-normalized probabilities $P(C' \mid C)$\;
\Return{$\mathcal{M}, P, \mathcal{A}$}
\end{algorithm}

The driving dynamics are modeled as a Markov chain over the clusters. The transition probability is defined as the probability of moving from cluster $C$ to cluster $C'$ in one time step:
\begin{equation}
\scalebox{0.87}{$\displaystyle P(C' \mid C) = \frac{\sum_{(s_t,s_{t+1})\in\mathcal{D}} \mathbb{I}\bigl(\mathcal{M}(s_t) = C \,\land\, \mathcal{M}(s_{t+1}) = C'\bigr)}{\sum_{(s_t,s_{t+1})\in\mathcal{D}} \mathbb{I}\bigl(\mathcal{M}(s_t) = C\bigr)}.$}
\end{equation}

Simultaneously, we construct an empirical acceleration distribution $\mathcal{A}(C)$ for each cluster using the pooled acceleration samples from the merged bins. Once the sample set is gathered, it is refined by removing outliers using the interquartile range (IQR) rule, retaining only values within the standard IQR bounds (Q1-1.5$\cdot \mathrm{IQR}$ to Q3+1.5$\cdot \mathrm{IQR}$). This process yields a robust, localized acceleration distribution \(\mathcal{A}(C)\) for each cluster, ensuring statistical validity while removing sensor noise. One limitation of this construction is that $\mathcal{A}(C)$ captures population-level stochasticity, as observations from all drivers encountered in cluster $C$ are pooled. While this is well suited for microscopic simulation aimed at reproducing macroscopic traffic dynamics, it does not retain individual driver fingerprints. Extending the model to represent individual-level heterogeneity is a promising direction for future work, for example by adding a driver identity dimension to the state space, constructing separate empirical distributions for behaviorally defined driver subgroups, or conditioning transition probabilities on driver-level features derived from trajectory history.

In the inference phase, at time $t$, we determine the current cluster $C_t$ for state $s_t$ using the mapping function $\mathcal{M}(s_t)$. However, to handle unseen states that were not populated during training, we employ a nearest-neighbor fallback mechanism: $C_t$ is assigned to the cluster with the closest centroid in the normalized feature space. This ensures that a valid current cluster for the current state is always identified, preventing model failure in edge cases.

We utilize this framework in two modes. In the deterministic version, namely, MC-CF (det), the most probable next cluster is selected, and the acceleration is the conditional mean:
\begin{equation}
C_{t+1} = \operatorname*{arg\,max}_{C'} P(C' \mid C_t), \quad a_t = \mathbb{E}[\mathcal{A}(C_{t+1})].
\end{equation}

In the stochastic version, denoted as MC-CF (stoch), the next cluster is sampled from the transition distribution, and the acceleration is drawn from the empirical distribution:
\begin{equation}
C_{t+1} \sim P(\cdot \mid C_t), \quad a_t \sim \mathcal{A}(C_{t+1}),
\end{equation}

\noindent allowing the model to capture both the stochasticity of state evolution and the variability of driver responses within a given traffic state. Note that the sampled acceleration may not deterministically place the vehicle into the sampled next cluster, due to state space discretization and clustering as well as the stochastic behavior of the leader vehicle. A consistency analysis showing that realized states remain spatially close to the intended cluster in nearly all cases is provided in Appendix~\ref{app:state_consistency}.

We also introduce a conservative inference variant, denoted MC-CF (cons, det), which applies a safety-oriented restriction to the acceleration sampling. The strategy is governed by two TTC thresholds, $\text{TTC}_\text{danger}$ and $\text{TTC}_\text{caution}$, and two corresponding percentile cutoffs, $p_\text{danger}$ and $p_\text{caution}$. Specifically, the model computes the time-to-collision $\text{TTC} = d_t / |\Delta v_t|$ when the follower is approaching the leader. If $\text{TTC} < \text{TTC}_\text{danger}$, the acceleration is sampled only from the bottom $p_\text{danger}$-th percentile of $\mathcal{A}(C_{t+1})$. If $\text{TTC}_\text{danger} \le \text{TTC} < \text{TTC}_\text{caution}$, it is sampled from the bottom $p_\text{caution}$-th percentile. Otherwise, the full distribution is used. The default configuration used throughout this paper sets $\text{TTC}_\text{danger} = 3.0$ s, $\text{TTC}_\text{caution} = 10.0$ s, $p_\text{danger} = 5$, and $p_\text{caution} = 30$. This restriction is motivated by the fact that naturalistic trajectory data inevitably contains accelerations from lane-changing maneuvers that superficially resemble aggressive car-following behavior, and state space discretization can pool borderline and safe conditions into the same cluster, making some sampled accelerations overly aggressive for the follower's actual position. Without this safeguard, such samples can induce artificial crashes during closed-loop simulation. Importantly, the sampled acceleration under conservative inference remains drawn from empirical data rather than being hard-coded. The conservative variant can be applied to either the deterministic or stochastic mode, yielding MC-CF (cons, det) and MC-CF (cons, stoch), respectively.

We evaluate the proposed models in two usage modes: one-step prediction and open-loop prediction. In one-step prediction, the update is performed for a single step and compared directly with the observed states.
\begin{equation}
\begin{aligned}
v_{t+1} &= \max\!\big( v_t + a_t \Delta t , 0 \big) \\
x_{t+1} &= x_t + \tfrac{1}{2}(v_t + v_{t+1}) \Delta t \\
d_{t+1} &= x^{\text{lead}}_{t+1} - x_{t+1}-l
\end{aligned}
\end{equation}

In open-loop prediction, the procedure is applied recursively to generate an entire follower trajectory given the sequence of ground truth leader states and the initial follower state, thereby assessing long-term predictive consistency.

\subsection{Baseline Model Calibration and Implementation}\label{sec:baseline_calib}
To benchmark the proposed MC-CF model against classical physics-based approaches, six representative car-following models are considered: IDM, SIDM, Van Arem~\cite{van2006impact}, FVDM-CTH~\cite{jiang2001full,punzo2021calibration}, FVDM-Sigmoid~\cite{jiang2001full,punzo2021calibration}, and Gipps. Their formulations, calibration procedure, and parameter bounds are detailed in Appendix~\ref{app:baseline}. To provide a broader comparison against modern data-driven methods, three additional baselines are included: a feedforward neural network regressor (FNN-reg) and a gated recurrent unit network (GRU-reg) that conditions on 1-second trajectory history, both of which are deterministic models; and a Bayesian Gaussian Mixture Model (BayesGMM) that also uses 1-second history to produce probabilistic acceleration predictions. Their architectures, training procedures, and implementation details are also described in Appendix~\ref{app:dd_baselines}.

% ---------- Results and Discussions ----------
\section{Results and Discussion}\label{sec:results}
In this section, we compare the trajectory prediction accuracy of the proposed MC-CF (det) and MC-CF (stoch) models against six physics-based baselines (IDM, Van Arem, FVDM-CTH, FVDM-Sigmoid, Gipps, and SIDM) and three data-driven baselines (FNN-reg, GRU-reg, and BayesGMM) across three interaction types: AV-following-HDV, HDV-following-AV, and HDV-following-HDV. All models are trained using the WOMD training dataset and evaluated on the WOMD test dataset. We then compare the distribution of state-transition probabilities of generated trajectories against ground truth trajectories using probabilities estimated from the WOMD training dataset. Next, we examine zero-shot generalization using the PHX dataset. Finally, we demonstrate the framework's scalability through microscopic ring road simulations.

\subsection{Trajectory Prediction Analysis}\label{sec:trajectory_prediction}
We present the trajectory prediction performance of the proposed MC-CF (det, stoch) models against nine baseline models: six physics-based models (IDM, Van Arem, FVDM-CTH, FVDM-Sigmoid, Gipps, and SIDM) and three data-driven models (FNN-reg, GRU-reg, and BayesGMM). All models are trained on the WOMD training dataset and tested on the WOMD test dataset separately for three interaction types: AV-following-HDV, HDV-following-AV, and HDV-following-HDV.

For deterministic models, including MC-CF (det), we generate a single trajectory ($K=1$) since they produce identical follower trajectories for a given leader trajectory. In contrast, stochastic models such as SIDM and MC-CF (stoch) can yield multiple plausible realizations conditioned on the same leader behavior. To capture this variability, we generate $K \in \{1, 3, 6, 10, 15\}$ follower trajectories for each leader trajectory in the test dataset.

\subsubsection{Evaluation Metrics}
We employ ten complementary metrics spanning two evaluation regimes. For one-step prediction, we use $RMSE(s)$, $RMSE(v)$, and $RMSE(a)$, which quantify instantaneous accuracy in spacing, speed, and acceleration, respectively; stochastic models are evaluated with $K=1$ here to allow direct comparison with deterministic models. For open-loop prediction, we use $minDTW(s)$, $minDTW(v)$, $minADE$, $minFDE$, $avgADE$, $avgFDE$, and the overlapping rate ($OR$), which together capture long-horizon trajectory fidelity, average predictive accuracy, and safety. All metrics except $OR$ are computed only over non-crashing trajectories to ensure a fair comparison across models. Formal definitions of all ten metrics are provided in Appendix~\ref{app:metrics}.

\subsubsection{Prediction Results}

Table~\ref{tab:HDV_following_HDV_results} presents the trajectory prediction performance for the HDV-following-HDV case, which is the most behaviorally rich interaction type and our primary focus, covering both one-step and open-loop results for all deterministic and stochastic models. All performance measures are averaged over the evaluated car-following pairs in the WOMD test dataset. Results for AV-following-HDV and HDV-following-AV, which show consistent trends, are provided in Appendix~\ref{app:av_results}.

\begin{table}[tb!]
\centering
\caption{Trajectory prediction performance for HDV-following-HDV.}
\vspace{0.2cm}
\label{tab:HDV_following_HDV_results}
\resizebox{\textwidth}{!}{
\begin{tabular}{lcccccccccc}
\hline
Model & \multicolumn{3}{c}{One-Step Prediction} & \multicolumn{7}{c}{Open-Loop Prediction} \\
 & $RMSE(s)$ & $RMSE(v)$ & $RMSE(a)$ & $minDTW(s)$ & $minDTW(v)$ & $minADE$ & $minFDE$ & $avgADE$ & $avgFDE$ & $OR$\\
\hline
\multicolumn{11}{l}{Deterministic} \\
IDM          & 0.0327 & 0.2126 & 2.5041 & 16.4298 & 3.3743 & 1.9564 & 3.9350 & 1.9564 & 3.9350          & 0.0015 \\
Van-Arem     & 0.0314 & 0.0797 & 0.7931 & 14.0278 & 3.1531 & 1.6164 & 3.6775 & 1.6164 & 3.6775          & 0.0082 \\
FVDM-CTH     & 0.0314 & 0.0883 & 0.8840 & 13.5194 & 3.0829 & 1.5557 & 3.6063 & 1.5557 & 3.6063          & 0.0134 \\
FVDM-Sigmoid & 0.0314 & 0.0887 & 0.8880 & 13.4735 & 3.0927 & 1.5512 & 3.5954 & 1.5512 & \textbf{3.5954*} & 0.0131 \\
Gipps        & 0.0316 & 0.1108 & 1.1092 & 17.3475 & 3.4963 & 2.0350 & 4.0746 & 2.0350 & 4.0746          & 0.0015 \\
FNN-reg      & 0.0314 & 0.0626 & 0.6136 & 16.6319 & 4.0000 & 1.6915 & 4.6599 & 1.6915 & 4.6599          & 0.0177 \\
GRU-reg      & 0.0318 & \textbf{0.0272*} & \textbf{0.1379*} & 19.6509 & 4.7659 & 1.9655 & 5.0526 & 1.9655 & 5.0526 & 0.1648 \\
MC-CF (det)  & 0.0314 & 0.0647 & 0.6357 & 15.1386 & 3.4969 & 1.5937 & 4.1493 & 1.5937 & 4.1493          & 0.0235 \\
MC-CF (cons, det) & 0.0313 & 0.0729 & 0.7216 & 14.6730 & 3.4680 & 1.5473 & 4.0156 & \textbf{1.5473*} & 4.0156 & 0.0043 \\
\hline
\multicolumn{11}{l}{Stochastic} \\
SIDM (1)                & 0.0204 & 0.1749 & 1.9229 & 16.5428 & 3.4063 & 1.9585 & 3.9548 & --     & --      & 0.0015 \\
SIDM (3)                & --     & --     & --     & 16.3052 & 3.3443 & 1.9331 & 3.8899 & --     & --      & --     \\
SIDM (6)                & --     & --     & --     & 16.1839 & 3.3117 & 1.9205 & 3.8555 & --     & --      & --     \\
SIDM (10)               & --     & --     & --     & 16.1065 & 3.2901 & 1.9123 & 3.8333 & --     & --      & --     \\
SIDM (15)               & --     & --     & --     & 16.0548 & 3.2759 & 1.9067 & 3.8182 & 1.9573 & 3.9540  & --     \\
BayesGMM (stoch, 1)     & \textbf{0.0174*} & 0.0466 & 0.4496 & 38.0453 & 8.3331 & 3.8675 & 8.2975 & --     & --      & \textbf{0.0000*} \\
BayesGMM (stoch, 3)     & --     & --     & --     & 18.9826 & 4.5306 & 2.1670 & 4.1429 & --     & --      & --     \\
BayesGMM (stoch, 6)     & --     & --     & --     & 12.3951 & 3.3007 & 1.5510 & 2.5854 & --     & --      & --     \\
BayesGMM (stoch, 10)    & --     & --     & --     &  9.0565 & 2.6667 & 1.2242 & 1.8178 & --     & --      & --     \\
BayesGMM (stoch, 15)    & --     & --     & --     & \textbf{7.2526*} & 2.3072 & 1.0369 & \textbf{1.3552*} & 3.8907 & 8.3514 & -- \\
MC-CF (stoch, 1)        & 0.0183 & 0.0825 & 0.8243 & 15.4709 & 3.5719 & 1.6225 & 4.2172 & --     & --      & 0.0256 \\
MC-CF (stoch, 3)        & --     & --     & --     & 11.7852 & 2.7482 & 1.2902 & 3.2159 & --     & --      & --     \\
MC-CF (stoch, 6)        & --     & --     & --     & 10.3477 & 2.4710 & 1.1589 & 2.8118 & --     & --      & --     \\
MC-CF (stoch, 10)       & --     & --     & --     &  9.4870 & 2.2970 & 1.0799 & 2.5616 & --     & --      & --     \\
MC-CF (stoch, 15)       & --     & --     & --     &  8.8799 & 2.1777 & 1.0222 & 2.3829 & 1.6048 & 4.1620  & --     \\
MC-CF (cons, stoch, 1)  & 0.0182 & 0.0850 & 0.8515 & 14.8540 & 3.4997 & 1.5677 & 4.0614 & --     & --      & 0.0043 \\
MC-CF (cons, stoch, 3)  & --     & --     & --     & 11.5162 & 2.7230 & 1.2697 & 3.1219 & --     & --      & --     \\
MC-CF (cons, stoch, 6)  & --     & --     & --     & 10.0223 & 2.4296 & 1.1387 & 2.6969 & --     & --      & --     \\
MC-CF (cons, stoch, 10) & --     & --     & --     &  9.2684 & 2.2714 & 1.0676 & 2.4650 & --     & --      & --     \\
MC-CF (cons, stoch, 15) & --     & --     & --     &  8.7199 & \textbf{2.1612*} & \textbf{1.0166*} & 2.3012 & 1.5588 & 4.0317  & --     \\
\hline
\multicolumn{11}{l}{\footnotesize \textit{Note}: * indicates the minimum (best) value within each column. In case of ties, multiple entries are marked.}\\
\multicolumn{11}{l}{\footnotesize -- denotes not applicable. $OR$: $K=1$ only; $avgADE$/$avgFDE$: $K=15$ for stochastic models; equals $minADE$/$minFDE$ for deterministic. $n_\text{viable}/n_\text{total} = 2704/3276$.}
\end{tabular}
}
\end{table}

The results in Table~\ref{tab:HDV_following_HDV_results} reveal several important patterns across the model families. For one-step prediction, GRU-reg achieves the lowest $RMSE(v)$ (0.0272) and $RMSE(a)$ (0.1379) by a wide margin, more than halving the error of the next-best model. BayesGMM also achieves the best $RMSE(s)$ (0.0174) among all models. This advantage is partly attributable to the use of 1-second trajectory history by both GRU-reg and BayesGMM, which provides access to recent acceleration patterns unavailable to instantaneous-state models such as MC-CF, FNN-reg, and the physics-based baselines. However, this short-term accuracy advantage does not carry over to open-loop rollout. GRU-reg produces the highest overlapping rate among all models ($OR = 0.1648$), and BayesGMM's open-loop errors at $K=1$ far exceed those of all other models ($minDTW(s) = 38.05$). Historical context that improves per-step accuracy does not guarantee stable long-horizon behavior.

In the open-loop regime at $K=15$, BayesGMM and MC-CF (cons, stoch) exhibit complementary strengths. BayesGMM achieves the best $minDTW(s)$ (7.2526) and $minFDE$ (1.3552), reflecting its ability to generate spatially diverse samples that cover the tails of the trajectory distribution. MC-CF (cons, stoch, 15) achieves the best $minDTW(v)$ (2.1612) and $minADE$ (1.0166), indicating superior average-position accuracy over the full trajectory horizon. BayesGMM's $avgADE$ at $K=15$ is 3.8907, more than twice that of MC-CF (cons, stoch, 15) at 1.5588, indicating that while BayesGMM can produce an occasional near-ground-truth sample, its typical trajectory quality is substantially lower. Sensitivity analyses of the conservative inference hyperparameters and the minimum cluster sample threshold ($N_{\min}$), confirming that these conclusions are robust to reasonable parameter choices, are provided in Appendix~\ref{app:sensitivity}. 

Among deterministic models, MC-CF (cons, det) achieves the best $avgADE$ (1.5473), outperforming all physics-based and neural-network baselines in this metric. The conservative inference strategy also reduces the overlapping rate from 2.35\% of MC-CF (det) to 0.43\%, which is essential for closed-loop simulation fidelity. As discussed further in Section~\ref{sec:simulation}, residual collision rates can be substantially reduced through the inclusion of additional trajectory data from diverse driving conditions.

It should be also noted that both MC-CF (stoch) and MC-CF (cons, stoch) show consistent improvement in all open-loop metrics as $K$ increases from 1 to 15, with the largest gains occurring between $K=1$ and $K=6$ and diminishing returns thereafter. SIDM shows negligible improvement with increasing $K$, reflecting its limited trajectory diversity.

\subsection{Transition Probability Analysis}

In this subsection, we present a statistical analysis to examine whether the distribution of the probabilities of observing follower trajectories generated by the calibrated models aligns with that of the ground truth follower trajectories. As previously discussed, the probability of observing the ground truth trajectory given a leader trajectory is not one, owing to the inherently stochastic nature of car-following behavior. Therefore, we evaluate and compare the distributions of trajectory probabilities from the generated and ground truth data. Following \citep{zhang2025waymo}, we use the geometric mean probability to represent the likelihood of a follower trajectory. Specifically, for each generated or ground truth follower trajectory with $T$ time steps (i.e., $T - 1$ transitions) in the WOMD test dataset, we compute:
\begin{equation}
\label{eq:geom_mean_combined}
\scalebox{0.92}{$\displaystyle\text{GeomMeanProb} = \exp\!\Bigl( \tfrac{1}{T-1}\sum_{t=1}^{T-1} \log P(C_{t+1} \mid C_t) \Bigr),$}
\end{equation}
\noindent where the transition probability $P(C_{t+1} \mid C_t)$ is estimated from the WOMD training dataset.

We then compare the distributions of these geometric mean probabilities using the Mann–Whitney test. The null hypothesis of the Mann–Whitney test assumes that the model-generated and ground truth transition probability distributions come from the same underlying population, that is, there is no significant difference between them. Note that for stochastic models, we generate one trajectory per leader trajectory since the WOMD test set already provides a sufficiently large number of car-following pairs for statistical comparison, as shown in Table \ref{tab:trajectory_seconds_WOMD}.

The Mann–Whitney test results in Tables~\ref{tab:transition_probability_comparison}a-c provide insight into how well each model's transition dynamics align with ground truth data. We caution that failure to reject the null hypothesis under this test is only one scalar summary of trajectory distribution similarity. Nonetheless, the p-value comparison across models is informative.

\begin{table}[tb!]
\centering
\caption{Comparison of model-generated and ground truth geometric mean probabilities using the Mann–Whitney test across three interaction types.}
\label{tab:transition_probability_comparison}
\small
\resizebox{0.6\textwidth}{!}{
\begin{tabular}{lrrrrr}
\\[-2mm]
\multicolumn{6}{l}{(a) AV-following-HDV} \\
\hline
Model & Test Statistic & $p$-Value & Significant & Mean & Median \\
\hline
Ground Truth         & –       & –     & –              & 0.388 & 0.367 \\
IDM                  & 96189   & 0.020 & True           & 0.367 & 0.329 \\
Van-Arem             & 86605   & 0.693 & \textbf{False} & 0.391 & 0.348 \\
FVDM-CTH             & 85696   & 0.514 & \textbf{False} & 0.396 & 0.354 \\
FVDM-Sigmoid         & 86296   & 0.629 & \textbf{False} & 0.394 & 0.355 \\
Gipps                & 108647  & 0.000 & True           & 0.337 & 0.281 \\
SIDM                 & 97552   & 0.006 & True           & 0.365 & 0.319 \\
FNN-reg              & 77885   & 0.004 & True           & 0.428 & 0.391 \\
GRU-reg              & 87414   & 0.870 & \textbf{False} & 0.396 & 0.344 \\
BayesGMM             & 84830   & 0.368 & \textbf{False} & 0.407 & 0.376 \\
MC-CF (det)          & 83461   & 0.197 & \textbf{False} & 0.410 & 0.369 \\
MC-CF (stoch)        & 90865   & 0.413 & \textbf{False} & 0.382 & 0.354 \\
MC-CF (cons, det)    & 85759   & 0.525 & \textbf{False} & 0.402 & 0.362 \\
MC-CF (cons, stoch)  & 92512   & 0.198 & \textbf{False} & 0.377 & 0.342 \\
\hline
\\[-2mm]
\multicolumn{6}{l}{(b) HDV-following-AV} \\
\hline
Model & Test Statistic & $p$-Value & Significant & Mean & Median \\
\hline
Ground Truth         & –       & –     & –              & 0.367 & 0.339 \\
IDM                  & 166801  & 0.000 & True           & 0.340 & 0.298 \\
Van-Arem             & 149486  & 0.770 & \textbf{False} & 0.359 & 0.333 \\
FVDM-CTH             & 144624  & 0.519 & \textbf{False} & 0.371 & 0.340 \\
FVDM-Sigmoid         & 147478  & 0.925 & \textbf{False} & 0.366 & 0.329 \\
Gipps                & 174002  & 0.000 & True           & 0.331 & 0.274 \\
SIDM                 & 179829  & 0.000 & True           & 0.319 & 0.266 \\
FNN-reg              & 138528  & 0.069 & \textbf{False} & 0.387 & 0.339 \\
GRU-reg              & 162739  & 0.004 & True           & 0.346 & 0.299 \\
BayesGMM             & 142508  & 0.292 & \textbf{False} & 0.373 & 0.351 \\
MC-CF (det)          & 134397  & 0.009 & True           & 0.388 & 0.366 \\
MC-CF (stoch)        & 146431  & 0.767 & \textbf{False} & 0.361 & 0.338 \\
MC-CF (cons, det)    & 135684  & 0.018 & True           & 0.387 & 0.358 \\
MC-CF (cons, stoch)  & 147750  & 0.967 & \textbf{False} & 0.361 & 0.334 \\
\hline
\\[-2mm]
\multicolumn{6}{l}{(c) HDV-following-HDV} \\
\hline
Model & Test Statistic & $p$-Value & Significant & Mean & Median \\
\hline
Ground Truth         & –        & –     & –              & 0.211 & 0.180 \\
IDM                  & 6116751  & 0.000 & True           & 0.193 & 0.162 \\
Van-Arem             & 5199860  & 0.039 & True           & 0.211 & 0.183 \\
FVDM-CTH             & 5039080  & 0.000 & True           & 0.218 & 0.186 \\
FVDM-Sigmoid         & 5246143  & 0.144 & \textbf{False} & 0.213 & 0.182 \\
Gipps                & 6568550  & 0.000 & True           & 0.182 & 0.153 \\
SIDM                 & 6481683  & 0.000 & True           & 0.185 & 0.154 \\
FNN-reg              & 4273979  & 0.000 & True           & 0.243 & 0.210 \\
GRU-reg              & 5949275  & 0.000 & True           & 0.202 & 0.164 \\
BayesGMM             & 5847285  & 0.000 & True           & 0.201 & 0.169 \\
MC-CF (det)          & 4570529  & 0.000 & True           & 0.232 & 0.201 \\
MC-CF (stoch)        & 5201238  & 0.041 & True           & 0.209 & 0.185 \\
MC-CF (cons, det)    & 4741486  & 0.000 & True           & 0.226 & 0.196 \\
MC-CF (cons, stoch)  & 5315410  & 0.579 & \textbf{False} & 0.205 & 0.183 \\
\hline
\multicolumn{6}{l}{\footnotesize \textit{Note}: A $p$-value below 0.1 indicates a statistically significant difference.}
\end{tabular}}
\end{table}

In the AV-following-HDV case (Table~\ref{tab:transition_probability_comparison}a), both MC-CF (stoch) and MC-CF (cons, stoch) produce transition dynamics broadly consistent with the ground truth ($p>0.1$), comparable to most physics-based models and several data-driven baselines. In the HDV-following-AV scenario (Table~\ref{tab:transition_probability_comparison}b), MC-CF (stoch) and MC-CF (cons, stoch) show high p-values ($p=0.767$ and $p=0.967$, respectively), while IDM, Gipps, SIDM, and GRU-reg show statistically significant deviations ($p<0.1$). In the HDV-following-HDV case (Table~\ref{tab:transition_probability_comparison}c), results are more discriminating: MC-CF (cons, stoch) ($p=0.579$) and FVDM-Sigmoid ($p=0.144$) are the only models without significant deviations, while all data-driven baselines show highly significant deviations ($p<0.001$), and the unconstrained MC-CF (stoch) yields a marginal result ($p=0.041$).

Overall, the proposed MC-CF variants, particularly MC-CF (stoch) and MC-CF (cons, stoch), produce transition dynamics that are broadly comparable to the ground truth across all three interaction types, performing at least as well as the baseline models. While FVDM-Sigmoid also does not show significant deviations in any of the three cases, it exhibits substantially higher prediction errors across all accuracy metrics compared to MC-CF (stoch) and MC-CF (cons, stoch) (Table~\ref{tab:HDV_following_HDV_results}), suggesting that the MC-CF framework offers a more favorable balance between transition-level behavioral consistency and kinematic prediction accuracy.

\subsection{Zero-Shot Generalization Analysis}
In this subsection, we evaluate the models' zero-shot generalization capability. Specifically, we test the models trained on the HDV-following-HDV subset of the WOMD training dataset on 106 HDV-following-HDV car-following pairs from the PHX dataset, as summarized in Table \ref{tab:trajectory_seconds_PHX}, considering both one-step and open-loop performance metrics. We focus on the HDV-following-HDV case because (i) AV behaviors may differ between the two datasets, as noted by \citep{zhang2025waymo}, and (ii) the WOMD training dataset contains substantially more HDV-following-HDV samples than the AV-included cases, providing a robust basis for cross-dataset evaluation.

The zero-shot results in Table~\ref{tab:HDV_following_HDV_zero_shot_results} reveal several patterns that are broadly consistent with, but also deepen, the findings from the WOMD test dataset. Models with access to 1-second trajectory history again dominate one-step prediction: GRU-reg achieves the lowest RMSE(a) (0.1887) and BayesGMM achieves the lowest RMSE(v) (0.0907), while MC-CF (cons, stoch) achieves the lowest RMSE(s) (0.0312). The performance advantage of history-conditioned models on one-step metrics, already observed in the WOMD evaluation, thus generalizes to an unseen dataset collected in a different geographic environment.

\begin{table}[tb!]
\centering
\caption{Trajectory prediction performance for HDV-following-HDV (Zero-Shot Generalization on PHX Dataset).}
\vspace{0.2cm}
\label{tab:HDV_following_HDV_zero_shot_results}
\resizebox{\textwidth}{!}{
\begin{tabular}{lcccccccccc}
\hline
Model & \multicolumn{3}{c}{One-Step Prediction} & \multicolumn{7}{c}{Open-Loop Prediction} \\
 & $RMSE(s)$ & $RMSE(v)$ & $RMSE(a)$ & $minDTW(s)$ & $minDTW(v)$ & $minADE$ & $minFDE$ & $avgADE$ & $avgFDE$ & $OR$\\
\hline
\multicolumn{11}{l}{Deterministic} \\
IDM          & 0.0701 & 0.2246 & 2.8060 &  81.3912 & 10.2871 & 6.7568 & 10.9818 & 6.7568 & 10.9818          & 0.0377 \\
Van-Arem     & 0.0694 & 0.1611 & 1.1125 &  69.9647 &  8.5555 & 5.7128 & 10.2753 & 5.7128 & 10.2753          & 0.0377 \\
FVDM-CTH     & 0.0694 & 0.1600 & 1.0900 &  62.0843 &  7.6615 & 4.9357 &  9.3843 & 4.9357 &  9.3843          & 0.0566 \\
FVDM-Sigmoid & 0.0694 & 0.1596 & 1.0844 &  61.2854 &  7.5394 & 4.8679 &  9.2701 & 4.8679 &  9.2701          & 0.0660 \\
Gipps        & 0.0696 & 0.1660 & 1.1813 &  88.3806 & 11.2156 & 7.4035 & 11.5607 & 7.4035 & 11.5607          & 0.0377 \\
FNN-reg      & 0.0693 & 0.1314 & 0.5749 &  52.0997 &  6.9482 & 4.1934 &  8.4532 & \textbf{4.1934*} & \textbf{8.4532*} & 0.0566 \\
GRU-reg      & 0.0700 & 0.1263 & \textbf{0.1887*} & 76.9325 & 11.6635 & 4.7039 & 12.4549 & 4.7039 & 12.4549 & 0.5472 \\
MC-CF (det)  & 0.0693 & 0.1324 & 0.5983 &  58.7282 &  7.6271 & 4.7146 &  9.8219 & 4.7146 &  9.8219          & 0.1038 \\
MC-CF (cons, det) & 0.0693 & 0.1340 & 0.6474 &  63.4022 &  7.6784 & 4.6404 &  9.3383 & 4.6404 &  9.3383          & 0.0472 \\
\hline
\multicolumn{11}{l}{Stochastic} \\
SIDM (1)                & 0.0345 & 0.2124 & 2.1237 & 81.8708 & 10.1084 & 6.7458 & 10.9890 & --     & --       & 0.0377 \\
SIDM (3)                & --     & --     & --     & 81.4790 & 10.0443 & 6.7084 & 10.9390 & --     & --       & --     \\
SIDM (6)                & --     & --     & --     & 81.3253 & 10.0135 & 6.6976 & 10.9202 & --     & --       & --     \\
SIDM (10)               & --     & --     & --     & 81.2543 &  9.9932 & 6.6913 & 10.9037 & --     & --       & --     \\
SIDM (15)               & --     & --     & --     & 81.1866 &  9.9698 & 6.6873 & 10.8930 & 6.7389 & 10.9788  & --     \\
BayesGMM (stoch, 1)     & 0.0316 & \textbf{0.0907*} & 0.6349 & 88.6017 & 14.2300 & 6.4184 & 14.3376 & --    & --      & \textbf{0.0000*} \\
BayesGMM (stoch, 3)     & --     & --     & --     & 42.6005 &  7.8788 & 3.4551 &  7.0603 & --     & --       & --     \\
BayesGMM (stoch, 6)     & --     & --     & --     & 28.4748 &  5.9406 & 2.5146 &  4.5114 & --     & --       & --     \\
BayesGMM (stoch, 10)    & --     & --     & --     & 21.1819 &  5.5200 & 2.0992 &  3.5138 & --     & --       & --     \\
BayesGMM (stoch, 15)    & --     & --     & --     & \textbf{18.0591*} & 5.0159 & \textbf{1.8955*} & \textbf{2.9033*} & 5.9235 & 12.4517 & -- \\
MC-CF (stoch, 1)        & 0.0314 & 0.1109 & 0.8036 & 55.1877 &  7.5953 & 4.4626 &  9.3564 & --     & --       & 0.0849 \\
MC-CF (stoch, 3)        & --     & --     & --     & 44.7221 &  6.2083 & 3.7144 &  7.6153 & --     & --       & --     \\
MC-CF (stoch, 6)        & --     & --     & --     & 41.3934 &  5.7185 & 3.4800 &  7.0230 & --     & --       & --     \\
MC-CF (stoch, 10)       & --     & --     & --     & 38.5924 &  5.1675 & 3.2051 &  6.4480 & --     & --       & --     \\
MC-CF (stoch, 15)       & --     & --     & --     & 37.2994 & \textbf{5.0151*} & 3.0988 &  6.1147 & 4.5604 &  9.3871 & --     \\
MC-CF (cons, stoch, 1)  & \textbf{0.0312*} & 0.1119 & 0.8285 & 60.2077 &  7.3054 & 4.5274 &  9.1226 & --    & --      & 0.0472 \\
MC-CF (cons, stoch, 3)  & --     & --     & --     & 52.7713 &  6.4777 & 3.9694 &  7.7423 & --     & --       & --     \\
MC-CF (cons, stoch, 6)  & --     & --     & --     & 49.6128 &  6.1408 & 3.7170 &  7.1841 & --     & --       & --     \\
MC-CF (cons, stoch, 10) & --     & --     & --     & 39.4399 &  5.4592 & 3.1978 &  6.2504 & --     & --       & --     \\
MC-CF (cons, stoch, 15) & --     & --     & --     & 37.6755 &  5.0434 & 3.0393 &  5.9018 & 4.7517 &  9.7022  & --     \\
\hline
\multicolumn{11}{l}{\footnotesize \textit{Note}: * indicates the minimum (best) value within each column. In case of ties, multiple entries are marked.}\\
\multicolumn{11}{l}{\footnotesize -- denotes not applicable. $OR$: $K=1$ only; $avgADE$/$avgFDE$: $K=15$ for stochastic models; equals $minADE$/$minFDE$ for deterministic. $n_\text{viable}/n_\text{total} = 47/106$.}
\end{tabular}
}
\end{table}

However, the open-loop regime reveals a sharp contrast. GRU-reg, despite its strong one-step accuracy, incurs a catastrophic overlapping rate of 0.5472 on PHX, substantially worse than its already-elevated OR of 0.1648 on WOMD. This collapse illustrates the compounding nature of open-loop error: the 1-second history advantage that benefits one-step prediction becomes a liability as autoregressive rollouts accumulate errors across time steps, and this instability is more pronounced on an out-of-distribution dataset. BayesGMM, by contrast, sustains competitive open-loop performance and achieves the best results at $K=15$ for minDTW(s), minADE, and minFDE (18.0591, 1.8955, and 2.9033, respectively). MC-CF (stoch) achieves the best minDTW(v) at $K=15$ (5.0151) with an overlapping rate of 0.0849, which can be further reduced to 0.0472 with conservative inference. MC-CF (cons, stoch) maintains comparable accuracy across most metrics, with some improving (minADE: 3.0393 vs. 3.0988; minFDE: 5.9018 vs. 6.1147) and others slightly higher (avgADE: 4.7517 vs. 4.5604).

The degradation of conventional physics-based models, including IDM, Van Arem, Gipps, both FVDM variants, and SIDM, is also consistent with the WOMD findings, as these models exhibit substantial error across both short-term accuracy and long-horizon metrics. Taken together, the PHX zero-shot results reinforce that the MC-CF framework and BayesGMM generalize well to unseen driving environments, while also confirming that open-loop stability, rather than one-step accuracy, is the more demanding criterion for zero-shot evaluation.

\subsection{Microscopic Simulation and Model Scalability}\label{sec:simulation}

\begin{figure}[tb!]
    \centering
    \includegraphics[width=0.3\linewidth]{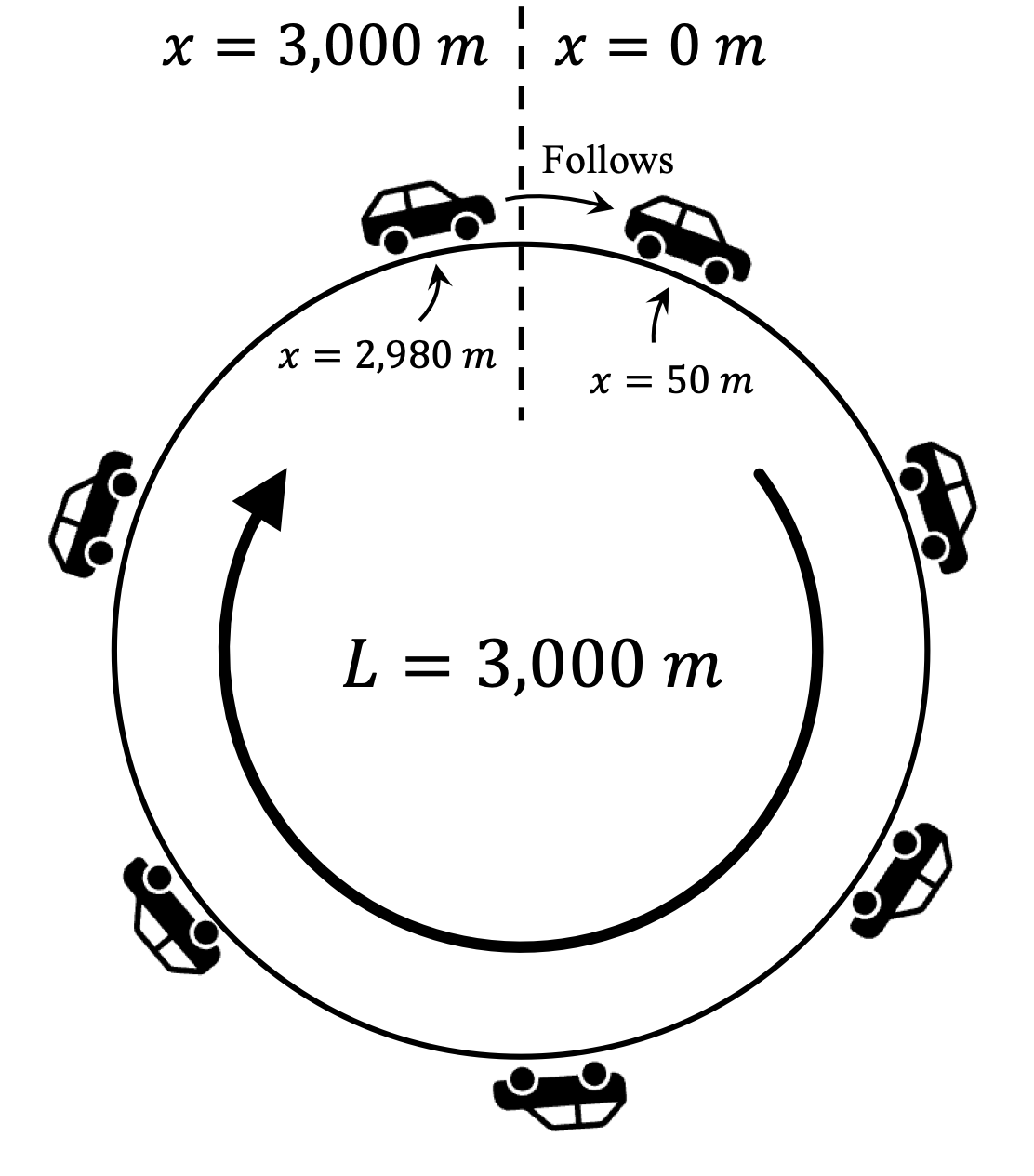}
    \caption{Schematic of the ring road simulation geometry ($L = 3,000$~m). The periodic boundary condition ensures that $x = 3,000~m$ and $x = 0~m$ represent the same physical location, creating a closed-loop system for evaluating long-term vehicle interactions.}
    \label{fig:ring_geometry}
\end{figure}

While the previous sections demonstrate the MC-CF model's predictive accuracy on real-world short-horizon trajectories, a robust car-following model must also perform reliably in microscopic simulations where errors can accumulate over time. In this section, we evaluate the performance of the proposed model in a ring road simulation, as illustrated in Figure~\ref{fig:ring_geometry}. This geometry represents a closed-loop system with a total length of $L = 3,000$ m (where the position $x = 3,000$ m is identical to $x = 0$ m), allowing us to assess long-term vehicle interactions and shockwave propagation over a 300-second horizon with a time step of $\Delta t = 0.1$ s. Since the WOMD contains the largest amount of HDV interaction data, we focus exclusively on a 100\% HDV simulation environment. We specifically utilize MC-CF (stoch) for the simulation, leveraging its superior open-loop trajectory prediction accuracy compared to MC-CF (det).

\subsubsection{Simulation Setup}

To demonstrate how incorporating diverse data and inference strategies systematically improves simulation stability and realism, we conduct an ablation study comparing the baseline physics-based models (IDM and SIDM) against four variants of the proposed MC-CF framework. The components of these variants are summarized in Table \ref{tab:ablation_setup}.

\begin{table}[tb!]
\centering
\caption{Ablation study configurations for the evaluated MC-CF models.}
\label{tab:ablation_setup}
\resizebox{0.8\columnwidth}{!}{
\begin{tabular}{lcccc}
\hline
\textbf{Model Name} & \textbf{\begin{tabular}[c]{@{}c@{}}WOMD CF \\ Data\end{tabular}} & \textbf{\begin{tabular}[c]{@{}c@{}}Free-flow (Solo) \\ Data\end{tabular}} & \textbf{\begin{tabular}[c]{@{}c@{}}Conservative \\ Inference\end{tabular}} & \textbf{\begin{tabular}[c]{@{}c@{}}Freeway (TGSIM) \\ Data\end{tabular}} \\ \hline
MC-CF (Original) & \checkmark & & & \\
MC-CF (Solo) & \checkmark & \checkmark & & \\
MC-CF (Solo+Conservative) & \checkmark & \checkmark & \checkmark & \\
MC-CF (Solo+Conservative+TGSIM) & \checkmark & \checkmark & \checkmark & \checkmark \\ \hline
\end{tabular}
}
\end{table}

The \textbf{MC-CF (Original)} model is trained purely on WOMD car-following pairs, which is identical to the MC-CF (stoch) model evaluated throughout the preceding sections.

The \textbf{MC-CF (Solo)} model is augmented with unconstrained solo driving trajectories (free-flow data) to provide appropriate acceleration distributions for large spacings. To incorporate free-flow behavior, we extracted trajectories of vehicles without leaders, or with spacings exceeding $45$ m for WOMD and $150$ m for TGSIM. For these cases, we assigned a ghost leader with $\Delta v = 0$ m/s and spacing $d = 45$ m for the MC-CF (Solo) and MC-CF (Solo+Conservative) models, and $d = 150$ m for the MC-CF (Solo+Conservative+TGSIM) model. To minimize the impact of urban noise in the WOMD, free-flow trajectories were exclusively sampled only from scenarios without traffic signals, stop signs, and crosswalks, which represented approximately 10\% of the total WOMD training set scenarios.

The \textbf{MC-CF (Solo+Conservative)} model applies the conservative inference strategy introduced in Section~\ref{sec:methodology} on top of the Solo data augmentation. As described there, this strategy restricts acceleration sampling to lower percentiles of $\mathcal{A}(C_{t+1})$ when TTC falls below safety thresholds, suppressing near-crash acceleration samples.

Finally, the \textbf{MC-CF (Solo+Conservative+TGSIM)} model is further augmented with high-speed freeway data from the TGSIM dataset. This addition extends the valid spacing threshold to $150$ m (originally $45$ m), the relative speed range to $[-30, 30]$ m/s (originally $[-10, 10]$ m/s), and the valid follower speed range to $[0, 40]$ m/s (originally $[0, 20]$ m/s). This model allows us to evaluate how expanding the state space with higher-speed samples impacts overall performance. Note that the solo driving data from the TGSIM dataset is added only to this specific model.

We designed four distinct simulation experiments to stress-test the models. The first three experiments stay within the speed range of the urban WOMD dataset, while the last experiment is introduced to test the models' capabilities in handling high-speed. The specific configurations, vehicle densities, and perturbation profiles for these scenarios are summarized in Table \ref{tab:sim_experiments}.

\begin{table}[tb!]
\centering
\caption{Configuration of the ring road simulation experiments.}
\label{tab:sim_experiments}
\resizebox{\textwidth}{!}{
\begin{tabular}{lccl}
\hline
\textbf{Experiment} & \textbf{Vehicles ($N$)} & \textbf{$v_{start}$ (m/s)} & \textbf{Perturbation Profile (Target Vehicle)} \\ \hline
Normal Equilibrium & 200 & 5.84 & - \\
Standard Shockwave & 200 & 5.84 & At $t=50$ s: $-1$ m/s$^2$ for $5$ s, hold speed for $10$ s, then $1$ m/s$^2$ for $5$ s \\
Severe Shockwave & 200 & 5.84 & At $t=50$ s: $-1$ m/s$^2$ for $10$ s, hold speed for $30$ s, then $1$ m/s$^2$ for $10$ s \\
High-Speed Shockwave & 40 & 30.00 & Same severe perturbation profile as the Severe Shockwave experiment \\ \hline
\end{tabular}
}
\end{table}

\subsubsection{Simulation Results}

For the stochastic models, $20$ independent simulation trials were conducted per scenario to evaluate the mean and standard deviation of collision events. Table \ref{tab:sim_crashes} summarizes these collision statistics. The results show that IDM and SIDM experience no crashes during the simulation. However, the MC-CF (Original) model clearly shows its limitations by resulting in numerous crashes, averaging $37.05$ crashes in the Severe Shockwave test. Incorporating the solo driving dataset significantly reduces the number of crashes in the urban scenarios, and adding the conservative sampling effectively eliminates crashes in both the Normal Equilibrium and Standard Shockwave experiments. In the Severe Shockwave experiment, MC-CF (Solo+Conservative) significantly reduces the number of collisions, though an average of $8.05$ crashes per simulation remains.

Notably, while adding the TGSIM dataset significantly reduces crashes in a \textbf{High-Speed Shockwave} scenario (a mean of $0.70$ crashes compared to $3.20$ without the TGSIM dataset), it slightly degrades performance in the lower-speed, dense urban scenarios compared to the MC-CF (Solo+Conservative) model. This can be attributed to the limited number of samples and the differing data quality in TGSIM compared to the massive WOMD dataset. However, as demonstrated in the high-speed experiment, adding even a small amount of freeway data substantially improves performance in high-speed regimes. If larger amounts of high-fidelity naturalistic freeway data (e.g., Waymo highway datasets) are collected in the future, we expect the model to organically self-correct across all speed regimes without such trade-offs.

\begin{table}[tb!]
\centering
\caption{Collision statistics (Mean $\pm$ Std) across 20 simulation trials for the four experiments.}
\label{tab:sim_crashes}
\resizebox{\textwidth}{!}{
\begin{tabular}{lcccccc}
\hline
\textbf{Experiment} & \textbf{IDM} & \textbf{SIDM} & \textbf{MC-CF (Original)} & \textbf{MC-CF (Solo)} & \textbf{MC-CF (Solo+Cons)} & \textbf{MC-CF (Solo+Cons+TGSIM)} \\ \hline
Normal Equilibrium & 0.00 $\pm$ 0.00 & 0.00 $\pm$ 0.00 & 12.05 $\pm$ 5.34 & 1.90 $\pm$ 1.48 & \textbf{0.00 $\pm$ 0.00} & 0.25 $\pm$ 0.55 \\
Standard Shockwave & 0.00 $\pm$ 0.00 & 0.00 $\pm$ 0.00 & 15.90 $\pm$ 5.30 & 6.00 $\pm$ 1.95 & \textbf{0.00 $\pm$ 0.00} & \textbf{0.00 $\pm$ 0.00} \\
Severe Shockwave & 0.00 $\pm$ 0.00 & 0.00 $\pm$ 0.00 & 37.05 $\pm$ 6.06 & 30.75 $\pm$ 4.83 & \textbf{8.05 $\pm$ 1.85} & 9.35 $\pm$ 2.94 \\
High-Speed Shockwave & 0.00 $\pm$ 0.00 & 0.00 $\pm$ 0.00 & 7.40 $\pm$ 3.08 & 11.10 $\pm$ 2.81 & 3.20 $\pm$ 1.47 & \textbf{0.70 $\pm$ 1.17} \\ \hline
\end{tabular}
}
\end{table}

\begin{figure}[tb!]
    \centering
    \begin{subfigure}[b]{0.48\textwidth}
        \includegraphics[width=\textwidth]{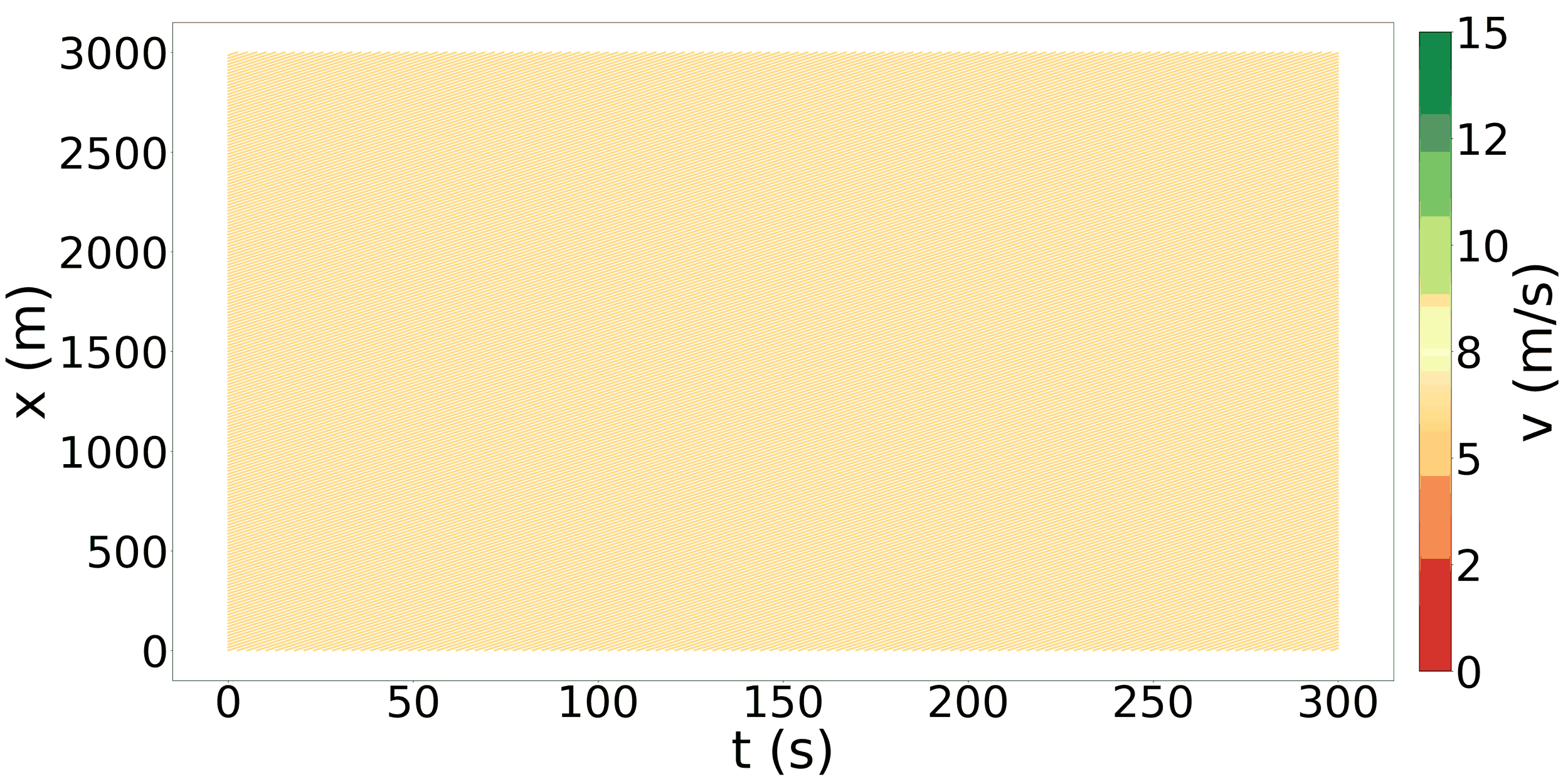}
        \caption{IDM}
    \end{subfigure}
    \hfill
    \begin{subfigure}[b]{0.48\textwidth}
        \includegraphics[width=\textwidth]{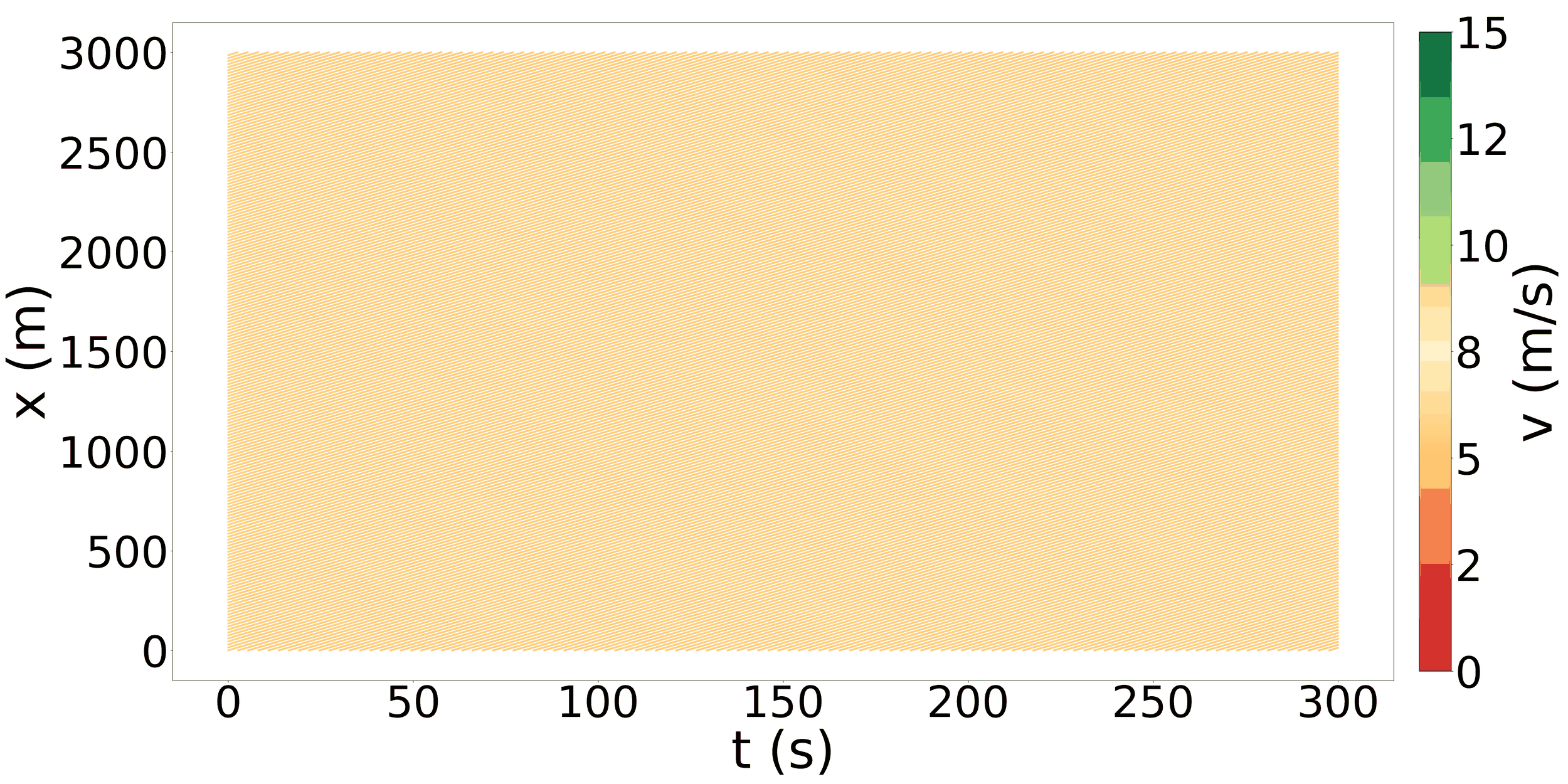}
        \caption{SIDM}
    \end{subfigure}
    \vskip\baselineskip
    \begin{subfigure}[b]{0.48\textwidth}
        \includegraphics[width=\textwidth]{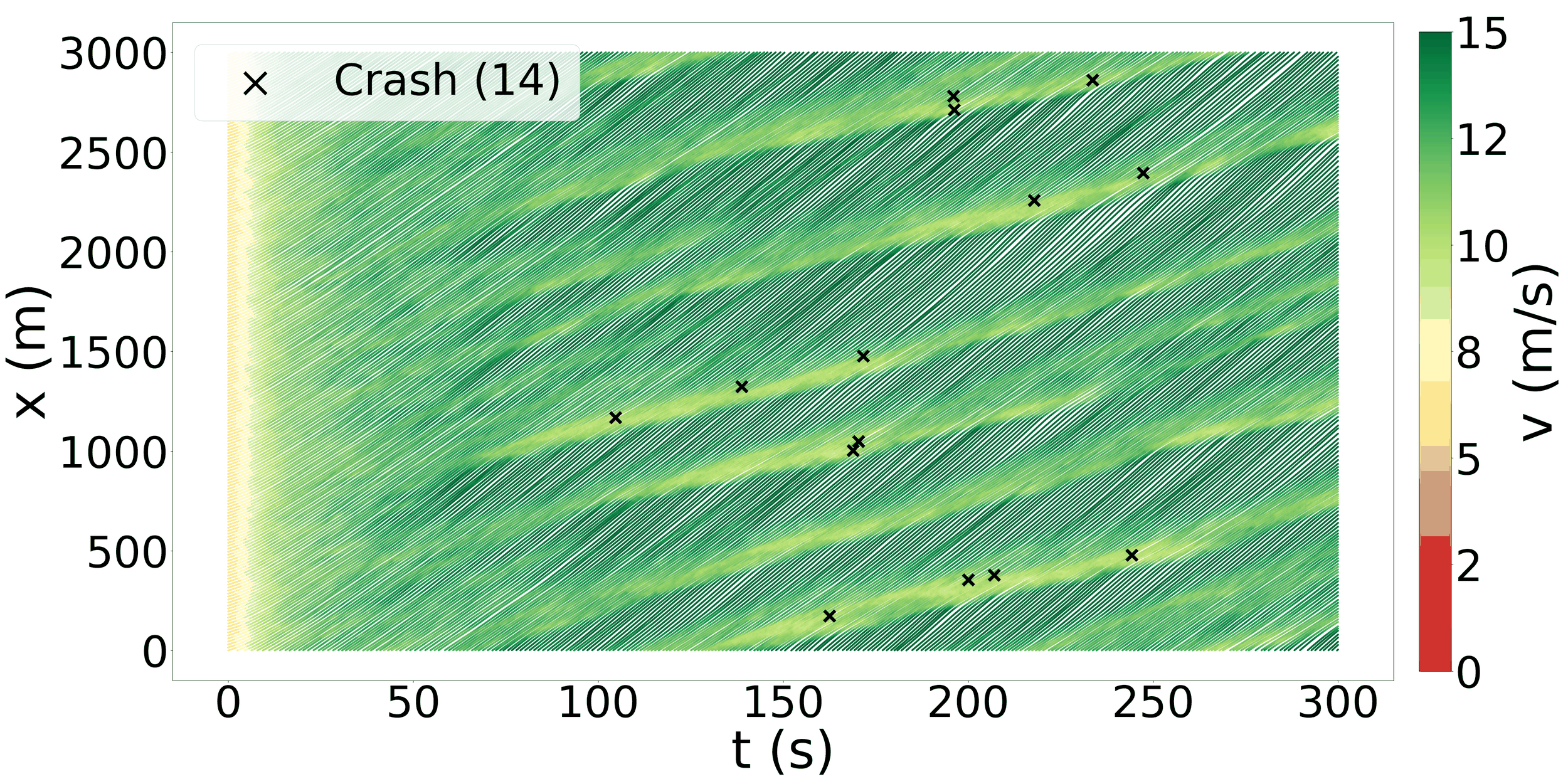}
        \caption{MC-CF (Original)}
    \end{subfigure}
    \hfill
    \begin{subfigure}[b]{0.48\textwidth}
        \includegraphics[width=\textwidth]{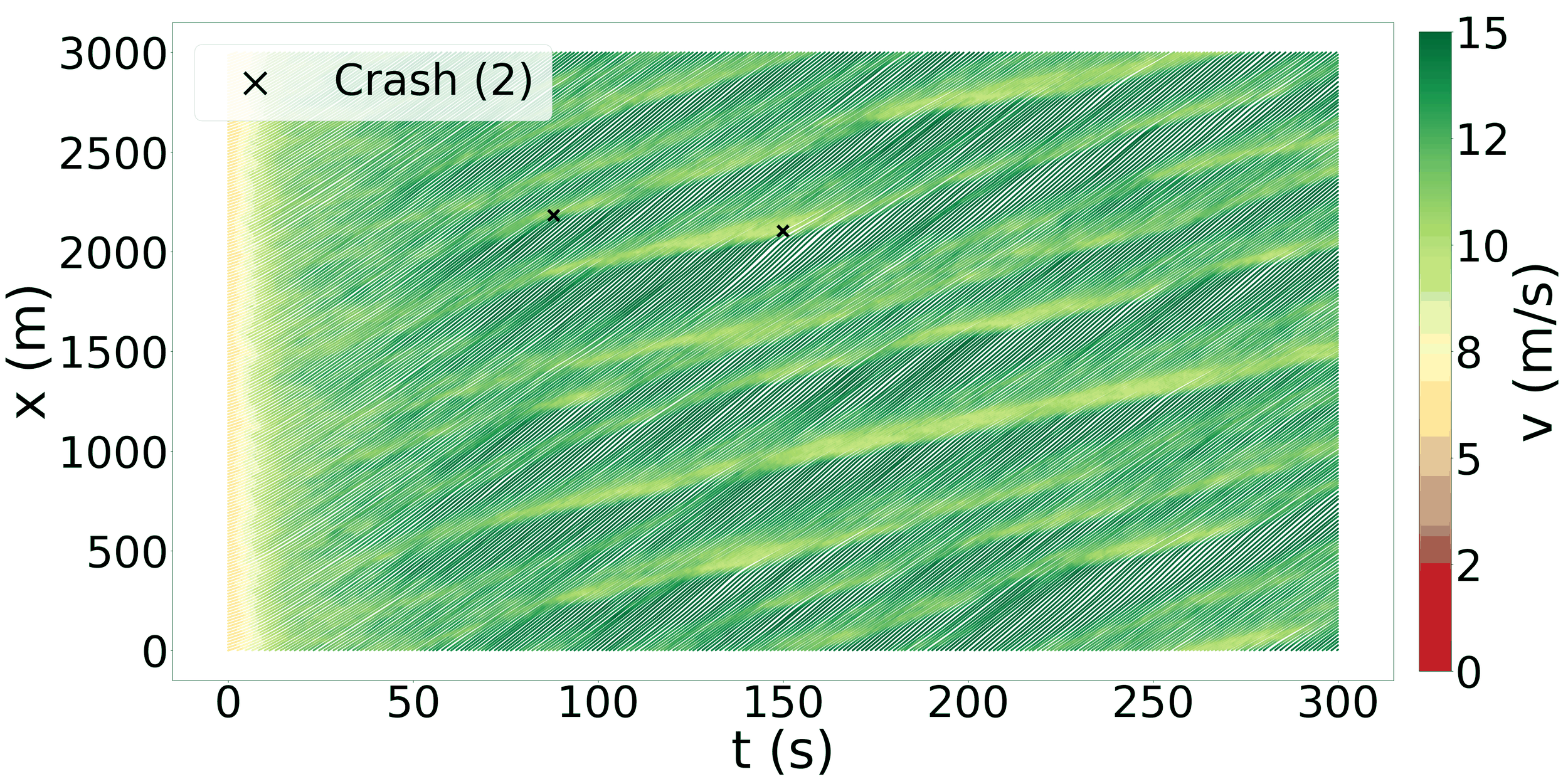}
        \caption{MC-CF (Solo)}
    \end{subfigure}
    \vskip\baselineskip
    \begin{subfigure}[b]{0.48\textwidth}
        \includegraphics[width=\textwidth]{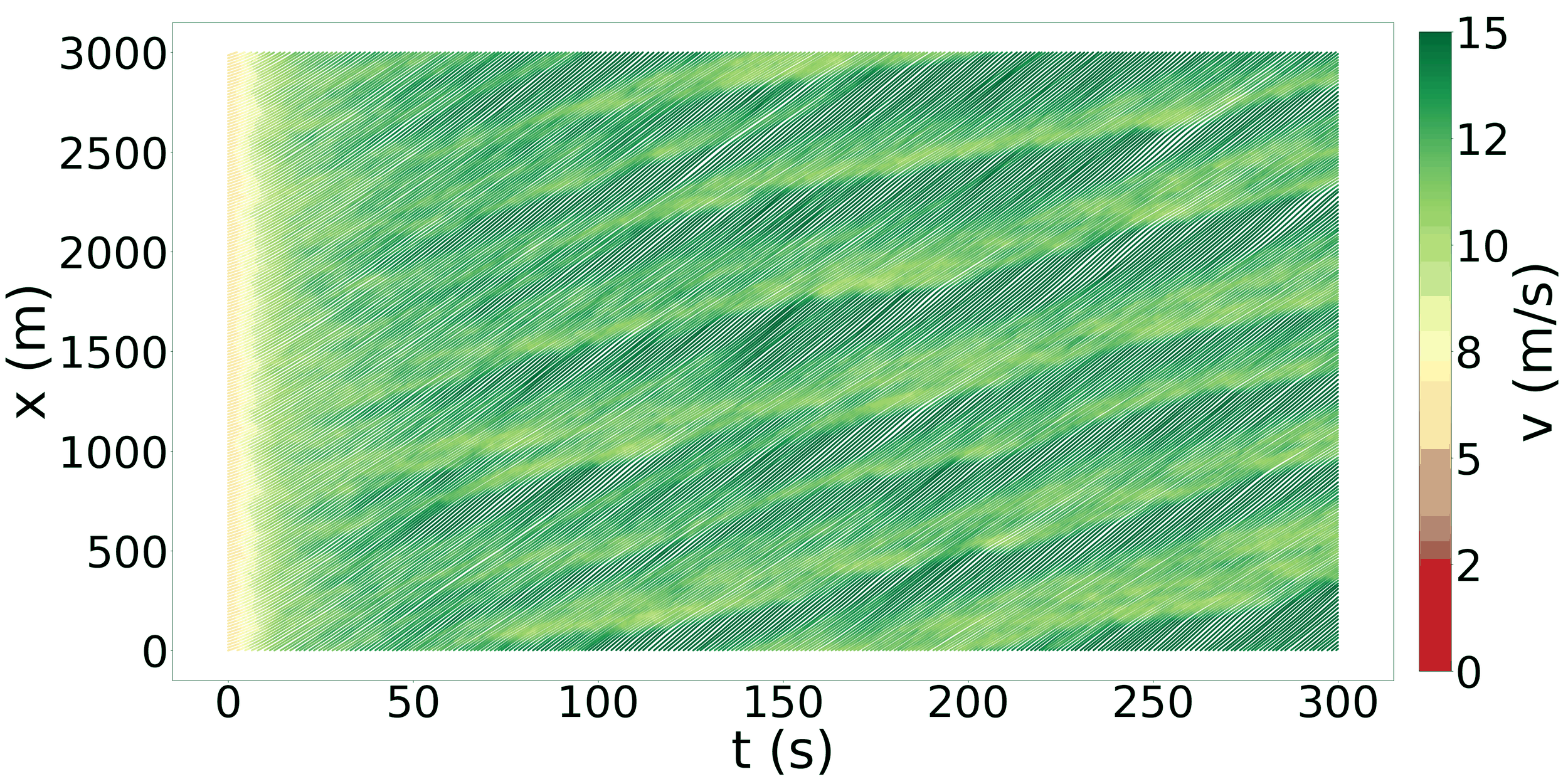}
        \caption{MC-CF (Solo+Conservative)}
    \end{subfigure}
    \hfill
    \begin{subfigure}[b]{0.48\textwidth}
        \includegraphics[width=\textwidth]{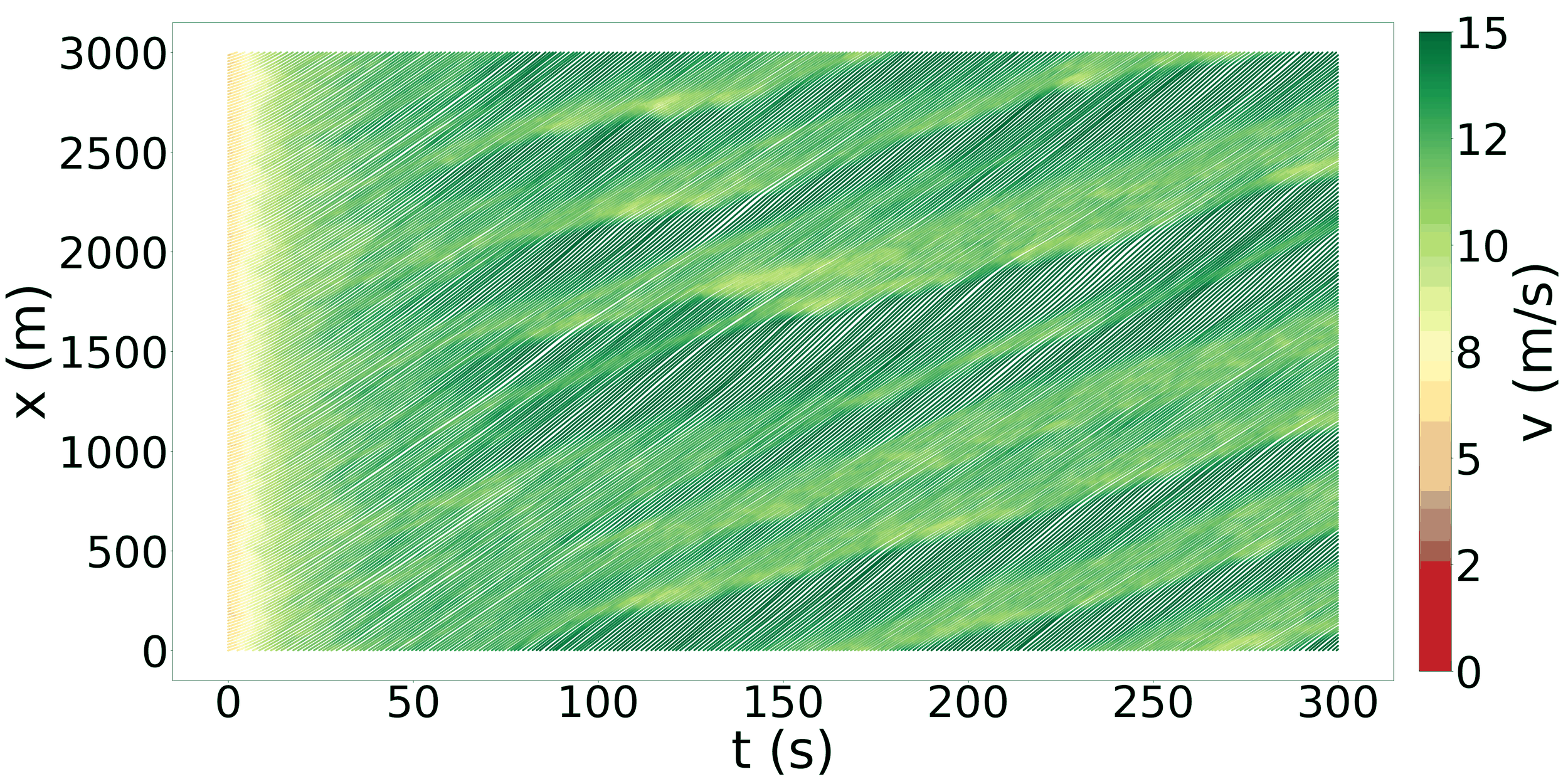}
        \caption{MC-CF (Solo+Conservative+TGSIM)}
    \end{subfigure}
    \caption{Trajectory plots for the Normal Equilibrium test ($N=200$, $v_{start}=5.84$ m/s).}
    \label{fig:sim_eq}
\end{figure}

Figure \ref{fig:sim_eq} illustrates the trajectories of the $200$ vehicles on a ring road for an arbitrarily selected random seed in the \textbf{Normal Equilibrium} test. We specifically start the simulation with an initial speed of $5.84$ m/s, which corresponds to the equilibrium speed obtained from the IDM simulation. As shown in Figures \ref{fig:sim_eq}a and \ref{fig:sim_eq}b, the IDM and SIDM maintain a perfectly uniform flow. However, we note that such perfect uniformity is highly unrealistic in real-world traffic consisting of heterogeneous human drivers.

The MC-CF (Original) model struggles with the urban noise embedded in its training data, resulting in $14$ crashes in the plotted equilibrium scenario. It is also interesting to observe that although both the IDM and MC-CF (Original) are calibrated/trained on the same dataset, they do not agree on the equilibrium speed. Considering that the MC-CF framework demonstrated significantly higher accuracy in both one-step and open-loop trajectory predictions (Table \ref{tab:HDV_following_HDV_results}), this discrepancy suggests that macroscopic conclusions (e.g., equilibrium speed or capacity) drawn from calibrated parametric models may be misleading. By adding solo driving data, the MC-CF (Solo) model effectively reduces the number of crashes. More importantly, applying the conservative inference mode, i.e., MC-CF (Solo+Conservative),  completely eliminates crashes in this experiment, matching the safety of the parametric models while simultaneously exhibiting realistic and stochastic speed fluctuations.

\begin{figure}[tb!]
    \centering
    \includegraphics[width=1.0\columnwidth]{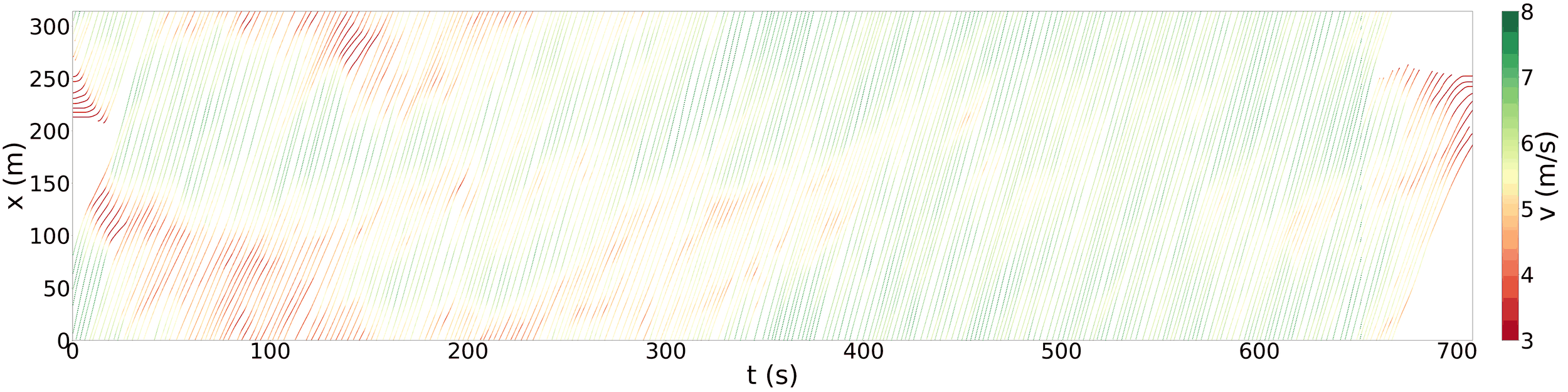 }
    \caption{Ground truth trajectories from the Nagoya Dome ring road experiment (Session 1520, $N=20$). The empirical data exhibits naturalistic speed fluctuations and forward moving shockwaves, contrasting with the rigid uniformity predicted by traditional parametric models.}
    \label{fig:nagoya_dome}
\end{figure}

To empirically validate the structural realism of these stochastic speed fluctuations, we compare them with real-world human driving data from the Nagoya Dome ring road experiment \cite{tadaki2013phase}. In this experiment, HDVs traveled continuously on a circular track with a radius of $50$ m (circumference approximately $314$ m). As illustrated by the ground truth trajectories in Figure \ref{fig:nagoya_dome}, naturalistic driving inherently exhibits continuous stochastic speed fluctuations and forward moving shockwaves. This empirical observation directly contradicts the perfectly homogeneous flow predicted by the IDM and SIDM (Figures \ref{fig:sim_eq}a–b). In contrast, the augmented MC-CF models (Figures \ref{fig:sim_eq}e–f) successfully reproduce this phenomenon, sustaining stable collision free flow while organically generating the speed variations observed in real-world experiment.

\begin{figure}[tb!]
    \centering
    \begin{subfigure}[b]{0.48\textwidth}
        \includegraphics[width=\textwidth]{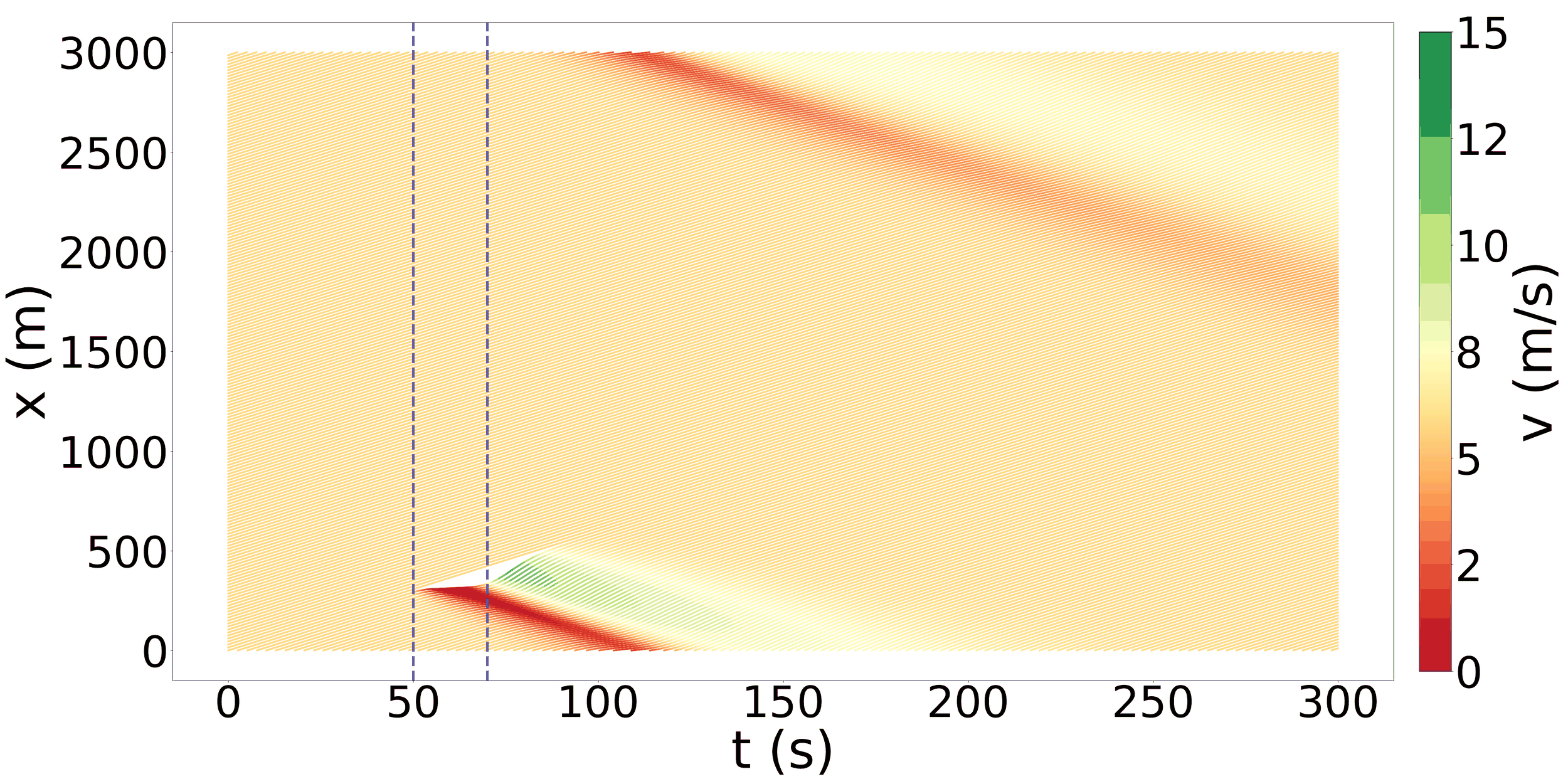}
        \caption{IDM}
    \end{subfigure}
    \hfill
    \begin{subfigure}[b]{0.48\textwidth}
        \includegraphics[width=\textwidth]{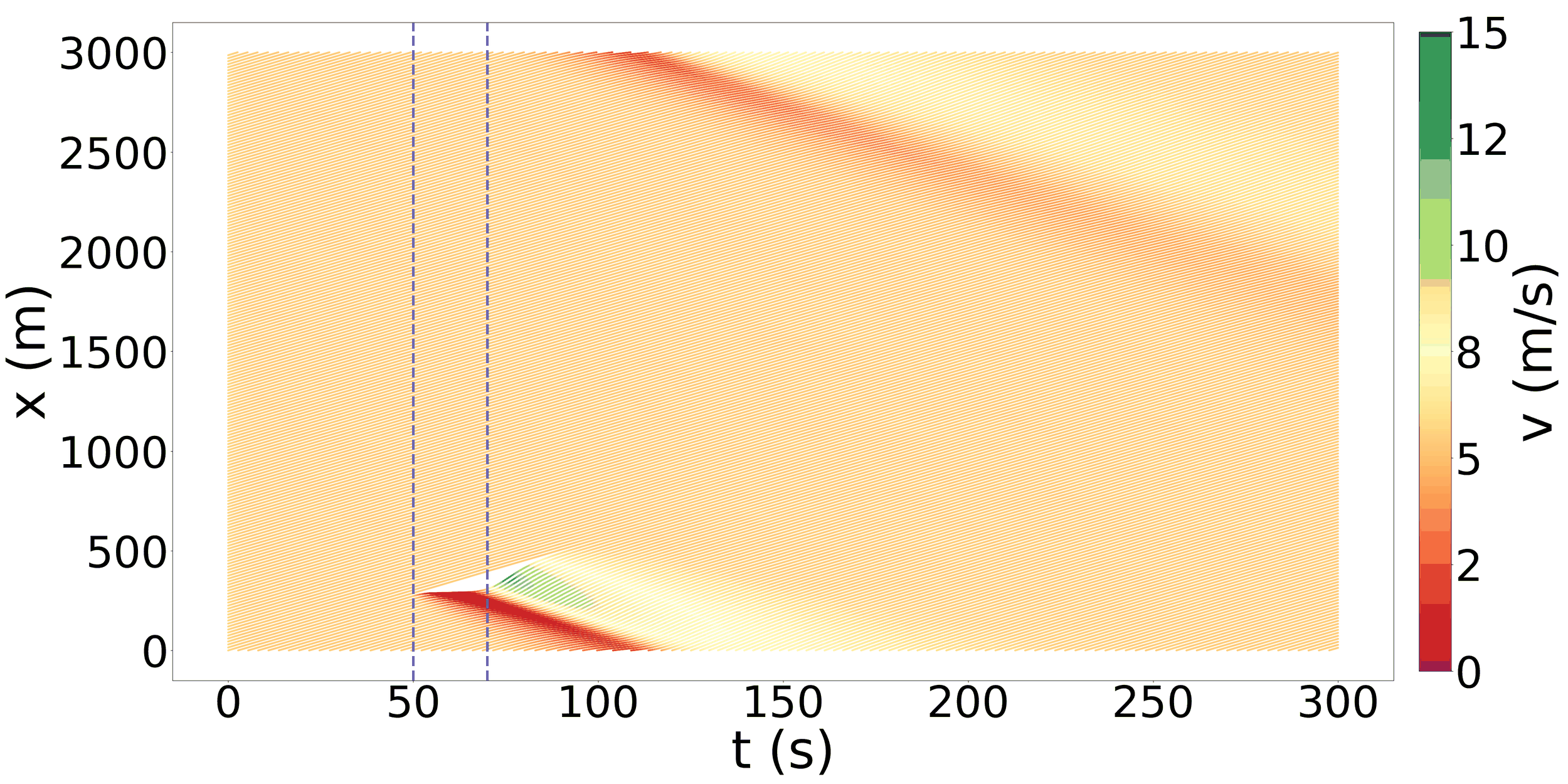}
        \caption{SIDM}
    \end{subfigure}
    \vskip\baselineskip
    \begin{subfigure}[b]{0.48\textwidth}
        \includegraphics[width=\textwidth]{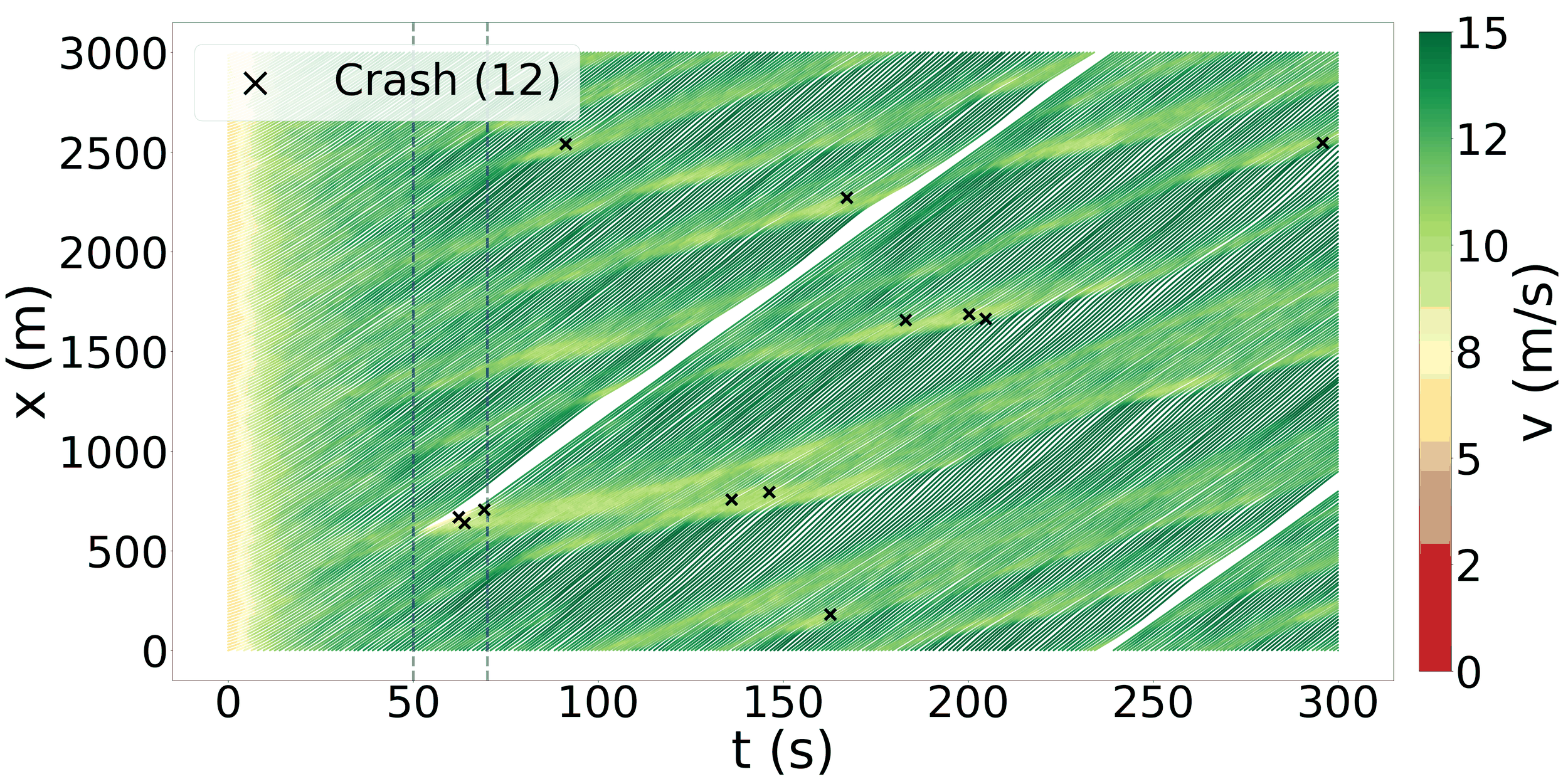}
        \caption{MC-CF (Original)}
    \end{subfigure}
    \hfill
    \begin{subfigure}[b]{0.48\textwidth}
        \includegraphics[width=\textwidth]{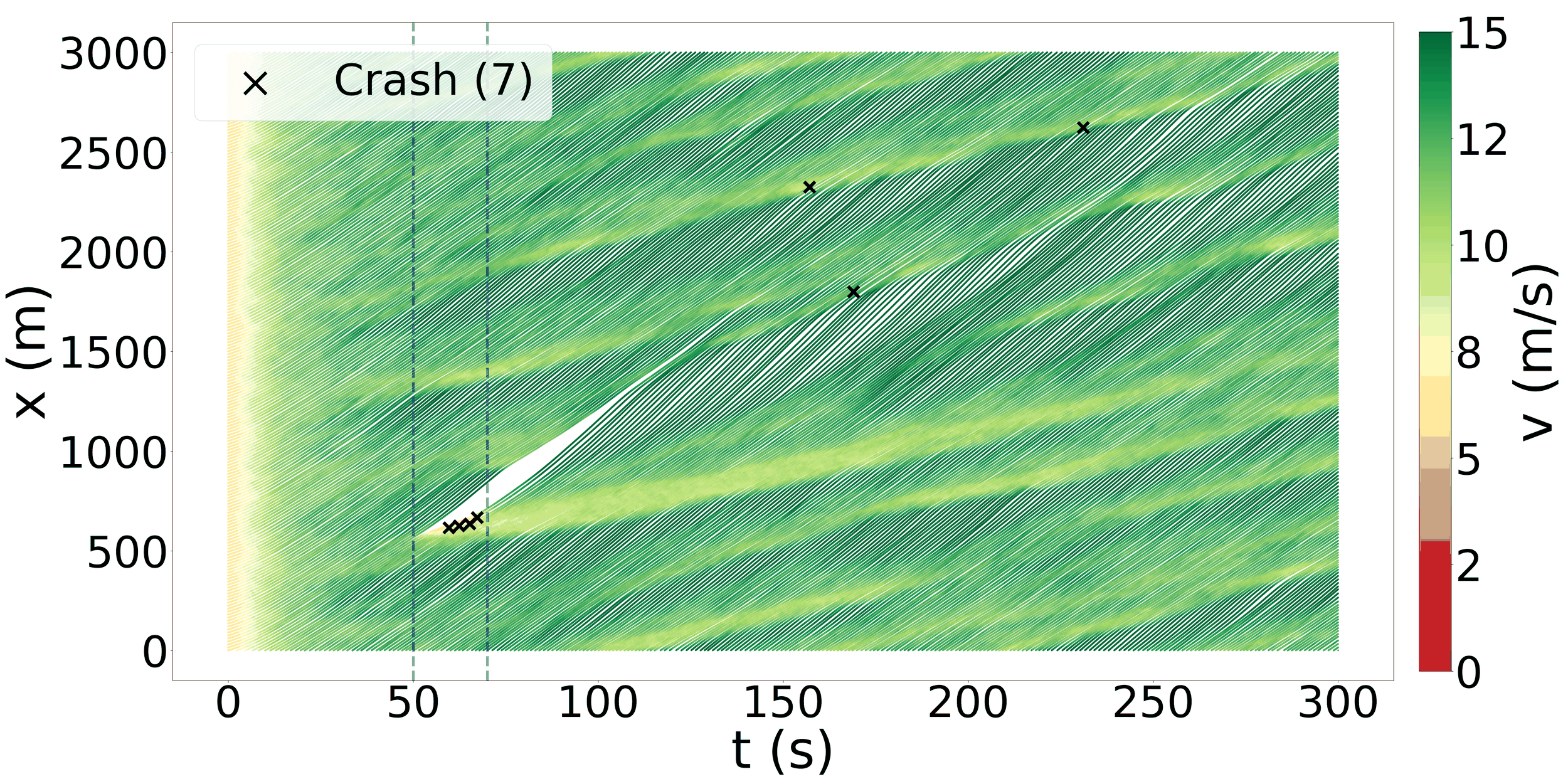}
        \caption{MC-CF (Solo)}
    \end{subfigure}
    \vskip\baselineskip
    \begin{subfigure}[b]{0.48\textwidth}
        \includegraphics[width=\textwidth]{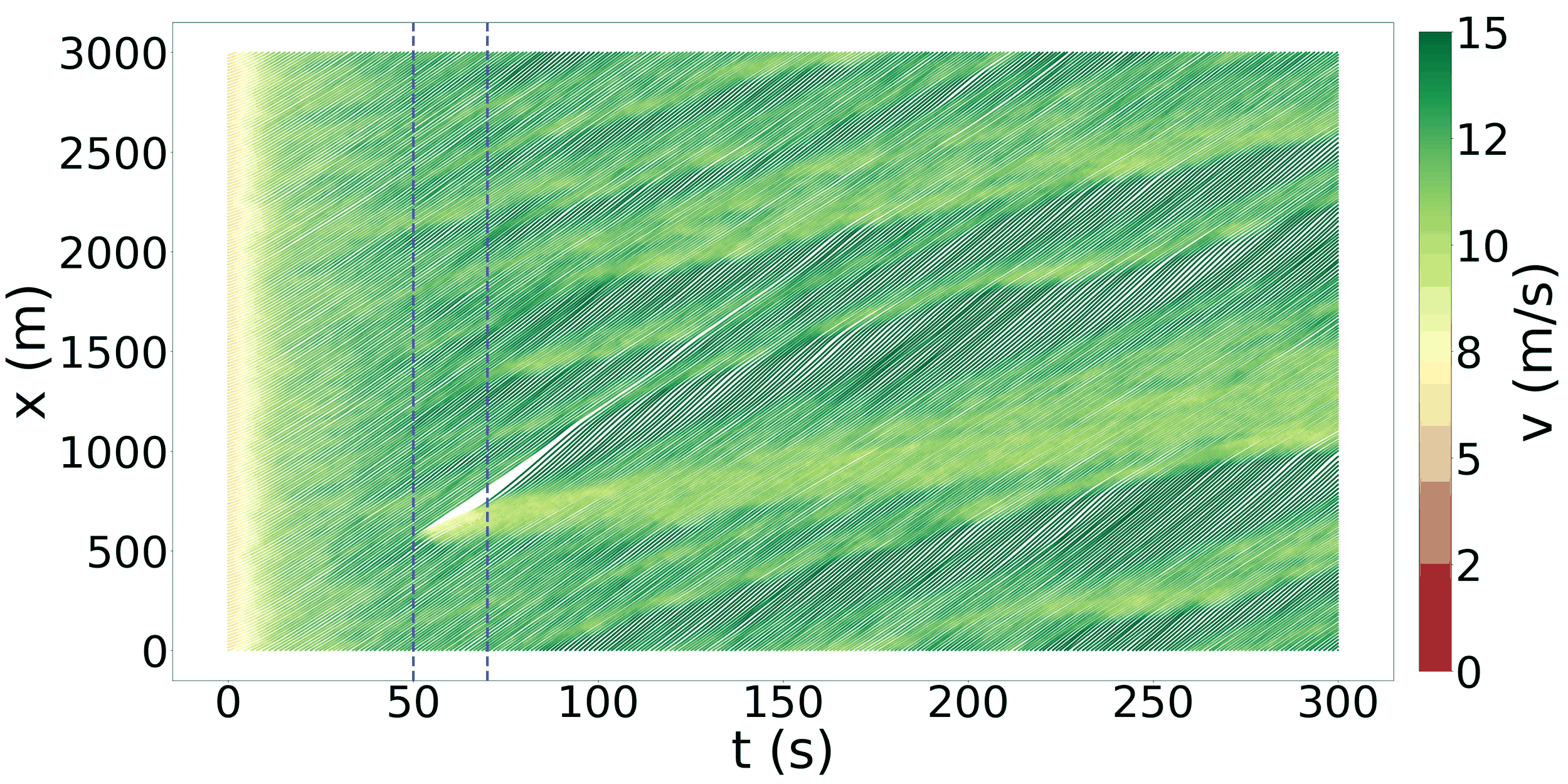}
        \caption{MC-CF (Solo+Conservative)}
    \end{subfigure}
    \hfill
    \begin{subfigure}[b]{0.48\textwidth}
        \includegraphics[width=\textwidth]{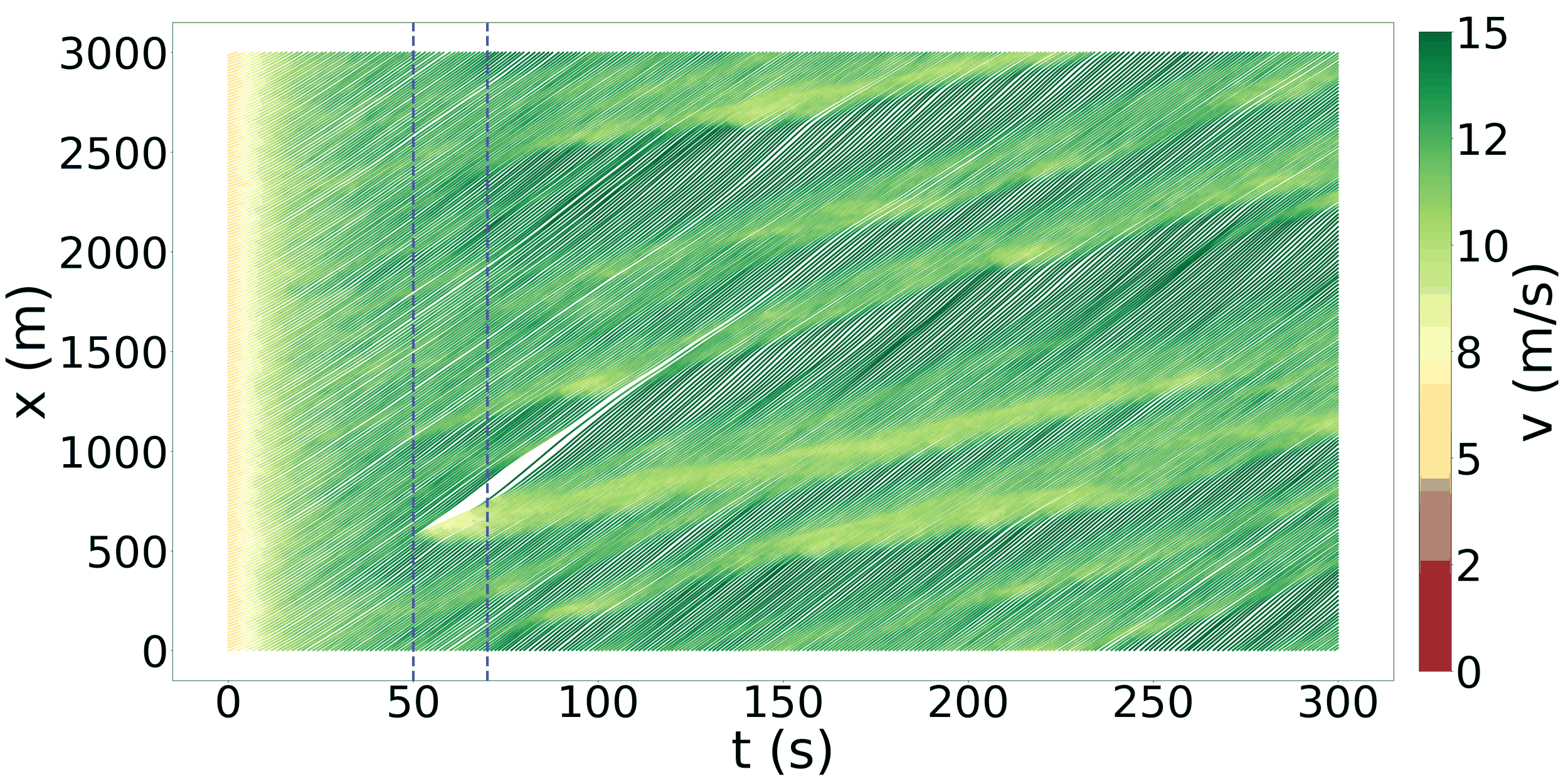}
        \caption{MC-CF (Solo+Conservative+TGSIM)}
    \end{subfigure}
    \caption{Trajectory plots for the Standard Shockwave test. MC-CF models with the conservative mode (e-f) eliminates collisions, while solo data (d-f) ensures proper gap reduction.}
    \label{fig:sim_std_sw}
\end{figure}

The \textbf{Standard Shockwave} test evaluates the response of following vehicles in a platoon with respect to a small perturbation introduced by the leader. As shown in Figures \ref{fig:sim_std_sw}a-b, the trajectories produced by the IDM and SIDM exhibit backward shockwave propagation, yet the system fails to recover to the equilibrium speed within the 300-second horizon. The MC-CF (Original) model (Figure \ref{fig:sim_std_sw}c) shows that vehicles recover their equilibrium speed in a shorter period of time. However, it fails to efficiently reduce the gap to the leader once the perturbation ends, leading to the formation of an unrealistic platoon.

The MC-CF (Solo) model (Figure \ref{fig:sim_std_sw}d) still exhibits collisions ($7$ crashes in the plotted trial) but successfully resolves the unrealistic platooning issue observed in the Original model. The inclusion of solo data enables followers to correctly recognize large gaps as free-flow conditions, allowing them to accelerate smoothly, close the gap, and return the system to equilibrium. When the conservative mode is applied alongside the solo data, collisions drop to zero (Figures \ref{fig:sim_std_sw}e and \ref{fig:sim_std_sw}f). This demonstrates the model's capability to maintain safety as long as the simulation remains within regions of the state space that are sufficiently represented in the training data.

\begin{figure}[tb!]
    \centering
    \begin{subfigure}[b]{0.48\textwidth}
        \includegraphics[width=\textwidth]{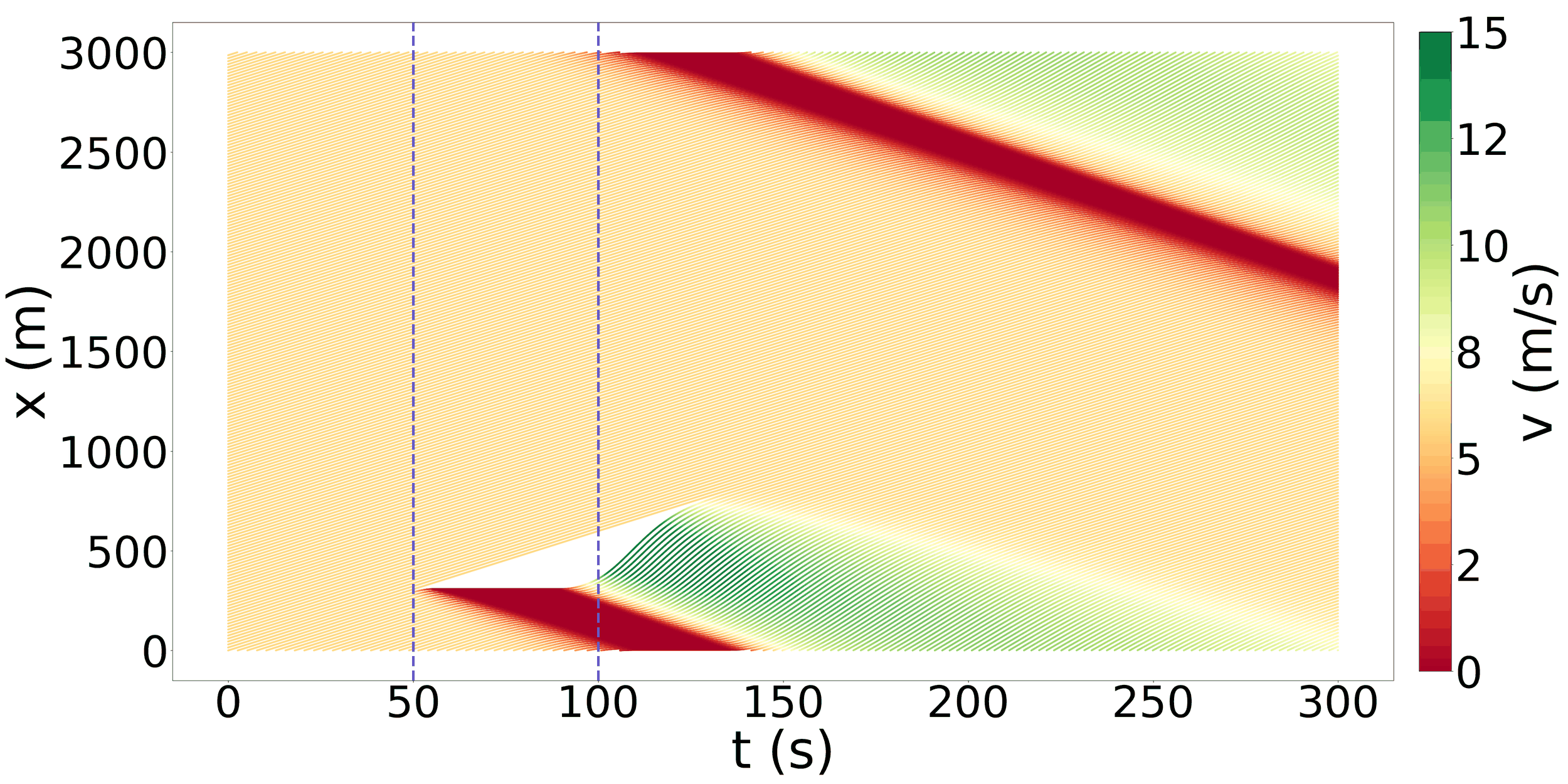}
        \caption{IDM}
    \end{subfigure}
    \hfill
    \begin{subfigure}[b]{0.48\textwidth}
        \includegraphics[width=\textwidth]{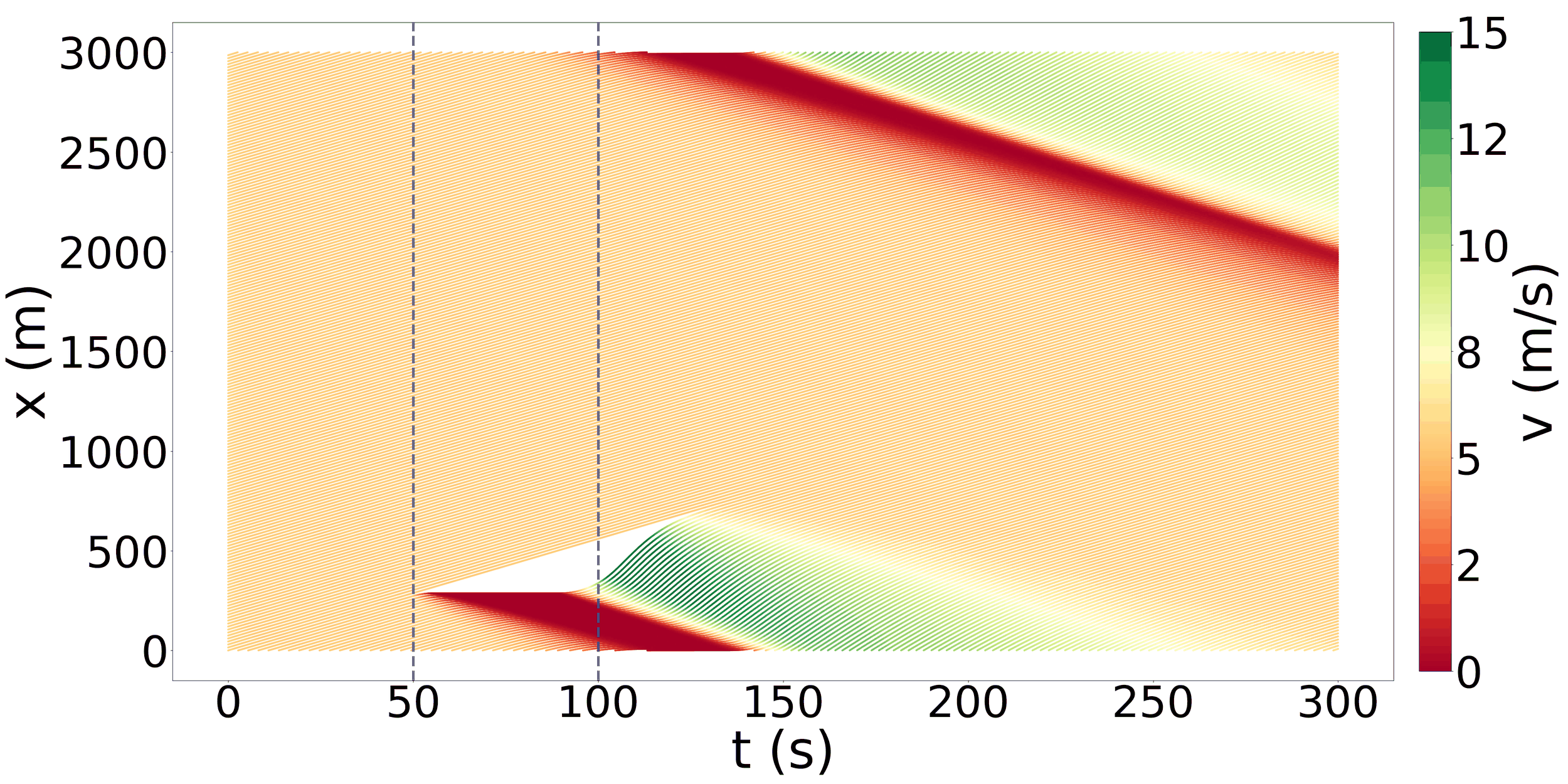}
        \caption{SIDM}
    \end{subfigure}
    \vskip\baselineskip
    \begin{subfigure}[b]{0.48\textwidth}
        \includegraphics[width=\textwidth]{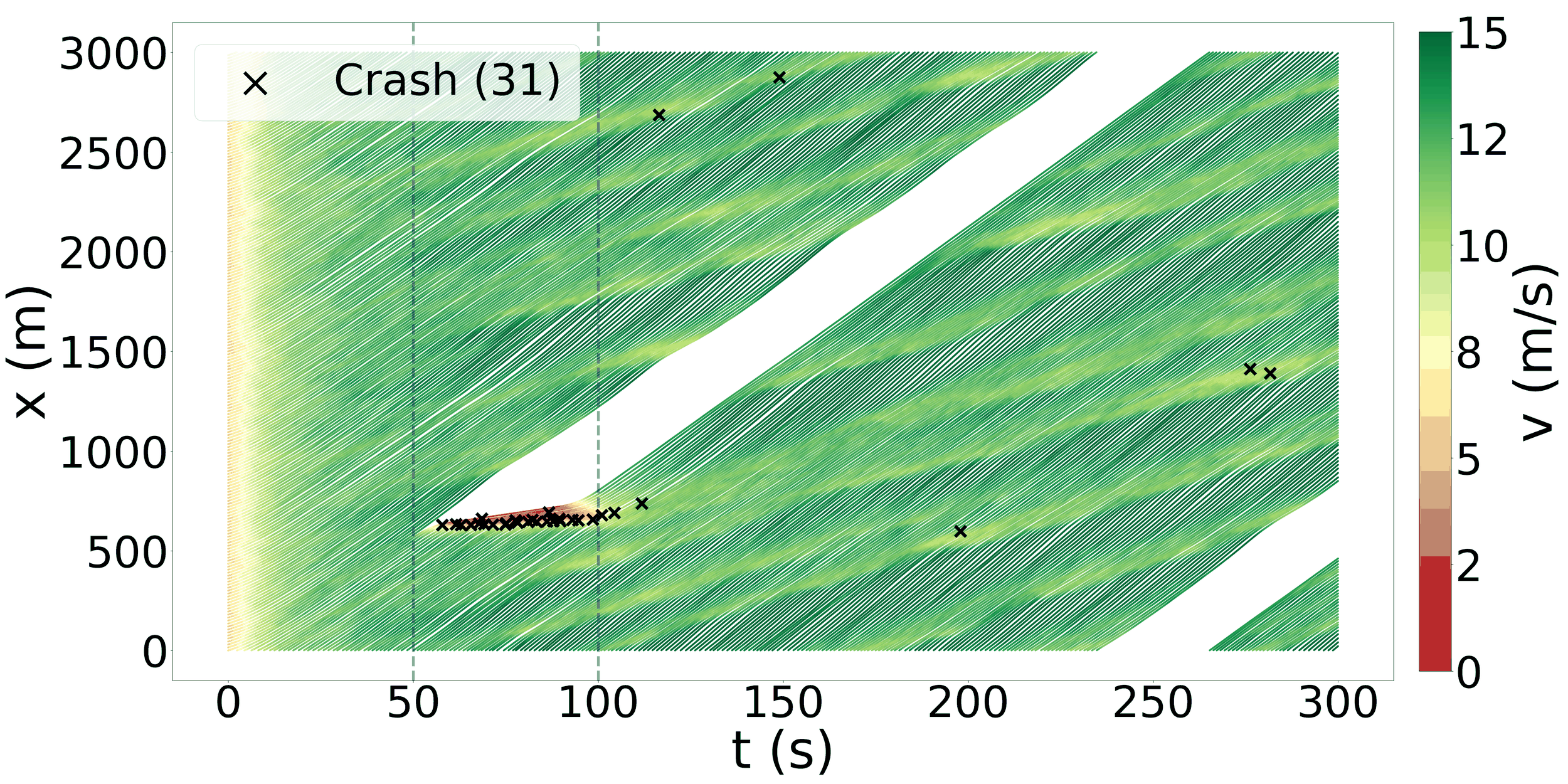}
        \caption{MC-CF (Original)}
    \end{subfigure}
    \hfill
    \begin{subfigure}[b]{0.48\textwidth}
        \includegraphics[width=\textwidth]{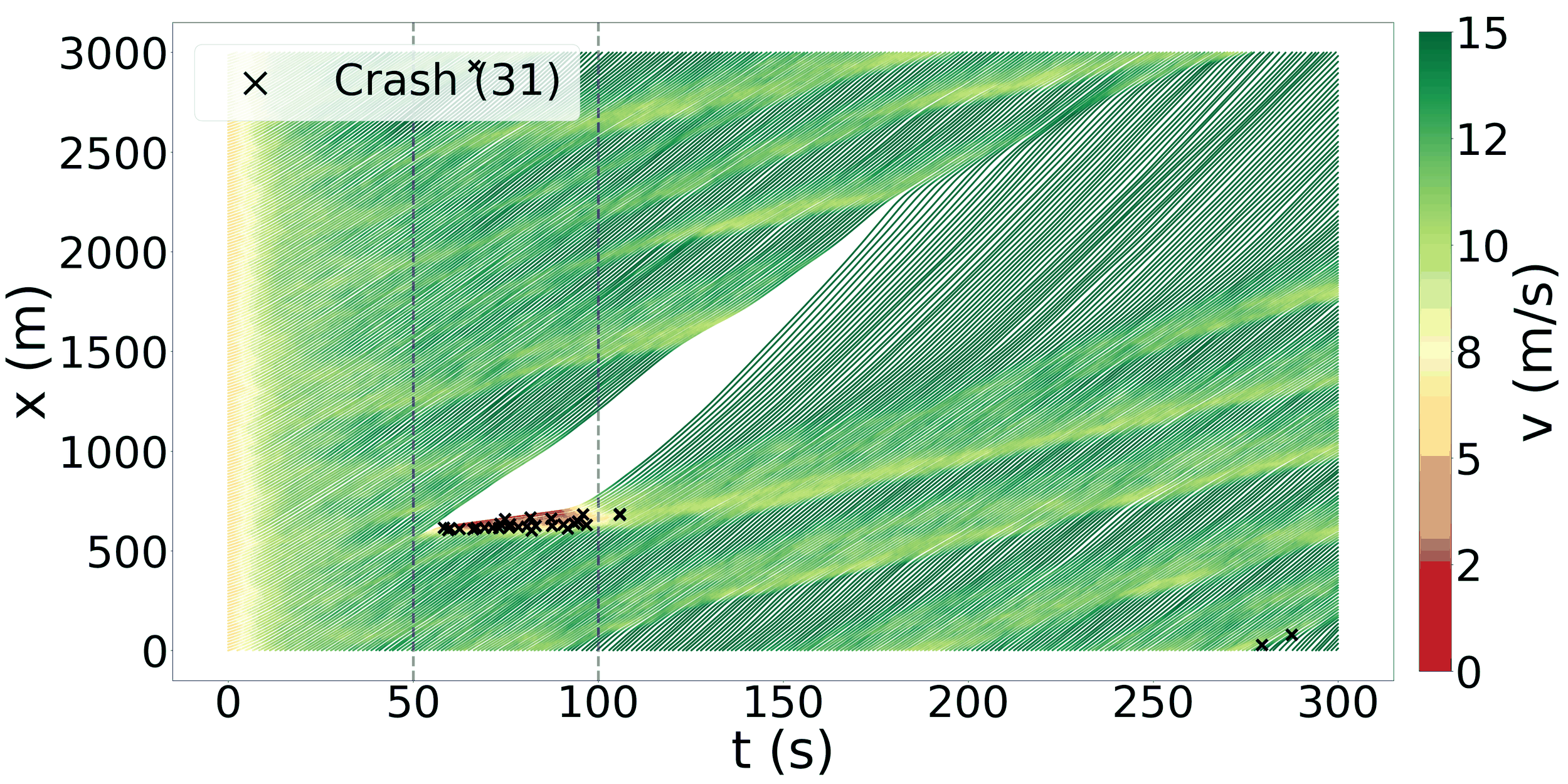}
        \caption{MC-CF (Solo)}
    \end{subfigure}
    \vskip\baselineskip
    \begin{subfigure}[b]{0.48\textwidth}
        \includegraphics[width=\textwidth]{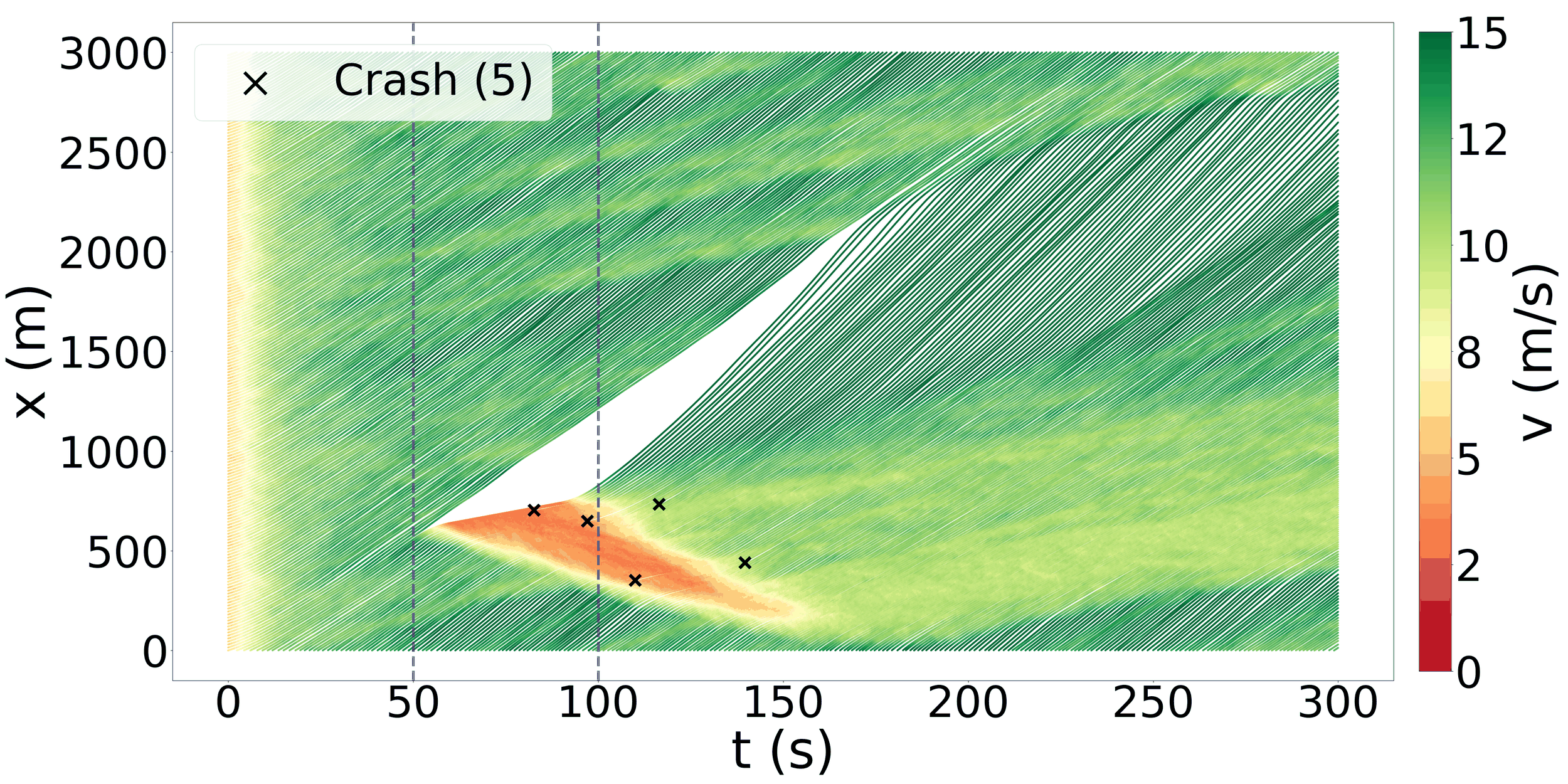}
        \caption{MC-CF (Solo+Conservative)}
    \end{subfigure}
    \hfill
    \begin{subfigure}[b]{0.48\textwidth}
        \includegraphics[width=\textwidth]{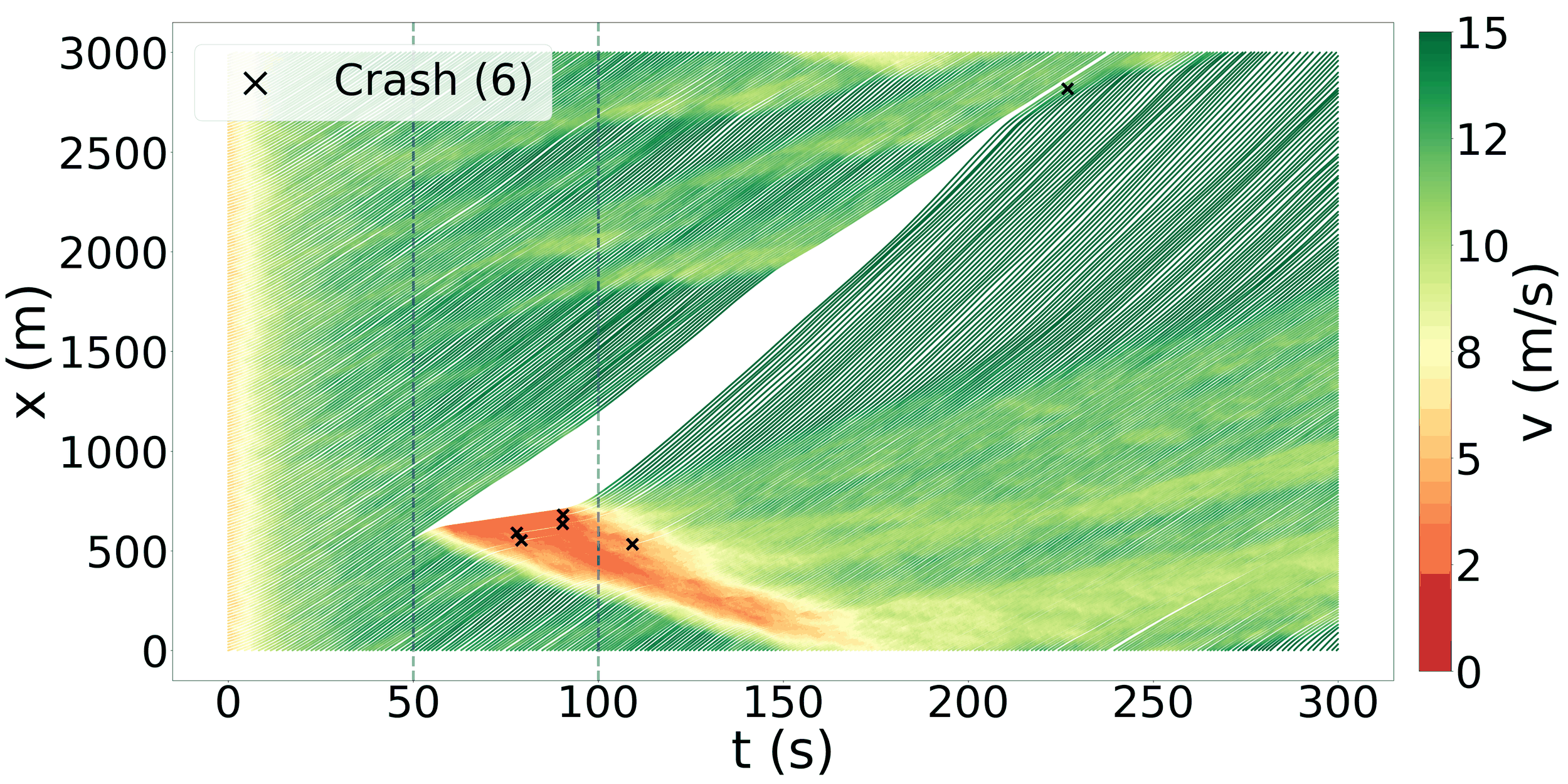}
        \caption{MC-CF (Solo+Conservative+TGSIM)}
    \end{subfigure}
    \caption{Trajectory plots for the Severe Shockwave test. IDM (a) and SIDM (b) show rigid shockwaves, while the MC-CF models with the conservative mode (e-f) display realistic backward and forward wave propagation.}
    \label{fig:sim_sev_sw}
\end{figure}

In the \textbf{Severe Shockwave} experiment (Figure \ref{fig:sim_sev_sw}), the IDM and SIDM produce mathematically perfect backward shockwaves that lack naturalistic variation. The MC-CF (Original) model (Figure \ref{fig:sim_sev_sw}c) fails entirely, resulting in heavy pile-ups ($31$ crashes in the plotted example) and failing once again to reduce the forward gap to the leader. The MC-CF (Solo) model (Figure \ref{fig:sim_sev_sw}d) solves the gap recovery issue but continues to suffer from major pile-ups.

In contrast, both the MC-CF (Solo+Conservative) and MC-CF (Solo+Conservative+TGSIM) models (Figures \ref{fig:sim_sev_sw}e-f) dramatically reduce the number of crashes ($5$ and $6$ crashes in their respective examples) and reproduce highly realistic traffic dynamics. Specifically, they generate a backward-propagating shockwave initiated by the controlled vehicle, followed by a forward-moving shockwave as vehicles dynamically accelerate to close the gap once the leader accelerates.

\begin{figure}[tb!]
    \centering
    \begin{subfigure}[b]{0.48\textwidth}
        \includegraphics[width=\textwidth]{Figures/N200_AV0_Spd100_SevereSW_T0_Solo_conservative.png}
        \caption{Trial 1}
    \end{subfigure}
    \hfill
    \begin{subfigure}[b]{0.48\textwidth}
        \includegraphics[width=\textwidth]{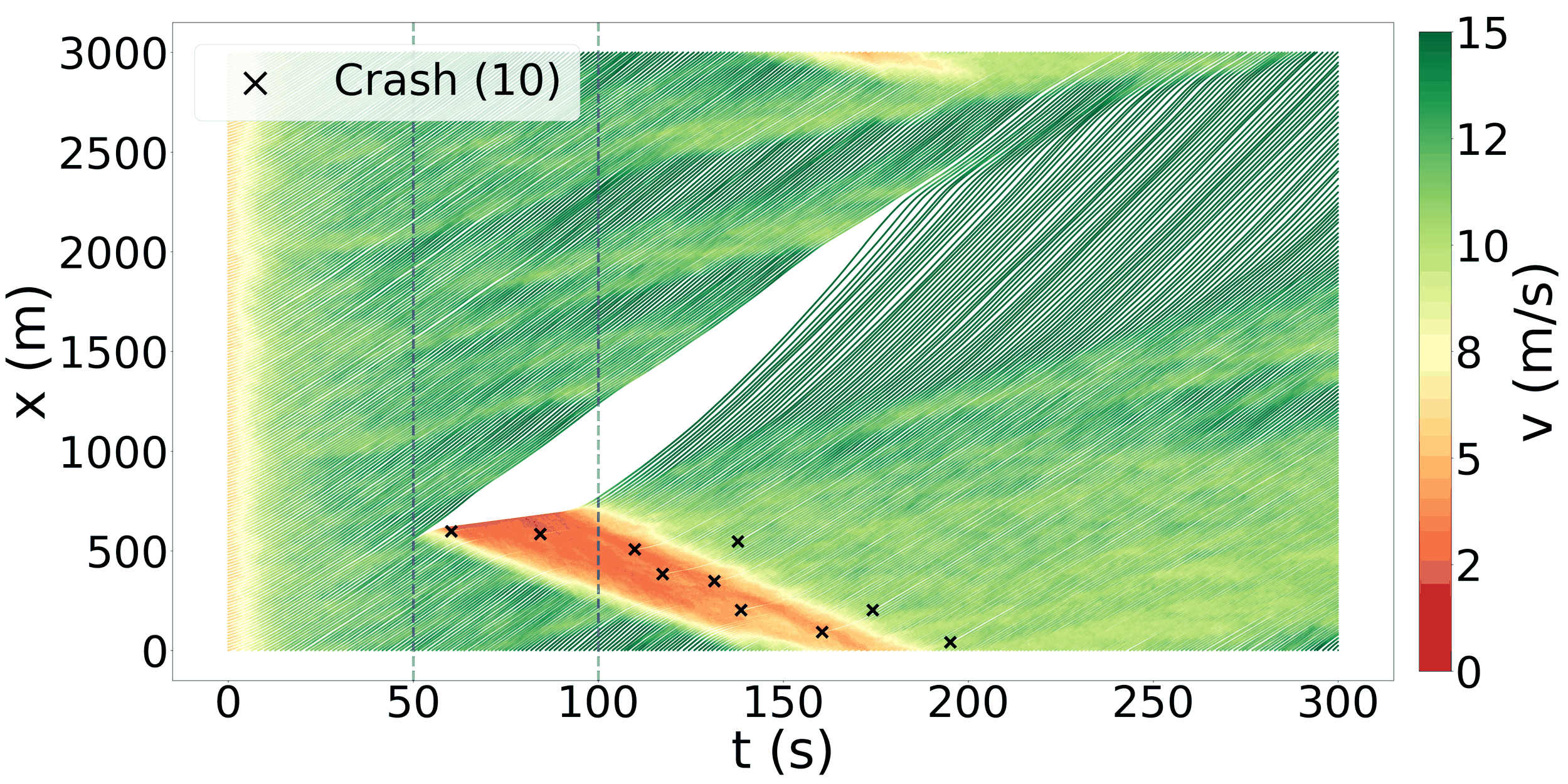}
        \caption{Trial 2}
    \end{subfigure}
    \caption{Two random seed evaluations of the MC-CF (Solo+Conservative) model under the Severe Shockwave test. Despite identical leader perturbations, shockwave propagation exhibits naturalistic, stochastic variability.}
    \label{fig:sim_stoch}
\end{figure}

A key advantage of the MC-CF framework over deterministic models is its ability to capture population-level stochasticity in human driving behavior. Figure \ref{fig:sim_stoch} displays the results of two different random seeds using the MC-CF (Solo+Conservative) model under the exact same severe shockwave perturbation. Visually, the trajectory plots demonstrate that the propagation speed, recovery time, and intensity of the shockwave vary noticeably between trials. This confirms that the model naturally replicates the stochasticity observed in real-world congested traffic, where identical leader behavior can trigger different macroscopic traffic outcomes depending on the probabilistic responses of the following vehicles.

\begin{figure}[tb!]
    \centering
    \begin{subfigure}[b]{0.48\columnwidth}
        \includegraphics[width=\textwidth]{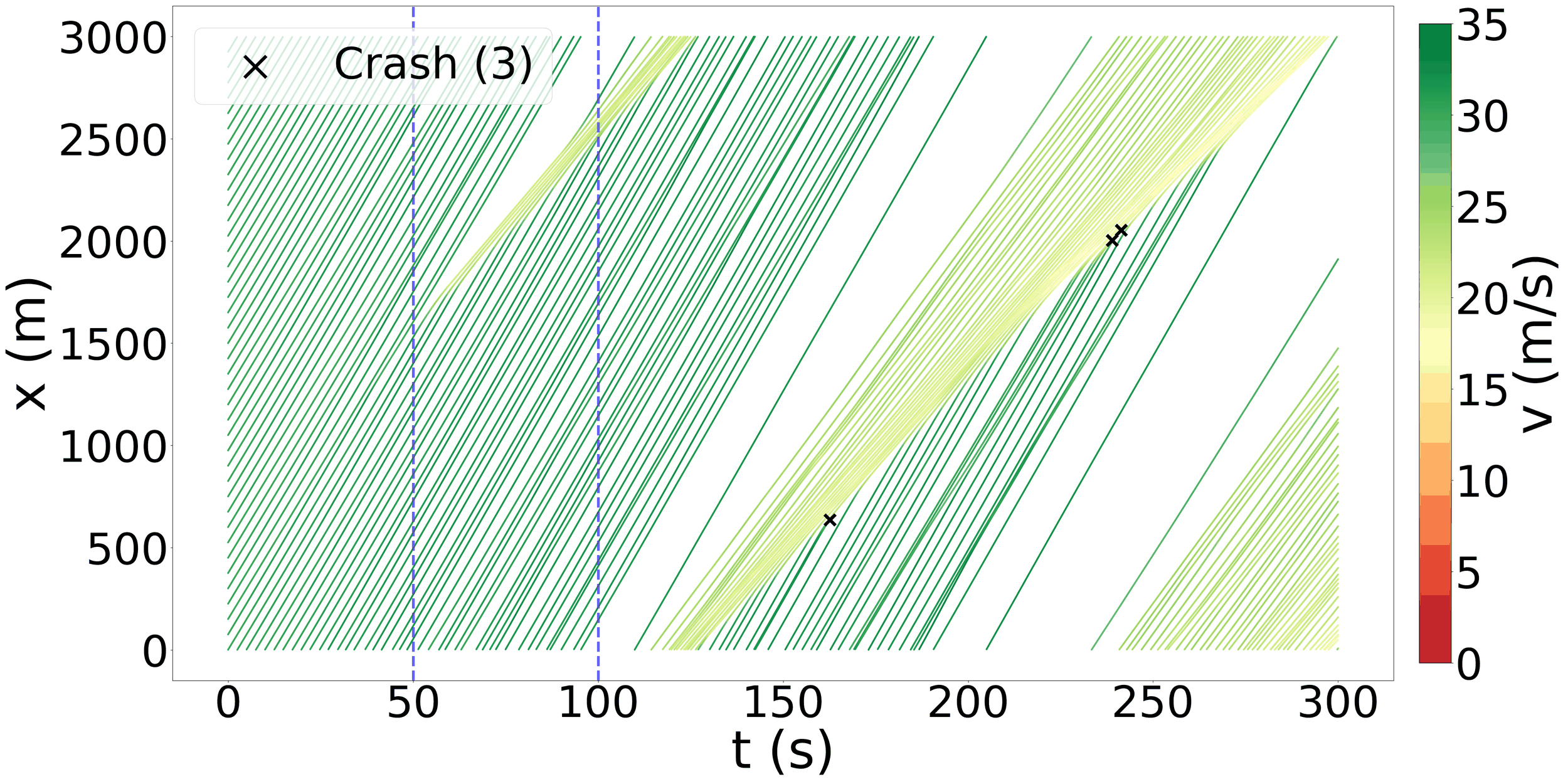}
        \caption{MC-CF (Solo+Conservative)}
    \end{subfigure}
    \hfill
    \begin{subfigure}[b]{0.48\columnwidth}
        \includegraphics[width=\textwidth]{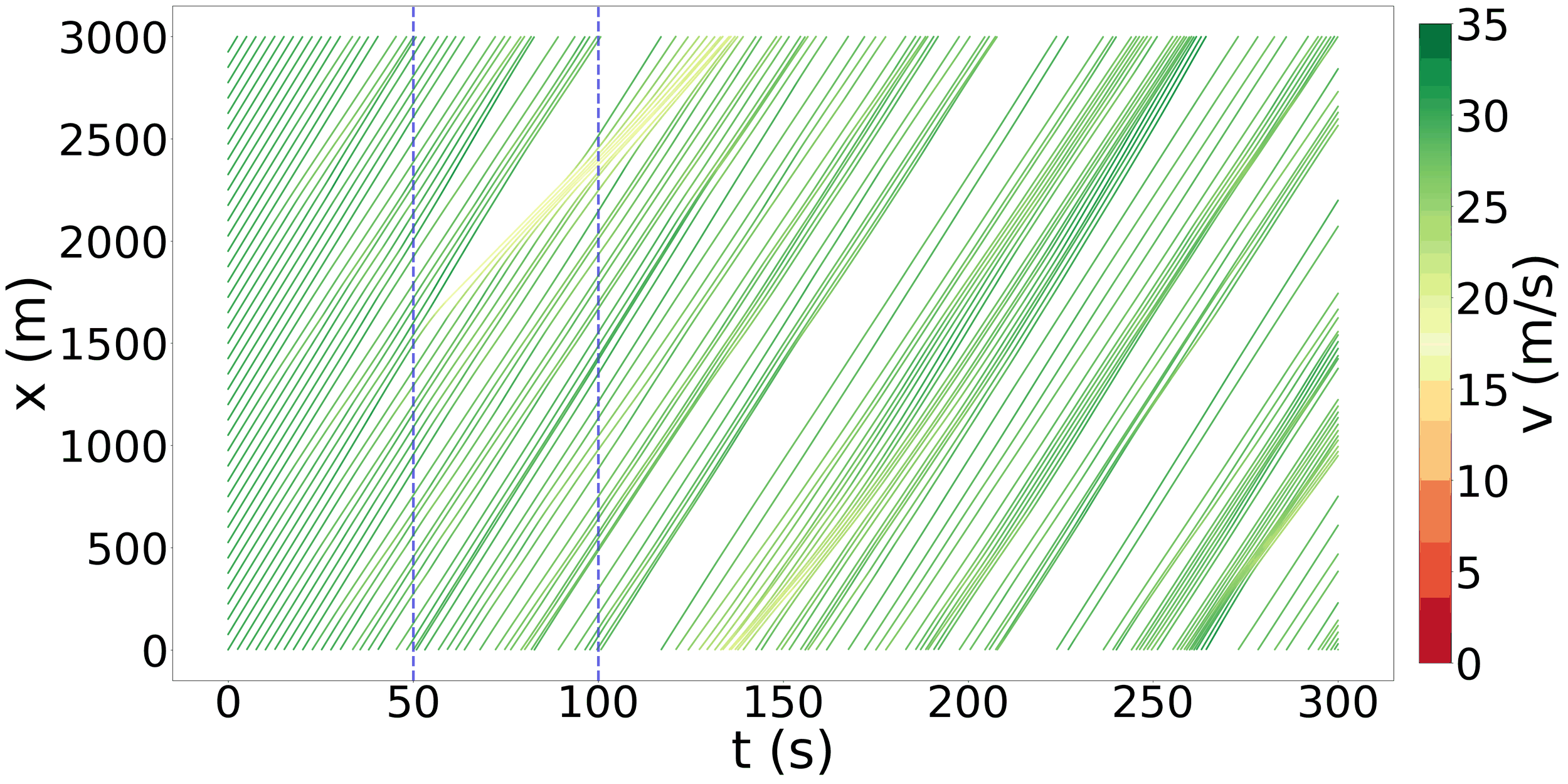}
        \caption{MC-CF (Solo+Cons+TGSIM)}
    \end{subfigure}
    \caption{Trajectory plots for the High-Speed Shockwave test. (a) MC-CF (Solo+Conservative) exhibits phantom braking due to a lack of high-speed training data, while (b) MC-CF (Solo+Conservative+TGSIM) successfully maintains high speeds and forms realistic platoons.}
    \label{fig:sim_hwy}
\end{figure}

Finally, we tested the framework's scalability to high-speed environments through the \textbf{High-Speed Shockwave} test. Because the WOMD dataset lacks high-speed freeway interactions, Figure \ref{fig:sim_hwy}a shows that the MC-CF (Solo+Conservative) model exhibits unrealistic decelerations and phantom braking, resulting in collisions ($3$ crashes in the plotted example). By integrating TGSIM data, the MC-CF (Solo+\allowbreak Conservative+\allowbreak TGSIM) model successfully populates the high-speed state space. As illustrated in Figure \ref{fig:sim_hwy}b, this addition allows vehicles to maintain high speeds reliably, form realistic platoons, and smoothly respond to the shockwave, dropping the crashes to $0$ in the visualized example. This highlights the scalability of the proposed framework, as its performance naturally improves with the incorporation of new and diverse data.

We note that the current variants of the MC-CF model are not perfect, as they still suffer from occasional crashes in the Severe Shockwave and High-Speed Shockwave scenarios. However, the presented framework achieves these realistic traffic dynamics without relying on predefined physical equations, instead sampling accelerations purely from empirical data distributions. As AVs collect increasingly vast, high-fidelity datasets across diverse operational design domains including freeways, the proposed model can organically ingest these new samples into its state bins without requiring structural recalibration. This inherent scalability ensures that the model will continuously improve, generating increasingly robust and realistic trajectories across both urban and freeway scenarios.

% ---------- Conclusion / Build ----------
\section{Conclusion}\label{sec:conclusion}

The enduring study of car-following behavior remains central to transportation engineering, informing critical applications from microscopic simulation to macroscopic traffic flow analysis. The emergence of extensive, high-fidelity datasets like the WOMD has the potential to fundamentally reshape our understanding of these dynamics. Building on this foundation, this paper proposes the MC-CF model as an empirical probabilistic sampling approach, which empirically derives state transitions and acceleration distributions directly from real-world trajectory data, thereby obviating the need for explicit calibration or restrictive parametric assumptions.

Our comprehensive evaluation reveals that the performance of the proposed MC-CF model is differentiated across evaluation regimes. In one-step prediction, data-driven models with access to 1-second trajectory history, namely GRU-reg and BayesGMM, achieve the lowest RMSE for speed and acceleration, reflecting the advantage of recency conditioning for instantaneous prediction. MC-CF (det) achieves competitive one-step performance without relying on any history input, and BayesGMM achieves the best RMSE(s) in the HDV-following-HDV setting. These patterns are consistent across both the WOMD test dataset and the entirely unseen PHX dataset, demonstrating the cross-domain transferability of the proposed approach.

In the open-loop regime, BayesGMM and MC-CF (cons, stoch) exhibit complementary strengths. BayesGMM achieves the best minDTW(s) and minFDE at $K=15$ on the WOMD HDV-following-HDV test set, while MC-CF (cons, stoch) achieves the best minDTW(v) and minADE and maintains an overlapping rate below 1\% across all interaction types. The consistently lower avgADE and avgFDE of MC-CF (stoch) and MC-CF (cons, stoch) compared to BayesGMM suggest that the typical trajectory quality of the MC-CF variants is better than that of BayesGMM.

Transition probability analysis further shows that MC-CF (stoch) and MC-CF (cons, stoch) produce transition dynamics broadly comparable to the ground truth across all three interaction types, with MC-CF (cons, stoch) showing no statistically significant deviation even in the more discriminating HDV-following-HDV case. Together, these results highlight that the MC-CF variants effectively capture the underlying probabilistic structure of real-world driving behavior with comparable accuracy to the best-performing baselines, and remain stable across long simulation horizons.

Beyond trajectory prediction, the proposed framework's inherent scalability allows it to evolve into a robust stochastic microscopic simulation tool. As demonstrated through the ring road ablation study, the straightforward aggregation of diverse data, such as unconstrained free-flow trajectories and high-speed freeway samples from TGSIM, alongside a safety-oriented conservative inference strategy, successfully mitigated the limitations of the baseline model. The enhanced MC-CF framework significantly reduced collision rates across equilibrium, standard shockwave, and high-speed scenarios, and accurately reproduced naturalistic and stochastic shockwave propagation. Nonetheless, an average of 8.05 crashes per simulation remains in the severe shockwave scenario, which represents an acknowledged limitation of the current model. As larger, high-fidelity datasets covering diverse operational design domains become available, the model can be continuously refined by incorporating new data without structural modification.

Future work will also leverage our model's robust representation of driving behavior for probabilistic stability analysis. This is a critical next step, given that many of the existing stability studies rely heavily on calibrated car-following models that we have demonstrated to deviate significantly from empirical ground truth.

% ---------- Declaration of COI ----------
\section*{DECLARATION OF CONFLICTING INTERESTS}
All other authors declare no potential conflicts of interest with respect to the research, authorship, and publication of this article.

\section*{Acknowledgments}
This work was supported by the National Science Foundation under Grant No. 2047937.

\section*{Declaration of Generative AI and AI-assisted Technologies in the Writing Process}
The authors acknowledge the use of AI-assisted tools (such as ChatGPT) for language editing and grammar refinement during manuscript preparation. No AI tool was used for generating novel content, data analysis, or drawing conclusions. All responsibility for the accuracy and integrity of the manuscript remains with the authors.

\section*{AUTHOR CONTRIBUTIONS}
\textbf{Sungyong Chung:} Conceptualization, data curation, formal analysis, investigation, methodology, software, visualization, Writing -- original draft. \textbf{Yanlin Zhang:} Conceptualization, methodology, validation, Writing -- review \& editing. \textbf{Nachuan Li:} Validation, visualization, Writing -- review \& editing. \textbf{Dana Monzer:} Validation, Writing -- review \& editing. \textbf{Alireza Talebpour:} Conceptualization, funding acquisition, methodology, project administration, supervision, validation, Writing -- review \& editing.

\newpage

\bibliographystyle{partc}
\bibliography{Draft}

% ---------- Appendix ----------
\appendix

\section{State Transition Consistency Analysis}\label{app:state_consistency}

An important conceptual question about the MC-CF framework is whether the kinematic update induced by the sampled acceleration actually places the follower vehicle into the sampled next cluster. That is, if the model samples cluster $C_{t+1}$ and then draws acceleration $a_t \sim \mathcal{A}(C_{t+1})$, the resulting vehicle state after the kinematic update may not fall within the boundaries of $C_{t+1}$. To quantify this consistency, we evaluate all transitions in the open-loop prediction trajectories across the WOMD test dataset (three interaction types) and the PHX zero-shot dataset.

For each transition, we record (i) the sampled (intended) next cluster and (ii) the cluster that the realized kinematic state actually maps to via $\mathcal{M}$. We then compute four statistics:
\begin{itemize}
    \item \textbf{Exact Match Rate}: the fraction of transitions where the realized cluster exactly matches the sampled cluster.
    \item \textbf{Near Match Rate}: the fraction of transitions where the realized cluster is either an exact match or a spatial neighbor (i.e., the normalized centroid distance is below 0.05).
    \item \textbf{Mean Centroid Distance}: the average normalized Euclidean distance between the centroids of the realized and sampled clusters, where each state dimension is scaled by its range.
    \item \textbf{Fallback Rate}: the fraction of inference steps that invoke the nearest-neighbor fallback mechanism (i.e., the current state does not directly map to any cluster seen during training).
\end{itemize}

Table~\ref{tab:state_consistency} summarizes these results across all four evaluation settings.

\begin{table}[tb!]
\centering
\caption{State transition consistency analysis: agreement between sampled next cluster and realized kinematic state across evaluation scenarios.}
\vspace{0.2cm}
\label{tab:state_consistency}
\resizebox{\textwidth}{!}{
\begin{tabular}{lrrrrrrrr}
\hline
Scenario & Steps & Pairs & Exact Match (\%) & Near Match (\%) & Mean Dist & Std Dist & $p_{75}$ Dist & $p_{95}$ Dist \\
\hline
AV-following-HDV      & 37,651 & 420 & 44.68 & 99.38 & 0.0116 & 0.0212 & 0.0169 & 0.0408 \\
HDV-following-AV      & 47,272 & 500 & 42.39 & 99.63 & 0.0116 & 0.0183 & 0.0181 & 0.0383 \\
HDV-following-HDV     & 46,106 & 500 & 22.81 & 99.88 & 0.0083 & 0.0115 & 0.0117 & 0.0240 \\
PHX HDV-following-HDV & 24,380 & 106 & 30.37 & 99.81 & 0.0071 & 0.0131 & 0.0092 & 0.0269 \\
\hline
\multicolumn{9}{l}{\footnotesize Fallback rates: AV-following-HDV 38.9\%, HDV-following-AV 42.9\%, HDV-following-HDV 43.4\%, PHX HDV-following-HDV 39.8\%.} \\
\multicolumn{9}{l}{\footnotesize Centroid distances are computed in normalized feature space (each dimension scaled by its range).}
\end{tabular}
}
\end{table}

The exact match rate ranges from 22.8\% (HDV-following-HDV, which has the finest discretization due to its large sample size) to 44.7\% (AV-following-HDV). The lower exact match rate in the HDV-following-HDV case is expected. Because the Freedman--Diaconis rule produces a finer grid for larger datasets (see Table~\ref{tab:fd_bins}), a single acceleration sample is less likely to land the vehicle in precisely the same narrow bin that was sampled. However, the near match rate is consistently 99.4\%--99.9\% across all four scenarios, indicating that the realized state is almost always within immediate spatial proximity of the sampled cluster. The mean normalized centroid distance between the realized and sampled clusters is 0.007--0.012, well within the typical bin width of 0.02--0.13 for each state dimension. Even at the 95th percentile, the centroid distance does not exceed 0.041, confirming that large discrepancies are effectively absent.

The fallback rate of approximately 39\%--43\% reflects the proportion of inference steps during open-loop prediction where the vehicle's state drifts beyond the range of clusters seen in training, triggering the nearest-neighbor mechanism. While this rate may initially appear high, it is a natural consequence of open-loop error accumulation. Once spacing or speed deviates slightly from the training distribution, nearby states are resolved via nearest-neighbor lookup rather than a direct map entry. The negligible centroid distances associated with these fallback steps confirm that the nearest-neighbor assignment does not introduce distributional artifacts; it recovers the closest empirically valid cluster. Taken together, these results validate the conceptual coherence of the MC-CF framework. The acceleration sampled from $\mathcal{A}(C_{t+1})$ almost invariably guides the vehicle into a state that is behaviorally consistent with $C_{t+1}$, even when an exact bin match does not occur.

\section{Baseline Model Calibration and Implementation}\label{app:baseline}
To benchmark the proposed MC-CF model, nine representative baselines are considered: six physics-based car-following models (IDM, SIDM, Van Arem's model, FVDM-CTH, FVDM-Sigmoid, and Gipps) and three data-driven models (FNN-reg, GRU-reg, and BayesGMM). The physics-based models span a diverse range of behavioral formulations from deterministic to stochastic dynamics. The data-driven baselines represent, respectively, an instantaneous-state feedforward regressor, a recurrent model with 1-second trajectory history \cite{wang2017capturing}, and a probabilistic Bayesian Gaussian mixture model \cite{chen2023bayesian}. Together, this collection provides a comprehensive comparison covering interpretability, stochasticity, and temporal conditioning.

\subsection{IDM}
The IDM proposed by \citep{treiber2000congested} specifies acceleration as
\begin{equation}
    a_t = a_{\max} \left[ 1 - \left(\frac{v_t}{v_0}\right)^{\delta} - \left( \frac{s^*(v_t, \Delta v_t)}{d_t} \right)^2 \right],
\end{equation}

\noindent where \(v_0\) is the desired speed, \(a_{\max}\) is the maximum acceleration, \(\delta\) is an acceleration exponent, \(\Delta v_t\) is the follower speed ($v_t$) minus the leader speed ($v_t^{lead}$), and \(d_t\) is the spacing.
The desired gap function is
\begin{equation}
    s^*(v_t, \Delta v_t) = s_0 + v_t T + \frac{v_t \, \Delta v_t}{2\sqrt{a_{\max} b}},
\end{equation}

\noindent with \(s_0\) the minimum gap, \(T\) the desired time headway, and \(b\) the comfortable deceleration.

\subsection{SIDM}

To capture oscillatory traffic dynamics, Treiber and Kesting~\cite{treiber2017intelligent} extended the IDM by introducing stochastic fluctuations (SIDM):
\begin{equation}
    a_t = a^{\text{IDM}}_t + \sigma \,\xi_t,
\end{equation}

\noindent where \(a^{\text{IDM}}_t\) is the deterministic IDM acceleration, \(\xi_t\) is a Gaussian random variable with zero mean and unit variance, and \(\sigma\) controls the strength of stochasticity.
This variant reflects heterogeneity in driver responses and reproduces traffic flow instabilities.

\subsection{Van Arem Model}

Van Arem et al.~\cite{van2006impact} proposed a control-oriented formulation for cooperative adaptive cruise control systems. The acceleration of the follower is determined as the minimum of a speed-based demand ($a_{t\_v}$) and a distance/speed-based demand ($a_{t\_d}$):
\begin{equation}
a_{t} = \min(a_{t\_v}, a_{t\_d}).
\end{equation}

The speed-based demand is given by
\begin{equation}
a_{t\_v} = k \cdot (v^{\text{int}} - v_t),
\end{equation}
\noindent where $v^{\text{int}}$ is the driver's intended speed and $k$ is a positive feedback gain.

The distance/speed-based demand is given by
\begin{equation}
a_{t\_d} = k_a a_t^{lead} - k_v \,\Delta v_t + k_d \,(d_t - d^{\text{ref}}_t),
\end{equation}

\noindent where \(a^{\text{lead}}_t\) is the acceleration of the leader, \(d^{\text{ref}}_t\) is a dynamic desired spacing derived from time headway and safety considerations, and \(k_a, k_v, k_d\) are positive feedback gains. The reference clearance $d^{\text{ref}}_t$ is defined as the maximum of three values:
\begin{equation}
d^{\text{ref}}_t = \max(r_{safe}, r_{system}, r_{min}),
\end{equation}
\noindent where $r_{min}$ is the minimum spacing.

The safe following distance ($r_{safe}$) is computed based on the speed of the follower and the deceleration capabilities of leader ($d_p$) and follower ($d$):
\begin{equation}
r_{safe} = \frac{(\Delta v_t)^2}{2} \left(\frac{1}{d_p} - \frac{1}{d}\right),
\end{equation}

\noindent and the system time-gap distance ($r_{system}$) is a function of the ego vehicle's speed and a system time headway setting $t_{system}$:
\begin{equation}
r_{system} = t_{system} \cdot v_t.
\end{equation}

\subsection{FVDM-CTH}

The FVDM by \citep{jiang2001full} combines a desired velocity rule with a velocity difference term.
In the constant time headway variant proposed by \citep{punzo2021calibration}, namely FVDM-CTH, the acceleration is
\begin{equation}
    a_t = K_1 \big( V(d_t) - v_t \big) - K_2 \,\Delta v_t,
\end{equation}

\noindent where \(K_1\) and \(K_2\) are positive feedback gains.
The desired velocity is given by
\begin{equation}
    V(d_t) =
\begin{cases}
0, & d_t \leq s_0, \\
\min\!\left( V_{\max}, \tfrac{d_t - s_0}{T} \right), & d_t > s_0,
\end{cases}
\end{equation}

\noindent with \(s_0\) the minimum spacing, \(T\) the desired time headway, and \(V_{\max}\) the maximum speed.

\subsection{FVDM-Sigmoid}

A smooth alternative, namely FVDM-Sigmoid, is to represent the desired velocity with a sigmoid profile \cite{punzo2021calibration}:
\begin{equation}
V(d_t) = \left\{\,
\begin{array}{@{}ll@{}}
0, & d_t \leq s_0, \\[4pt]
\multicolumn{2}{@{}l@{}}{\tfrac{V_{\max}}{2}
  \left[1 - \cos\!\left(\tfrac{\pi(d_t - s_0)}{T V_{\max}}\right)\right],} \\[2pt]
& s_0 < d_t < s_0 + T V_{\max}, \\[4pt]
V_{\max}, & d_t \geq s_0 + T V_{\max}.
\end{array}
\right.
\end{equation}

The acceleration structure remains the same as in the CTH formulation, with the sigmoid desired speed replacing the piecewise linear function.

\subsection{Gipps}

Finally, the Gipps model~\cite{gipps1981behavioural} balances free-flow acceleration with collision-avoidance constraints. The follower's speed at the next time step is determined by
\begin{equation}
\begin{aligned}
v_{t+1} &= \min\biggl(
v_t + 2.5\,a_{\max}\tau
\bigl(1 - \tfrac{v_t}{V_{\max}}\bigr)\\
&\qquad\quad \sqrt{0.025 + \tfrac{v_t}{V_{\max}}},\\
&\quad -b\bigl(\tfrac{\tau}{2} + \theta\bigr)
+ \Bigl(b^2\bigl(\tfrac{\tau}{2} + \theta\bigr)^{2}\\
&\qquad + b\Big[2(d_t - s_0) - \tau v_t
+ \tfrac{(v_t^{\mathrm{lead}})^2}{\hat{b}}\Big]\Bigr)^{\!1/2}
\biggr)
\end{aligned}
\end{equation}

\noindent where \(a_{\max}\) denotes the maximum acceleration, \(b\) is the comfortable deceleration, \(\tau\) is the reaction time, \(\theta\) is an additional anticipation parameter, \(s_0\) is the minimum spacing, \(V_{\max}\) is the desired speed, and \(\hat b\) represents the expected deceleration capability of the leader vehicle.
The resulting acceleration is approximated as \(a_t = (v_{t+1} - v_t)/\tau\).

\subsection{Data-Driven Baselines: FNN-reg, GRU-reg, and BayesGMM}\label{app:dd_baselines}

To provide a broader comparison against modern data-driven methods, three additional baselines are included alongside the six physics-based models.

\paragraph{FNN-reg.}
The feedforward neural network regressor (FNN-reg) is a three-layer multilayer perceptron with architecture $3 \to 20 \to 20 \to 1$ (ReLU activations), trained to directly predict the follower acceleration from the instantaneous kinematic state $(v_t, \Delta v_t, d_t)$ via mean squared error (MSE) regression. Inputs are min-max normalized to the 1st--99th percentile range of the training data. The model is trained for 500 epochs using the AdamW optimizer (learning rate $5 \times 10^{-3}$, weight decay $10^{-2}$) with mini-batch sampling of 512 samples, yielding approximately 521 trainable parameters. FNN-reg represents the simplest data-driven baseline: a compact, instantaneous-state regressor with no temporal context.

\paragraph{GRU-reg.}
The gated recurrent unit regressor (GRU-reg) follows the architecture proposed by \citep{wang2017capturing}: a single-layer GRU with 32 hidden units processes a 10-step (1-second) normalized history of $(v_t, \Delta v_t, d_t)$, and the final hidden state is projected to a scalar acceleration prediction via a linear head, yielding approximately 3,585 trainable parameters. The model is trained with the same AdamW setup as FNN-reg, sampling sequences with full real history from training pairs (pairs shorter than 10 steps are excluded). GRU-reg captures short-term temporal dependencies---a signal unavailable to instantaneous-state models---and consistently achieves the lowest one-step RMSE for speed and acceleration across interaction types. However, this history conditioning advantage does not translate to open-loop stability: autoregressive rollout compounding errors over long horizons leads to substantially elevated collision rates compared to MC-CF and physics-based models.

\paragraph{BayesGMM.}
The Bayesian Gaussian Mixture Model (BayesGMM) follows \citep{chen2023bayesian}. The model represents the joint distribution of a $T_{\text{past}}=10$-step kinematic history and a $P_{\text{future}}=3$-step prediction horizon as an augmented variable $\mathbf{x} \in \mathbb{R}^{(T_{\text{past}}+P_{\text{future}}) \times 3}$, with each row containing $(v, \Delta v, d)$ normalized to zero mean and unit variance. A $K=60$-component Gaussian mixture is fitted using a Gibbs sampler under a Gaussian-inverse-Wishart prior ($\alpha_0 = 1/K$, $\nu_0 = 39 + 10$, $\lambda_0 = 10^{-3}$), run for 200 burn-in and 100 collecting iterations. To keep the Gibbs sampler tractable, the training set is capped at 50,000 subsampled observations. At inference, the posterior predictive distribution over the future block is obtained by (i) drawing one posterior sample at random from the $Q_2=100$ collected samples, (ii) computing component responsibilities from the past-block likelihood, and (iii) sampling from the resulting conditional Gaussian. The implied acceleration is derived from the predicted next-step speed. BayesGMM achieves the best one-step RMSE for spacing and competitive one-step accuracy overall, and produces zero overlapping trajectories ($OR=0$). It also achieves strong oracle open-loop metrics at $K=15$ (best $minDTW(s)$ and $minFDE$ in the HDV-following-HDV setting). However, the independently drawn step-level samples under autoregressive rollout result in by far the highest jerk (see Appendix~\ref{app:jerk}), and the $avgADE$ and $avgFDE$ at $K=15$ are substantially higher than those of MC-CF variants (e.g., $avgADE$ of 3.89 vs. 1.56 for MC-CF (cons, stoch)), indicating that typical trajectory quality is considerably lower despite competitive best-case performance.

\subsection{Calibration Procedure}

The parameters for all six physics-based car-following models were calibrated using a trajectory-based optimization approach. This method minimizes the RMSE between the simulated follower speed ($v_{\text{sim}}$) and the actual follower speed ($v_{\text{true}}$) over the entire time series for all segments in the training dataset. The objective function for calibration is:

\begin{equation}
\begin{aligned}
\text{Minimize }& \text{RMSE}_v \\
={}& \sqrt{
  \tfrac{1}{\sum_{j=1}^{M} T_j}
  \sum_{j=1}^{M}\sum_{t=1}^{T_j}
  (v_{\text{sim}, j, t} - v_{\text{true}, j, t})^{2}
}
\end{aligned}
\end{equation}

\noindent where $M$ is the total number of distinct car-following pairs in the training dataset, $T_j$ is the number of time steps of the $j$-th car-following pair, and $v_{\text{sim}, j, t}$ is the simulated speed of the follower in pair $j$ at time $t$. The minimization is performed by simulating the follower's trajectory over the entire trajectory using only the true initial conditions (position and speed) of the follower vehicle, allowing simulation errors to propagate throughout the time series.

The optimization was performed using the Differential Evolution (DE) algorithm \cite{storn1997differential} to effectively explore the parameter space. Specifically, the DE algorithm was implemented using the SciPy optimization library (\texttt{scipy.optimize.differential\_evolution}). The DE parameters were configured as follows: strategy = \texttt{best1bin}, population size = 15, mutation factor in \([0.5, 1.0]\), recombination probability = 0.7, maximum iterations = 50, and convergence tolerance = 0.01. The calculated acceleration for all models was constrained within a practical range of \([-10, 5]~\text{m/s}^2\) to reflect real-world vehicle limitations. Table~\ref{tab:parameter_bounds} summarizes the parameter bounds used for each model. This rigorous setup ensures a fair and systematic baseline comparison against the proposed MC-CF models.

\begin{table}[tb!]
    \centering
    \caption{Parameter Bounds Applied for Baseline Models Calibration.}
    \vspace{0.2cm}
    \label{tab:parameter_bounds}
    \small
    \begin{tabular}{l l c c c c}
    \hline
     Model & Parameter & Symbol & Unit & [ & ] \\
    \hline
    IDM & Desired Speed & $v_0$ & m/s & 5.0 & 50.0 \\
    & Desired Time Headway & $T$ & s & 0.5 & 3.0 \\
    & Maximum Acceleration & $a_{\max}$ & m/s$^2$ & 0.1 & 5.0 \\
    & Comfortable Decel. & $b$ & m/s$^2$ & 0.1 & 10.0 \\
    & Minimum Spacing & $s_0$ & m & 0.5 & 10.0 \\
    & Acceleration Exponent & $\delta$ & --- & 1.0 & 10.0 \\
    \cline{1-6}
    SIDM & IDM Parameters & --- & --- & --- & --- \\
    & Noise Std Dev & $\sigma$ & m/s$^2$ & 0.01 & 2.0 \\
    \cline{1-6}
    Van Arem & Acceleration Gain & $k_a$ & --- & 0.1 & 5.0 \\
    & Speed Gain & $k_v$ & s$^{-1}$ & 0.1 & 5.0 \\
    & Spacing Gain & $k_d$ & s$^{-2}$ & 0.1 & 5.0 \\
    & System Time Headway & $t_{\text{system}}$ & s & 0.5 & 3.0 \\
    & Intended Speed & $v_{\text{int}}$ & m/s & 5.0 & 50.0 \\
    & Minimum Spacing & $r_{\min}$ & m & 0.1 & 5.0 \\
    & Leader Decel. Capability & $d_p$ & m/s$^2$ & 0.1 & 10.0 \\
    & Follower Decel. Capability & $d$ & m/s$^2$ & 0.1 & 10.0 \\
    & Intended Speed Gain & $k$ & --- & 0.1 & 1.0 \\
    \cline{1-6}
    FVDM-CTH & Spacing Gain & $K_1$ & s$^{-2}$ & 0.1 & 5.0 \\
    & Velocity Gain & $K_2$ & s$^{-1}$ & 0.1 & 5.0 \\
    & Minimum Spacing & $s_0$ & m & 0.1 & 10.0 \\
    & Desired Time Headway & $T$ & s & 0.5 & 3.0 \\
    & Maximum Speed & $V_{\max}$ & m/s & 5.0 & 50.0 \\
    \cline{1-6}
    FVDM-Sigmoid & Same as FVDM-CTH & --- & --- & --- & --- \\
    \cline{1-6}
    Gipps & Maximum Acceleration & $a_{\max}$ & m/s$^2$ & 0.5 & 3.0 \\
    & Comfortable Deceleration & $b$ & m/s$^2$ & 1.0 & 4.0 \\
    & Reaction Time & $\tau$ & s & 0.1 & 1.5 \\
    & Anticipation Time & $\theta$ & s & 0.3 & 1.0 \\
    & Minimum Spacing & $s_0$ & m & 0.1 & 10.0 \\
    & Desired Speed & $V_{\max}$ & m/s & 5.0 & 50.0 \\
    & Expected Leader Decel. & $\hat{b}$ & m/s$^2$ & 2.0 & 5.0 \\
    \hline
    \end{tabular}
\end{table}

\section{Evaluation Metrics}\label{app:metrics}

Model evaluation is conducted in two categories: one-step prediction and open-loop prediction. For the one-step prediction, we compute the RMSE for spacing ($RMSE(s)$), speed ($RMSE(v)$), and acceleration ($RMSE(a)$). These metrics quantify how accurately each model estimates the instantaneous spacing, speed, and acceleration of the follower vehicle. Given that each car-following pair contains, on average, more than 90 time steps with 420, 544, and 3,276 test pairs for AV-following-HDV, HDV-following-AV, and HDV-following-HDV cases, respectively (as shown in Table \ref{tab:trajectory_seconds_WOMD}), the stochastic models are evaluated with $K=1$ for the one-step prediction analysis. This provides a robust measure of their instantaneous predictive capability, allowing a direct comparison with the deterministic models.

Open-loop prediction, on the other hand, assesses long-term trajectory consistency and spatial accuracy. For stochastic models, evaluation is based on the best-matching realization among the $K$ generated trajectories, acknowledging that multiple follower trajectories may be plausible given the same leader trajectory in real-world driving.

We employ seven complementary open-loop metrics, each capturing different aspects of trajectory quality. For a fair comparison between models, all open-loop metrics other than the overlapping rate ($OR$) are computed using the car-following pairs without overlapping only. Specifically, to ensure valid pairwise comparison, given that shorter predicted trajectories starting from the true initial point will naturally be closer to the ground truth, the evaluation set includes only those car-following pairs in the test dataset that (i) do not result in a crash for any deterministic model, (ii) have at least one non-crashing trajectory among the $K$ SIDM trajectories, and (iii) have at least one non-crashing trajectory among the $K$ MC-CF (stoch) trajectories. As a result, all open-loop metrics are computed over the same time horizon for all models, ensuring a fair comparison.

First, the dynamic time warping (DTW) distance computes the minimal cumulative cost required to align two sequences by allowing non-linear time stretching. DTW is particularly effective for time series that are out of phase, as it warps the time axis to find an optimal alignment between two sequences \(X=[x_1, x_2, \ldots, x_m]\) and \(Y=[y_1, y_2, \ldots, y_n]\) \cite{zhang2024characterizing}. This capability is crucial in car-following scenarios where minor temporal shifts in driver behavior or model predictions can lead to large Euclidean distances, even if the overall shape of the trajectory is similar. The DTW distance between \(X\) and \(Y\) is given by
\begin{equation}
DTW(X,Y) \;=\; \min_{W} \sum_{(i,j)\in W} d(x_i,y_j),
\end{equation}
\noindent where \(W=[(i_1,j_1),\ldots,(i_M,j_M)]\) is a warping path that maps elements of \(X\) to elements of \(Y\), and \(d(\cdot,\cdot)\) denotes the local cost, taken in this study as squared Euclidean distance.

We compute DTW for each of the $K'$ $(\leq K)$ generated non-crashing trajectories and then take the best match realization among them, where $K'=1$ for deterministic models. The minimum DTW is calculated separately for spacing and speed, denoted as $minDTW(s)$ and $minDTW(v)$, respectively.

For spacing, let the ground truth spacing sequence be \(s=[s_1,\ldots,s_T]\) and the \(k^{\mathrm{th}}\) predicted spacing be \(\hat{s}_{k}=[\hat{s}_{1,k},\ldots,\hat{s}_{T,k}]\).
The DTW cost for the \(k^{\mathrm{th}}\) sample is
\begin{equation}
DTW_k\big(\hat{s}_{k},s\big) \;=\; \min_{W} \sum_{(i,j)\in W} d\big(\hat{s}_{i,k},s_j\big),
\end{equation}
\noindent and the minimum DTW for spacing is then defined as
\begin{equation}
minDTW(s) \;=\; \min_{k\in\{1,\ldots,K'\}} DTW_k\big(\hat{s}_{:,k},s\big).
\end{equation}

Similarly, for speed, let the ground truth speed be \(v=[v_1,\ldots,v_T]\) and the \(k^{\mathrm{th}}\) predicted speed be \(\hat{v}_{k}=[\hat{v}_{1,k},\ldots,\hat{v}_{T,k}]\).
We compute
\begin{equation}
minDTW(v) \;=\; \min_{k\in\{1,\ldots,K'\}} DTW_k\big(\hat{v}_{k},v\big).
\end{equation}

The third open-loop prediction metric, the minimum average displacement error (minADE), quantifies the mean spatial deviation between predicted and ground truth trajectories while accounting for stochastic predictions. For each non-crashing trajectory sample $k \in \{1, \ldots, K'\}$, the average displacement error (ADE) is computed as
\begin{equation}
ADE_k = \frac{1}{T} \sum_{t=1}^{T} \lVert \hat{y}_{t,k} - y_t \rVert_2,
\end{equation}
\noindent where $y_t$ denotes the ground truth position of the follower vehicle at time step $t$, $\hat{y}_{t,k}$ represents the predicted position at time step $t$ for the $k^{th}$ non-crashing trajectory, and $T$ represents the total number of time steps in the trajectory. For each car-following pair, the minADE across $K'$ non-crashing trajectories is then taken as
\begin{equation}
minADE = \min_{k \in \{1, \ldots, K'\}} ADE_k.
\end{equation}

Next, the minimum final displacement error ($minFDE$) measures the smallest final position error among the $K'$ non-crashing predicted trajectories:
\begin{equation}
minFDE = \min_{k \in \{1, \ldots, K'\}} \lVert \hat{y}_{T,k} - y_T \rVert_2.
\end{equation}

To complement the oracle min-based metrics, we also report the average ADE ($avgADE$) and average FDE ($avgFDE$), which average performance over all $K'$ non-crashing trajectories rather than selecting the best one:
\begin{equation}
avgADE = \frac{1}{K'} \sum_{k=1}^{K'} ADE_k, 
\end{equation}

\begin{equation}
avgFDE = \frac{1}{K'} \sum_{k=1}^{K'} \lVert \hat{y}_{T,k} - y_T \rVert_2.
\end{equation}

These non-oracle metrics provide a direct measure of average predictive accuracy and penalize models that generate a wide spread of trajectories with only occasional near-ground-truth samples. For deterministic models ($K'=1$), $avgADE$ and $avgFDE$ equal $minADE$ and $minFDE$, respectively. For stochastic models, they are reported at $K=15$.

Finally, the $OR$ evaluates safety-related consistency by computing the ratio of collision events to the total number of car-following pairs in the test dataset. A collision is counted when the follower's predicted position overlaps with the leader's position at any time step. To ensure a fair comparison between deterministic and stochastic models, the $OR$ is computed using a single generated trajectory ($K=1$) for each car-following pair. Given the large size of the test dataset as shown in Table \ref{tab:trajectory_seconds_WOMD}, using one realization for stochastic models still gives a statistically meaningful estimate of their safety, indicating how often the model produces collisions across the full dataset.

Together, the three one-step prediction metrics and the seven open-loop prediction metrics provide a comprehensive evaluation framework that captures instantaneous accuracy, long-horizon trajectory fidelity, average predictive quality, and simulation safety, allowing fair comparison across deterministic and stochastic car-following models.

% CASE 2: BiBTeX used to generate mypaper.bbl (to be further fine tuned)
%\input{mypaper.bbl} % outcomment this line in Case 2

%If you don't use BiBTex, you can manually itemize references as shown below.

\section{Trajectory Prediction Results: AV-Following-HDV and HDV-Following-AV}\label{app:av_results}

Tables~\ref{tab:AV_following_HDV_results} and~\ref{tab:HDV_following_AV_results} present trajectory prediction results for the AV-following-HDV and HDV-following-AV interaction types, respectively. The overall performance patterns are consistent with those reported in Table~\ref{tab:HDV_following_HDV_results}. GRU-reg achieves the best one-step $RMSE(v)$ and $RMSE(a)$ but at the cost of high open-loop collision rates, and as a deterministic model it performs worse than stochastic models in various open-loop accuracy metrics including $avgADE$ and $avgFDE$. BayesGMM achieves the best $minDTW(s)$, $minDTW(v)$, and $minFDE$ at $K=15$, while MC-CF (stoch) and MC-CF (cons, stoch) achieve comparable open-loop prediction accuracy with notably lower $avgADE$ and $avgFDE$ compared to BayesGMM. These consistent trends across interaction types support the generalizability of the MC-CF framework beyond the HDV-HDV setting.

\begin{table}[tb!]
\centering
\caption{Trajectory prediction performance for AV-following-HDV.}
\vspace{0.2cm}
\label{tab:AV_following_HDV_results}
\resizebox{\textwidth}{!}{
\begin{tabular}{lcccccccccc}
\hline
Model & \multicolumn{3}{c}{One-Step Prediction} & \multicolumn{7}{c}{Open-Loop Prediction} \\
 & $RMSE(s)$ & $RMSE(v)$ & $RMSE(a)$ & $minDTW(s)$ & $minDTW(v)$ & $minADE$ & $minFDE$ & $avgADE$ & $avgFDE$ & $OR$\\
\hline
\multicolumn{11}{l}{Deterministic} \\
IDM          & 0.0182 & 0.1576 & 1.6521 & 13.0913 & 3.8395 & 1.5523 & 3.3610 & 1.5523 & \textbf{3.3610*} & 0.0024 \\
Van-Arem     & 0.0172 & 0.0988 & 0.9839 & 14.0582 & 4.0063 & 1.6844 & 3.7955 & 1.6844 & 3.7955           & 0.0071 \\
FVDM-CTH     & 0.0170 & 0.0831 & 0.8190 & 12.6668 & 3.7184 & 1.4810 & 3.5016 & \textbf{1.4810*} & 3.5016 & 0.0048 \\
FVDM-Sigmoid & 0.0170 & 0.0839 & 0.8269 & 12.9033 & 3.7478 & 1.5131 & 3.5581 & 1.5131 & 3.5581           & 0.0048 \\
Gipps        & 0.0171 & 0.0945 & 0.9357 & 14.8513 & 3.9633 & 1.8108 & 3.7799 & 1.8108 & 3.7799           & 0.0024 \\
FNN-reg      & 0.0167 & 0.0613 & 0.5792 & 14.5544 & 4.4330 & 1.5635 & 4.1088 & 1.5635 & 4.1088           & 0.0190 \\
GRU-reg      & 0.0179 & \textbf{0.0335*} & \textbf{0.1498*} & 23.5561 & 8.6039 & 2.2154 & 6.8037 & 2.2154 & 6.8037 & 0.1024 \\
MC-CF (det)  & 0.0168 & 0.0641 & 0.6077 & 14.7558 & 4.6919 & 1.5404 & 4.2206 & 1.5404 & 4.2206           & 0.0452 \\
MC-CF (cons, det) & 0.0168 & 0.0700 & 0.6747 & 14.6505 & 4.6758 & 1.5158 & 4.1633 & 1.5158 & 4.1633           & 0.0048 \\
\hline
\multicolumn{11}{l}{Stochastic} \\
SIDM (1)                & 0.0076 & 0.1204 & 1.2403 & 13.1932 & 3.8252 & 1.5761 & 3.3855 & --     & --      & 0.0024 \\
SIDM (3)                & --     & --     & --     & 13.0665 & 3.7910 & 1.5618 & 3.3489 & --     & --      & --     \\
SIDM (6)                & --     & --     & --     & 12.9990 & 3.7747 & 1.5542 & 3.3304 & --     & --      & --     \\
SIDM (10)               & --     & --     & --     & 12.9554 & 3.7653 & 1.5493 & 3.3177 & --     & --      & --     \\
SIDM (15)               & --     & --     & --     & 12.9303 & 3.7572 & 1.5463 & 3.3098 & 1.5762 & 3.3871  & --     \\
BayesGMM (stoch, 1)     & \textbf{0.0044*} & 0.0508* & 0.5027* & 44.3739 & 10.6502 & 4.3215 & 10.0250 & -- & -- & \textbf{0.0000*} \\
BayesGMM (stoch, 3)     & --     & --     & --     & 20.9065 &  5.1812 & 2.2664 &  4.7910 & --     & --      & --     \\
BayesGMM (stoch, 6)     & --     & --     & --     & 13.3752 &  3.6300 & 1.6180 &  2.9311 & --     & --      & --     \\
BayesGMM (stoch, 10)    & --     & --     & --     &  9.5263 &  2.8240 & 1.2087 &  2.0194 & --     & --      & --     \\
BayesGMM (stoch, 15)    & --     & --     & --     & \textbf{7.9938*} & \textbf{2.4236*} & 1.0520 & \textbf{1.5258*} & 4.5166 & 10.6921 & -- \\
MC-CF (stoch, 1)        & 0.0055 & 0.0782 & 0.7726 & 14.3716 & 4.5351 & 1.5182 & 4.1102 & --     & --      & 0.0429 \\
MC-CF (stoch, 3)        & --     & --     & --     & 11.7876 & 3.7566 & 1.2793 & 3.3591 & --     & --      & --     \\
MC-CF (stoch, 6)        & --     & --     & --     & 10.2642 & 3.3125 & 1.1434 & 2.9338 & --     & --      & --     \\
MC-CF (stoch, 10)       & --     & --     & --     &  9.4995 & 3.0527 & 1.0762 & 2.6933 & --     & --      & --     \\
MC-CF (stoch, 15)       & --     & --     & --     &  8.9985 & 2.9191 & \textbf{1.0355*} & 2.5433 & 1.5484 & 4.1770 & -- \\
MC-CF (cons, stoch, 1)  & 0.0055 & 0.0791 & 0.7711 & 14.3946 & 4.6250 & 1.5084 & 4.1373 & --     & --      & 0.0071 \\
MC-CF (cons, stoch, 3)  & --     & --     & --     & 11.9166 & 3.7924 & 1.2695 & 3.4071 & --     & --      & --     \\
MC-CF (cons, stoch, 6)  & --     & --     & --     & 10.8728 & 3.4941 & 1.1745 & 3.1003 & --     & --      & --     \\
MC-CF (cons, stoch, 10) & --     & --     & --     &  9.8996 & 3.2271 & 1.0904 & 2.8228 & --     & --      & --     \\
MC-CF (cons, stoch, 15) & --     & --     & --     &  9.4915 & 3.0746 & 1.0509 & 2.6776 & 1.5107 & 4.1240  & --     \\
\hline
\multicolumn{11}{l}{\footnotesize \textit{Note}: * indicates the minimum (best) value within each column. In case of ties, multiple entries are marked.}\\
\multicolumn{11}{l}{\footnotesize -- denotes not applicable. $OR$: $K=1$ only; $avgADE$/$avgFDE$: $K=15$ for stochastic models; equals $minADE$/$minFDE$ for deterministic. $n_\text{viable}/n_\text{total} = 361/420$.}
\end{tabular}
}
\end{table}

\begin{table}[tb!]
\centering
\caption{Trajectory prediction performance for HDV-following-AV.}
\vspace{0.2cm}
\label{tab:HDV_following_AV_results}
\resizebox{\textwidth}{!}{
\begin{tabular}{lcccccccccc}
\hline
Model & \multicolumn{3}{c}{One-Step Prediction} & \multicolumn{7}{c}{Open-Loop Prediction} \\
 & $RMSE(s)$ & $RMSE(v)$ & $RMSE(a)$ & $minDTW(s)$ & $minDTW(v)$ & $minADE$ & $minFDE$ & $avgADE$ & $avgFDE$ & $OR$\\
\hline
\multicolumn{11}{l}{Deterministic} \\
IDM          & 0.0633 & 0.1837 & 2.0834 & 17.1651 & 3.5458 & 1.9593 & 4.0849 & 1.9593 & 4.0849           & \textbf{0.0000*} \\
Van-Arem     & 0.0628 & 0.1021 & 1.0211 & 16.4752 & 3.6460 & 1.8491 & 4.1421 & 1.8491 & 4.1421           & 0.0092 \\
FVDM-CTH     & 0.0628 & 0.0902 & 0.8982 & 15.4723 & 3.4688 & 1.7221 & 3.9475 & 1.7221 & 3.9475           & 0.0147 \\
FVDM-Sigmoid & 0.0628 & 0.0908 & 0.9038 & 15.4419 & 3.4930 & 1.7100 & 3.9406 & \textbf{1.7100*} & \textbf{3.9406*} & 0.0147 \\
Gipps        & 0.0628 & 0.1047 & 1.0450 & 16.9242 & 3.5538 & 1.9329 & 3.9883 & 1.9329 & 3.9883           & \textbf{0.0000*} \\
FNN-reg      & 0.0627 & 0.0659 & 0.6384 & 17.6969 & 4.1774 & 1.8172 & 4.6256 & 1.8172 & 4.6256           & 0.0184 \\
GRU-reg      & 0.0562 & \textbf{0.0265*} & \textbf{0.1317*} & 18.8850 & 5.1467 & 1.8441 & 5.1512 & 1.8441 & 5.1512 & 0.0147 \\
MC-CF (det)  & 0.0627 & 0.0692 & 0.6729 & 18.9825 & 4.5627 & 1.9040 & 5.0448 & 1.9040 & 5.0448           & 0.0221 \\
MC-CF (cons, det) & 0.0627 & 0.0769 & 0.7538 & 19.0030 & 4.5588 & 1.8986 & 5.0124 & 1.8986 & 5.0124           & \textbf{0.0000*} \\
\hline
\multicolumn{11}{l}{Stochastic} \\
SIDM (1)                & 0.0251 & 0.1482 & 1.6103 & 17.3199 & 3.5538 & 1.9798 & 4.1386 & --     & --      & \textbf{0.0000*} \\
SIDM (3)                & --     & --     & --     & 16.4765 & 3.3613 & 1.8949 & 3.9074 & --     & --      & --     \\
SIDM (6)                & --     & --     & --     & 16.1261 & 3.2641 & 1.8599 & 3.8095 & --     & --      & --     \\
SIDM (10)               & --     & --     & --     & 15.9052 & 3.2109 & 1.8388 & 3.7447 & --     & --      & --     \\
SIDM (15)               & --     & --     & --     & 15.7335 & 3.1663 & 1.8229 & 3.6966 & 1.9669 & 4.0971  & --     \\
BayesGMM (stoch, 1)     & \textbf{0.0219*} & 0.0558* & 0.5381* & 58.9775 & 11.7765 & 5.4715 & 12.2298 & -- & -- & \textbf{0.0000*} \\
BayesGMM (stoch, 3)     & --     & --     & --     & 28.3009 &  5.8193 & 2.9480 &  5.6197 & --     & --      & --     \\
BayesGMM (stoch, 6)     & --     & --     & --     & 17.2961 &  3.7059 & 1.9822 &  3.2185 & --     & --      & --     \\
BayesGMM (stoch, 10)    & --     & --     & --     & 12.7078 &  3.0443 & 1.5555 &  2.2751 & --     & --      & --     \\
BayesGMM (stoch, 15)    & --     & --     & --     & \textbf{10.2967*} & \textbf{2.5158*} & 1.3191 & \textbf{1.7787*} & 5.6249 & 12.8313 & -- \\
MC-CF (stoch, 1)        & 0.0239 & 0.0857 & 0.8588 & 19.0244 & 4.7908 & 1.8988 & 5.0747 & --     & --      & 0.0239 \\
MC-CF (stoch, 3)        & --     & --     & --     & 14.8034 & 3.5267 & 1.5709 & 3.9541 & --     & --      & --     \\
MC-CF (stoch, 6)        & --     & --     & --     & 13.2546 & 3.1601 & 1.4349 & 3.4830 & --     & --      & --     \\
MC-CF (stoch, 10)       & --     & --     & --     & 12.3001 & 2.9695 & 1.3451 & 3.2100 & --     & --      & --     \\
MC-CF (stoch, 15)       & --     & --     & --     & 11.6066 & 2.8000 & \textbf{1.2833*} & 3.0170 & 1.9034 & 5.0187 & -- \\
MC-CF (cons, stoch, 1)  & 0.0237 & 0.0879 & 0.8775 & 19.3137 & 4.6430 & 1.9154 & 5.0342 & --     & --      & \textbf{0.0000*} \\
MC-CF (cons, stoch, 3)  & --     & --     & --     & 15.4450 & 3.5574 & 1.6197 & 3.9918 & --     & --      & --     \\
MC-CF (cons, stoch, 6)  & --     & --     & --     & 13.4492 & 3.1632 & 1.4560 & 3.4559 & --     & --      & --     \\
MC-CF (cons, stoch, 10) & --     & --     & --     & 12.5846 & 2.9933 & 1.3768 & 3.2057 & --     & --      & --     \\
MC-CF (cons, stoch, 15) & --     & --     & --     & 11.9009 & 2.8465 & 1.3163 & 3.0262 & 1.9052 & 4.9897  & --     \\
\hline
\multicolumn{11}{l}{\footnotesize \textit{Note}: * indicates the minimum (best) value within each column. In case of ties, multiple entries are marked.}\\
\multicolumn{11}{l}{\footnotesize -- denotes not applicable. $OR$: $K=1$ only; $avgADE$/$avgFDE$: $K=15$ for stochastic models; equals $minADE$/$minFDE$ for deterministic. $n_\text{viable}/n_\text{total} = 526/544$.}
\end{tabular}
}
\end{table}

\section{Sensitivity Analysis}\label{app:sensitivity}

This appendix assesses the sensitivity of MC-CF performance to two key design choices: the minimum cluster sample threshold ($N_{\min}$) and the conservative inference TTC hyperparameters. All evaluations use MC-CF (cons, stoch) on the HDV-following-HDV test set with $K=15$.

\subsection*{Sensitivity to Minimum Cluster Sample Threshold ($N_{\min}$)}

The parameter $N_{\min}$ controls the minimum number of observations required per cluster before merging; it governs the resolution of the state space, with smaller values producing finer-grained clusters. Table~\ref{tab:sensitivity_nmin} reports results for $N_{\min} \in \{5, 10, 20\}$, corresponding to 247,702, 123,235, and 62,418 clusters respectively.

\begin{table}[tb!]
\centering
\caption{Sensitivity of MC-CF (cons, stoch) open-loop performance to minimum cluster sample threshold $N_{\min}$ (default TTC configuration, $K=15$).}
\vspace{0.2cm}
\label{tab:sensitivity_nmin}
\resizebox{0.6\textwidth}{!}{
\begin{tabular}{lcccccccc}
\hline
$N_{\min}$ & $n_\text{clusters}$ & $minADE$ & $minFDE$ & $avgADE$ & $avgFDE$ & $OR$ & $n_\text{viable}/n_\text{total}$ \\
\hline
5  & 247,702 & 1.0169 & 2.2419 & 1.5578 & 4.0480 & 0.0052 & 3264/3276 \\
10 & 123,235 & 1.0350 & 2.2627 & 1.5660 & 4.0511 & 0.0037 & 3266/3276 \\
20 &  62,418 & 1.0417 & 2.2775 & 1.5681 & 4.0409 & 0.0034 & 3267/3276 \\
\hline
\multicolumn{8}{l}{\footnotesize $OR$: $K=1$ first run only. ADE/FDE computed over viable pairs ($n_\text{viable}$).}
\end{tabular}
}
\end{table}

Performance is highly stable across this range. MinADE varies by at most 0.025 (2.4\% relative to the default), and all other ADE/FDE metrics vary by less than 1\%. The OR decreases slightly with larger $N_{\min}$ as coarser clustering reduces the chance of sampling from a sparsely populated cluster region. These results confirm that model performance is not sensitive to the specific choice of $N_{\min}$ within a practically reasonable range.

\subsection*{Sensitivity to Conservative Inference Hyperparameters}

The conservative inference strategy introduces two TTC thresholds ($\text{TTC}_\text{danger}$ and $\text{TTC}_\text{caution}$) and two corresponding percentile cutoffs ($p_\text{danger}$ and $p_\text{caution}$). We evaluate three configurations: a \textbf{loose} configuration with lower TTC thresholds and broader percentile windows, the \textbf{default} configuration used throughout the paper, and a \textbf{strict} configuration with higher TTC thresholds and narrower percentile windows.

\begin{table}[tb!]
\centering
\caption{Sensitivity of MC-CF (cons, stoch) open-loop performance on the HDV-following-HDV test set to conservative inference hyperparameters ($K=15$).}
\vspace{0.2cm}
\label{tab:sensitivity}
\resizebox{\textwidth}{!}{
\begin{tabular}{lcccccccccc}
\hline
Config & $\text{TTC}_\text{danger}$ (s) & $\text{TTC}_\text{caution}$ (s) & $p_\text{danger}$ (\%) & $p_\text{caution}$ (\%) & $minADE$ & $minFDE$ & $avgADE$ & $avgFDE$ & $OR$ & $n_\text{viable}/n_\text{total}$ \\
\hline
Loose   & 2 &  5 & 10 & 40 & 1.0128 & 2.2897 & 1.5645 & 4.0802 & 0.0058 & 3263/3276 \\
Default & 3 & 10 &  5 & 30 & 1.0350 & 2.2627 & 1.5660 & 4.0511 & 0.0037 & 3266/3276 \\
Strict  & 5 & 15 &  5 & 20 & 1.0996 & 2.3650 & 1.6239 & 4.1442 & 0.0037 & 3266/3276 \\
\hline
\multicolumn{11}{l}{\footnotesize $p_\text{danger}$: percentile cap when TTC $<$ TTC$_\text{danger}$; $p_\text{caution}$: cap when TTC$_\text{danger}$ $\leq$ TTC $<$ TTC$_\text{caution}$.} \\
\multicolumn{11}{l}{\footnotesize $OR$: $K=1$ first run only. ADE/FDE computed over viable pairs ($n_\text{viable}$).}
\end{tabular}
}
\end{table}

The results in Table~\ref{tab:sensitivity} reveal a modest and monotone sensitivity to the conservative inference threshold. Moving from the loose to the default configuration reduces the overlapping rate from 0.58\% to 0.37\% while incurring only a marginal change in trajectory error (minADE: $+0.022$, minFDE: $-0.027$). Tightening further to the strict configuration offers no additional safety benefit (OR remains at 0.37\%) but degrades trajectory accuracy more noticeably (minADE: $+0.065$, minFDE: $+0.102$), suggesting that over-restricting the acceleration distribution reduces behavioral realism without improving safety. The default configuration therefore represents a practical optimum: it achieves the safety benefit of the conservative mode with minimal impact on open-loop trajectory quality. Across both sweeps, the maximum variation in any open-loop metric relative to the default configuration is less than 7\%, confirming that the overall conclusions of this paper are robust to reasonable choices of these hyperparameters.

\section{Jerk Analysis}\label{app:jerk}

A natural concern with the discrete bin-sampling mechanism of the MC-CF model is whether it produces jerky acceleration profiles. Jerk, defined as the rate of change of acceleration ($da/dt$), is a key indicator of ride comfort and behavioral smoothness. Because MC-CF samples acceleration independently at each 0.1-second time step without explicit temporal correlation between consecutive samples, abrupt changes in acceleration are structurally possible. We compute jerk root mean square (jerkRMS) over all one-step prediction transitions in the HDV-following-HDV test set, defined as:
\begin{equation}
\text{jerkRMS} = \sqrt{ \frac{1}{T-1} \sum_{t=1}^{T-1} \left( \frac{a_{t+1} - a_t}{\Delta t} \right)^2 },
\end{equation}
where $\Delta t = 0.1$ s.

\begin{table}[tb!]
\centering
\caption{Jerk root mean square (jerkRMS, $m/s^3$) for all models on the HDV-following-HDV test set ($K=1$).}
\vspace{0.2cm}
\label{tab:jerk}
\resizebox{0.45\textwidth}{!}{
\begin{tabular}{lcc}
\hline
Model & jerkRMS (mean) & jerkRMS (std) \\
\hline
\multicolumn{3}{l}{Deterministic} \\
IDM               & 0.8291 & 1.0812 \\
Van-Arem          & 0.4102 & 0.1940 \\
FVDM-CTH          & 0.3662 & 0.2021 \\
FVDM-Sigmoid      & 0.3643 & 0.2031 \\
Gipps             & 0.4750 & 0.2235 \\
FNN-reg           & 0.3316 & 0.1873 \\
GRU-reg           & 0.4296 & 0.3374 \\
MC-CF (det)       & 2.1508 & 0.8697 \\
MC-CF (cons, det)      & 2.3770 & 1.1105 \\
\hline
\multicolumn{3}{l}{Stochastic ($K=1$)} \\
SIDM              & 2.1224 & 1.0024 \\
BayesGMM          & 25.4478 & 21.1850 \\
MC-CF (stoch)     & 7.6637 & 2.6503 \\
MC-CF (cons,stoch) & 7.2973 & 2.4609 \\
\hline
\end{tabular}
}
\end{table}

Table~\ref{tab:jerk} reveals a clear divide between model classes. Physics-based models (IDM, Van Arem, FVDM variants, Gipps) and neural network baselines (FNN-reg, GRU-reg) produce low jerkRMS values of 0.33--0.83 $m/s^3$, reflecting smooth analytical acceleration functions or temporally correlated hidden state representations. The deterministic MC-CF variants exhibit substantially higher jerk (2.15--2.38 $m/s^3$), attributable to discontinuous transitions between cluster centroids whose mean accelerations can differ markedly. MC-CF (stoch) shows further elevation (7.66 $m/s^3$) because both the cluster transition and the acceleration draw within the cluster are independently stochastic at every step. It should be noted that BayesGMM exhibits by far the highest jerk (25.45 $m/s^3$), confirming that high jerk is a general characteristic of nonparametric samplers that independently draw accelerations at each time step, not a unique failure mode of the MC-CF architecture. The conservative inference strategy provides a modest jerk reduction for MC-CF (cons, stoch) (7.30 vs.\ 7.66 $m/s^3$) by restricting the tail of the acceleration distribution in near-collision states.

The elevated jerk of MC-CF models reflects a design trade-off inherent to time-step-level empirical sampling. Each step is independent, which maximizes behavioral diversity and enables faithful reproduction of the transition probability structure, but at the cost of temporal smoothness. Addressing this trade-off is a natural direction for future work. Candidate approaches include (i) temporal correlation in the sampling process (e.g., sampling acceleration increments rather than absolute values), (ii) post-processing trajectory smoothing filters, and (iii) finer state discretization as data volume increases, which would naturally reduce inter-cluster acceleration jumps.

\end{document}